\documentclass[preprint,prd,tightenlines,nofootinbib,eqsecnum]{revtex4-1}

\usepackage{amsmath}
\usepackage{amsthm}
\usepackage{amsfonts}
\usepackage{amssymb}
\usepackage{mathrsfs}
\usepackage{bm}
\usepackage{graphicx}
\usepackage{hyperref}
\usepackage{color}
\usepackage{array}
\usepackage{booktabs}
\usepackage{verbatim}
\usepackage{addfont}
\usepackage{enumerate}
\usepackage[modulo]{lineno}
\usepackage[scr=boondox,scrscaled=1.05]{mathalfa}

\newcolumntype{C}{>{$}c<{$}}
\AtBeginDocument{
\heavyrulewidth=.08em
\lightrulewidth=.05em
\cmidrulewidth=.03em
\belowrulesep=.65ex
\belowbottomsep=0pt
\aboverulesep=.4ex
\abovetopsep=0pt
\cmidrulesep=\doublerulesep
\cmidrulekern=.5em
\defaultaddspace=.5em
}

\definecolor{CiteColor}{rgb}{0,0.6,0.1}
\definecolor{URLColor}{rgb}{1,0,0.7}
\definecolor{LinksColor}{rgb}{1,0,0.1}

\hypersetup{colorlinks,linkcolor=LinksColor,citecolor=CiteColor,urlcolor=URLColor}

\newcommand{\beq}{\begin{equation}}
\newcommand{\eeq}{\end{equation}}
\newcommand{\ud}{\mathrm{d}}

\newcommand{\scL}{\mathscr{L}}
\newcommand{\scC}{\mathscr{C}}

\newcommand{\scP}{\mathscr{P}}
\newcommand{\scT}{\mathscr{T}}
\newcommand{\scS}{\mathscr{S}}
\newcommand{\scA}{\mathscr{A}}
\newcommand{\scO}{\mathscr{O}}

\newcommand{\RR}{\mathbb{R}}

\newtheorem*{theorem*}{Theorem}

\begin{document}

\setcounter{tocdepth}{1}

\title{ The Geometry of Isochrone Orbits  \texorpdfstring{\\}{}
        \normalfont{ \normalsize{\textit{from Archimedes' parabolae to Kepler's third law}}} \texorpdfstring{\\}{} \vspace{1cm}}

\author{Paul Ramond$^{\dag \ddag}$\footnote{Email: paul.ramond@obspm.fr (contact author)} and J\'er$\hat{\text{o}}$me Perez$^{\ddag}$\footnote{Email: jerome.perez@ensta-paris.fr}\vspace{0.5cm}}
\affiliation{$^{\dag}$Laboratoire Univers et TH\'eories\\Observatoire de Paris, PSL Research University, CNRS, Universit\'e Paris Diderot, Sorbonne Paris Cit\'e\\5 place Jules Janssen 92190 Meudon, France}
\author{}
\affiliation{$^{\ddag}$Laboratoire de Math\'ematiques Appliqu\'ees\\ENSTA Paris, Institut Polytechnique de Paris\\828 Boulevard des Mar\'echaux 91120 Palaiseau, France \vspace{2cm}}

\begin{abstract}
The Kepler potential $\propto{-1/r}$ and the Harmonic potential $\propto r^2$ share the following remarkable property: In either of these potentials, a bound test particle orbits with a radial period that is independent of its angular momentum. For this reason, the Kepler and harmonic potentials are called \textit{isochrone}. In this paper, we solve the following general problem: Are there any other isochrone potentials, and if so, what kind of orbits do they contain? To answer these questions, we adopt a geometrical point of view initiated by Michel H\'enon in 1959, in order to explore and classify exhaustively the set of isochrone potentials and isochrone orbits. In particular, we provide a geometrical generalization of Kepler's third law, and give a similar law for the apsidal angle, of any isochrone orbit. We also relate the set of isochrone orbits to the set of parabolae in the plane under linear transformations, and use this to derive an analytical parameterization of any isochrone orbit. Along the way we compare our results to known ones, pinpoint some interesting details of this mathematical physics problem, and argue that our geometrical methods can be exported to more generic orbits in potential theory. \\

\vspace{1cm}

\textit{Keywords : Classical gravity; Potential theory; Orbital mechanics; Isochrony; Kepler's laws }
\end{abstract}

\date{\today}

\pacs{}

\maketitle

\clearpage


\clearpage


\section*{Introduction}

The concept of \textit{isochrony} in physics can be traced back to Galileo's pendulum and his discovery of isochrone oscillations: The period of (small) oscillations of a simple pendulum is independent of its initial conditions. In modern classical mechanics, the one-dimensional motion of an oscillator or small oscillations of a pendulum are characterized by harmonic potentials  $\psi(q)=\omega^2 q^2/2$ where $\omega$ is the common pulsation to all orbits and $q$ represents the varying amplitude of the oscillation through time. In general or theoretical physics the concept of isochrony is crystallized around this fundamental potential: This notion is reducted to potentials $V$ which offer constant periods for all solutions of the ordinary differential equation $\ddot{q}+\partial_q V(q)=0$ (see \cite{Sfec.15} and reference therein). Applications for such problems range from scalar field cosmologies \cite{HaLi.02} to quantum mechanics \cite{Do.05}; in the former the isochronous property is often associed to regularly spaced discrete spectra. In all cases isochrony is appreciated for providing exact models and explicit analytical formulae. \\

In this paper we are interested in a more general paradigm which include the historical and classical previous one. It comes from gravitational potential theory applied to astrophysics. It was coined by Michel H\'enon in 1959 to qualify a gravitational potential meant to describe globular clusters. As the core of such spherical clusters of stars is roughly homogeneous, their mean field potential is harmonic at small radial distances $r\ll 1$. By opposition, stars confined to the outer parts only feel a Kepler potential $\psi(r)=-\mu/r$ associated with a point mass distribution, seeing the cluster from far away $r\gg 1$. In these two celebrated potentials, bound test particles orbit along ellipses; and their associated orbital period exhibit the striking feature of being independent of the angular momentum of the particle. Michel H\'enon then proposed looking for a general potential characterized by this property, in order to describe globular clusters as a whole. \\

In his seminal paper \cite{HeI.59} (in French, for an English version see \cite{Bi.14}) he succeeded in solving this ambitious problem and found what he called \textit{isochrone} potential: $\psi(r)=-\mu/s$, where $s:=b^2+\sqrt{b^2+r^2}$ and $b$ is a size parameter closely related to the half-mass radius of the system. While having the requested dynamical properties, the corresponding mass density distribution, obtained by solving the Poisson equation, was in good agreement with some of the observed globular clusters available in 1959. Although the recent refinement of observations has actually revealed a wider diversity, H\'enon's isochrone model remains at the center of cluster modeling for at least two reasons. As the harmonic and Kepler potentials, this potential is fully integrable and its action-angle formalism provides a fundamental basis for both the modeling and simulations of stellar systems (see e.g., \cite{McGBi.90}). More recently, a detailed numerical analysis \cite{SPDnum} showed that the isochrone model could be associated with the initial state of the evolution of singular stellar systems (e.g., globular clusters and/or Low Surface Brightness Galaxies); a result that followed an involved extension of many aspects of H\'enon's work on isochrone potentials \cite{SPD}. \\

The modern version of the isochrony proposed by \cite{SPD} extended many mathematical aspects of the work pioneered by H\'enon on isochrone potentials. In particular, other kinds of potentials with the isochrone property were found and classified using elements of group theory and Euclidean geometry. In the present paper, building on these results, we go a step further in two directions. On the one hand, we provide a fully geometrical treatment of the problem first posed by H\'enon, namely: Finding all isochrone potentials. We shall see that with a geometrical treatment, one family of potentials was left aside in \cite{SPD}. Therefore, we complete and exhaustively classify all isochrone potentials, based on their physical properties. On the other hand, we study in details the shape, properties and conditions of existence of isochrone orbits, i.e., bounded orbits in isochrone potentials. In particular we generalize Kepler’s third law to all isochrone orbits, providing a synthetic analytic formula for both the radial period and the apsidal angle. We also detail and fulfill a geometrical program that leads to an analytic parameterization of any isochrone orbit, completing the program started in \cite{SPD}. \\

This paper’s main content is the solution to a problem of mathematical physics: Finding the complete set of isochrone potentials and describing the isochrone orbits. It is remarkable that it can be solved analytically and that everything is expressible in terms of elementary functions. Furthermore, these solutions can be obtained using elementary Euclidean geometry. We stress that, physically speaking, the isochrone potentials with interesting properties are the Kepler, the harmonic and the H\'enon one, as was already found by H\'enon. All other potentials are necessary to get the complete picture of isochrony, but present somewhat unfamiliar physical properties that shall be discussed. They may nonetheless be of some interest as toy-models for astrophysics or electrodynamics, and also for academic purposes. Many of our results and geometrical methods are relevant to orbits in any central potential, as shall be pointed out in the text. Throughout the paper, we emphasize on the geometry of the problem, fill in some gaps that may be found in \cite{SPD} and pinpoint some interesting mathematical physics details. Computations that are not central to the results are left in the appendices, while the main text is organized in four main sections, as follows:\\

\textbullet \, In Sec.~\ref{sec:iso} we briefly mention well-known results about bounded orbits in central potentials, along with our notations and conventions (Sec.~\ref{sec:defs}). We define the notion of isochrony for potentials and the H\'enon variables (Sec.~\ref{sec:isopot}) that shall be used throughout the paper. \\

\textbullet \, The aim of Sec.~\ref{sec:geo} is twofold: First we derive an explicit formula for the radial period in an arbitrary isochrone potential in terms of geometrical quantities (Sec.~\ref{sec:Henonform}), and second, we use this formula to give a geometrical proof that isochrone potentials are parabolae in H\'enon's variable (Sec.~\ref{sec:isopara}). This proof is inspired by the findings of Archimedes. \\

\textbullet \, Based on these results, in Sec.~\ref{sec:sec3} we first sum up some generalities on parabolae (Sec.~\ref{sec:genpara}) in the plane. We then discuss the physical and mathematical properties of the associated potentials (Sec.~\ref{sec:portion}) and draw the bifurcation diagram that ensures the existence of periodic orbits, in terms of the energy and angular momentum of the test particle. This is necessary to give an exhaustive classification of isochrone potentials (Sec.~\ref{sec:complete}). \\

\textbullet \, This leads naturally to Sec.~\ref{sec:Kep} where our main and new results are stated. We provide a generalization of Kepler's third law (Sec.~\ref{sec:T&Theta}) for all isochrone orbits, both for the radial period and the apsidal angle. We discuss their geometrical meaning in various context. We then show how to geometrically derive an analytic parameterization of any orbit in any isochrone potential (Sec.~\ref{sec:param}). Lastly we depict some isochrone orbits, analyze their properties and classify them (Sec.~\ref{sec:classorbits}). \\

\clearpage
\section{Periodic orbits in central potentials} \label{sec:iso}
In this first section, the aim is to lay down the definitions and notations that shall be used in this paper. First, in Sec.~\ref{sec:defs}, we derive some standard results regarding periodic orbits of test particles in a given central potential. In Sec.~\ref{sec:isopot}, we define the qualifier \textit{isochrone} for a central potential, as well as the H\'enon variables that shall be used throughout the paper.
\subsection{Basic definitions} \label{sec:defs}
Let us consider the three-dimensional Euclidean space and an inertial frame of reference equipped with the usual spherical coordinates $(r,\theta,\varphi)$ and the associated natural basis $(\vec{e}_r,\vec{e}_\theta,\vec{e}_\varphi)$. We assume that around the origin $O=(0,0,0)$ lies a spherically symmetric distribution of matter with mass density $\rho(r)$. This system generates a gravitational potential, denoted $\psi(r)$, that obeys Poisson's equation
\beq \label{Poisson}
\Delta \psi (r) = \frac{1}{r^2} \frac{\ud (r^2 \psi^{\prime})}{\ud r} = 4 \pi G \rho (r) \, ,
\eeq
where a prime $^{\prime}$ denotes a differentiation with respect to $r$ and $G$ is the universal gravitational constant\footnote{In electrostatics, $G$ must be replaced by the Coulomb constant $-1/4\pi\varepsilon_0$, where $\varepsilon_0$ is the vacuum electric pe rmittivity. In this context, $\rho(r)$ is the charge density and is not required to be positive as in the gravitational case.}. We shall also use the usual dot $\dot{r}$ for the time derivative $\ud r / \ud t$. \\

Let us now consider a test particle of mass $m$ orbiting this system, with position vector $\vec{r}$ and velocity vector $\vec{v}:=\ud \vec{r} / \ud t$. From the spherical symmetry, the angular momentum $\vec{L}:=m \vec{r}\times\vec{v}$ of the particle is conserved. Its norm can be computed explicitly and is given by $|\vec{L}|=mr^2\dot{\theta}$, with the usual notation $\dot{\theta}=\ud \theta / \ud t$ for the time derivative. The total energy $E$ of the particle, sum of a kinetic term $m|\vec{v}|^2/2$ and a potential term $m\psi$, is conserved as well. Let us introduce $\xi := E/m$, the (total) energy of the particle per unit mass ; and $\Lambda := |\vec{L}|/m$, the (norm of the) angular momentum per unit mass. The explicit computation of the energy in terms of $r$ yields the following energy conservation equation
\beq \label{eomr}
\xi = \frac{1}{2} \biggl( \frac{\ud r}{\ud t} \biggr)^2 + \frac{\Lambda^2}{2r^2} + \psi(r) \, .
\eeq

Since $E, \vec{L}$ and $m$ are conserved quantities, $\xi$ and $\Lambda$ are two constants of motion for the particle. In a given potential $\psi$, the quantities $(\xi,\Lambda)$ are sufficient to know everything about the dynamics of a particle, up to initial conditions. Accordingly, we may abuse notation and speak of $(\xi,\Lambda)$ as \textit{a particle}. Along with some initial conditions, Eq.~\eqref{eomr} is a nonlinear ordinary differential equation for the function $t\mapsto r(t)$. We are interested in orbits and therefore will consider bounded solutions to Eq.~\eqref{eomr}. \\

Since the orbit is bounded and the function $r$ continuous, we may define $r_P$ and $r_A$ as the minimum and maximum values of $r(t)$. Accordingly, we will sometimes use the notation $[r_P,r_A]$ for an orbit since, physically, $r_P$ is (the radius of) the \textit{periapsis}, i.e., the point along the orbit closest to $O$ ; and $r_A$ that of the \textit{apoapsis}, the one farthest from $O$. At these turning points, the radial velocity $\dot{r}\vec{e}_r$ vanishes and $\dot{r}$ changes sign. Consequently, by Eq.~\eqref{eomr}, $r_P$ and $r_A$ are two solutions to the following algebraic equation\footnote{When there are more than two solutions to Eq.~\eqref{rarp}., say $(r_1,\ldots,r_n)$ for some $n \geq 3$, the orbit is selected on the graph by the initial radius $r_0$: its periapsis and apoapsis being $r_P=r_i$ and $r_A=r_{i+1}$, where $i$ is such that $r_0\in[r_i,r_{i+1}]$.}
\beq \label{rarp}
\xi = \frac{\Lambda^2}{2r^2} + \psi(r) =: \psi_e(r) \, ,
\eeq
where we have introduced the \textit{effective potential} $\psi_e(r)$, sum of the potential $\psi(r)$ and the centrifugal term $\Lambda^2/2r^2$. Note that when there is unique solution $r_C$ to Eq.~\eqref{rarp}, the associated orbit is circular, of radius $r=r_C$. This can always be seen as the degenerate case $r_P\rightarrow r_A$. \\

\begin{figure}[!htbp]
	\includegraphics[width=0.8\linewidth]{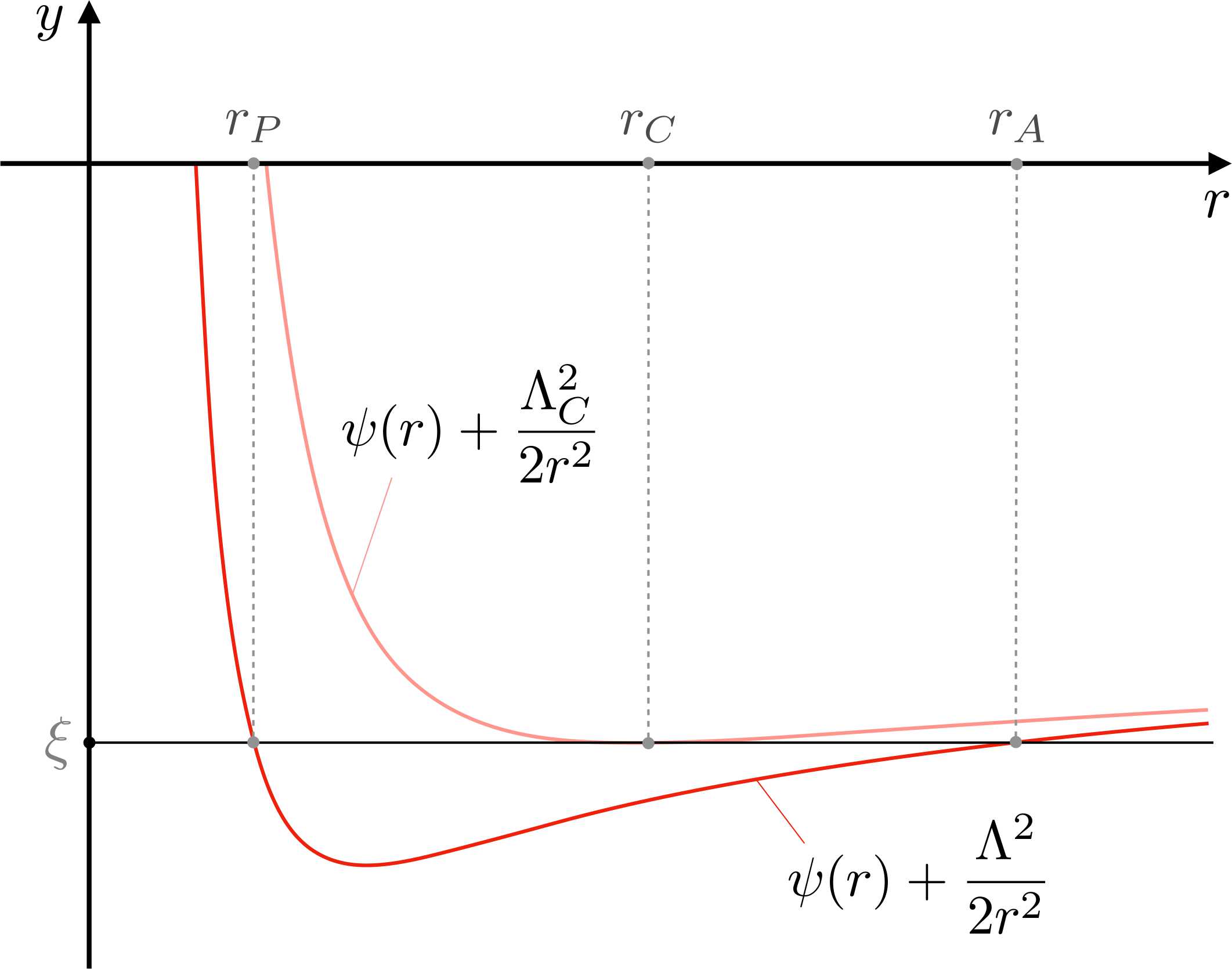}
	\caption{The graph $y=\psi_e(r)$ corresponds to the effective potential $\psi(r)+\Lambda^2/2r^2$. Two $\psi_e$ are depicted, associated with two particles with different angular momenta : $\Lambda$ (bottom curve, red) and $\Lambda_C>\Lambda$ (top curve, light red). The vertical line $y=\xi$ defines two orbits associated with the same energy $\xi$. Particle $(\xi,\Lambda)$ is on a generic, non-circular orbit $[r_P,r_A]$ and particle $(\xi,\Lambda_C)$ is on a circular orbit of radius $r_C$. Note that they both orbit in the same potential $\psi$. \label{fig:newphie}}
\end{figure}

It is customary to use the effective potential to study geometrically the orbit of a particle, depending on its energy. As depicted in Fig.~\ref{fig:newphie}, one plots the function $\psi_e$ for a given value of $\Lambda$, and then draws a line of height $\xi$. By construction, any choice of initial conditions will yield $\xi\geq \min{\psi_e}$. When there are two intersections between the line $y=\xi$ and the curve $y=\psi_e(r)$, the orbit is non-circular and one can read the periapsis $r_P$ and apoapsis $r_A$ as the abscissae of the intersection points. When there is only one intersection, its abscissa is the orbital radius $r_C$ and the orbit is circular. Furthermore, at $r=r_C$ the tangent to the curve is the horizontal line $y=\xi$, and therefore, $\psi_e^{\prime}(r_C)=0$. Two exemples of this well-known construction are depicted in Fig.~\ref{fig:newphie} for two particles with same energy, but different angular momenta.
 \\

We stress that, by virtue of Eq.~\eqref{eomr}, the quantity $\xi-\psi_e(r)\propto \dot{r}^2$ should always be strictly positive when $r(t)\in]r_P,r_A[$, and vanish at $r_P$ and $r_A$, by definition. This remark is important for later, so we summarize it geometrically as
\beq \label{positivity}
	\text{On an orbit $[r_P,r_A]$} \, \, : \, \,
		\begin{cases}
			\, \, y=\xi \, \, \, \text{intersects} \, \, \, y=\psi_e(r) \, \, \, \text{at} \, \, \, r=r_P, r_A \, , \\
			\, \, y=\xi \, \, \, \text{is above} \, \, \, y=\psi_e(r) \, \, \, \text{when} \, \, \, r\in ]r_P,r_A[ \, .
		\end{cases}
\eeq
\subsubsection{Radial Period}
It is a remarkable result of Hamiltonian dynamics that any bounded and continuous solution to Eq.~\eqref{eomr} must be periodic \cite{Arn}. In other words, if an orbit is bounded in a central potential, it is necessarily radially periodic. We shall denote by $T$ the radial period (\textit{the period} in short hereafter), i.e., the smallest $T\in\mathbb{R}^\star_+$ such that $r(t+T)=r(t)$ for all $t\geq0$. Note that $T$ always exists for bound orbits, and it should not be confused with the period of motion of the particle (i.e., the period of $t\mapsto \vec{r}(t)\in\mathbb{R}^3$), which only exists if the orbit is closed\footnote{For example, in a harmonic potential the period of motion is twice the radial period, and in a Kepler potential, both periods coincide.} in real-space (to be discussed below).   \\

For a generic, non-circular orbit\footnote{We do not define $T$ for circular orbits, since the radial motion for the latter is $r(t)=r_C$ where $r_C$ is a mere constant, therefore any $T\in\mathbb{R}$ is a radial period. }, one can get a formula for $T$ by first isolating the variables $t$ and $r$ in Eq.~\eqref{eomr}. This yields
\beq \label{dT}
\ud t = \pm \frac{\ud r}{\sqrt{2\xi - 2\psi(r) - \Lambda^2/{r^2}}} \, .
\eeq
In this formula, the $+$ sign corresponds to an increasing radius $r(t)$, i.e., when the particle goes from $r_P$ to $r_A$, whereas the $-$ sign corresponds to a decreasing radius, i.e., when the particle comes from $r_A$ back to $r_P$. Integrating Eq.~\eqref{dT} over a full period and taking into account the two different signs provides the following integral formula for the period
\beq \label{defT}
T := 2 \int_{r_P}^{r_A} \frac{\ud r}{\sqrt{2\xi - 2\psi(r) - \Lambda^2/{r^2}}} \, .
\eeq
Notice that the bounds of the integral $r_P$ and $r_A$ are precisely the values making the denominator vanish, by virtue of Eq.~\eqref{rarp}. The fact that $x\mapsto1/\sqrt{x}$ is integrable near $0$ ensures the convergence of the integral.\footnote{Indeed, using Eq.~\eqref{rarp}, we have the Taylor expansion $\xi-\psi_e(r)=\psi_e^\prime(r_P)(r_P-r)+o(r_P-r)$, and $\psi_e^\prime(r_P)\neq0$ since the orbit is non-circular. The integrand in Eq.~\eqref{defT} is thus equivalent to $(r_P-r)^{-1/2}$, which is integrable at $r_P$. The same holds at $r_A$.}.
\subsubsection{Apsidal angle}
\begin{figure}[!htbp]
	\includegraphics[width=1.0\linewidth]{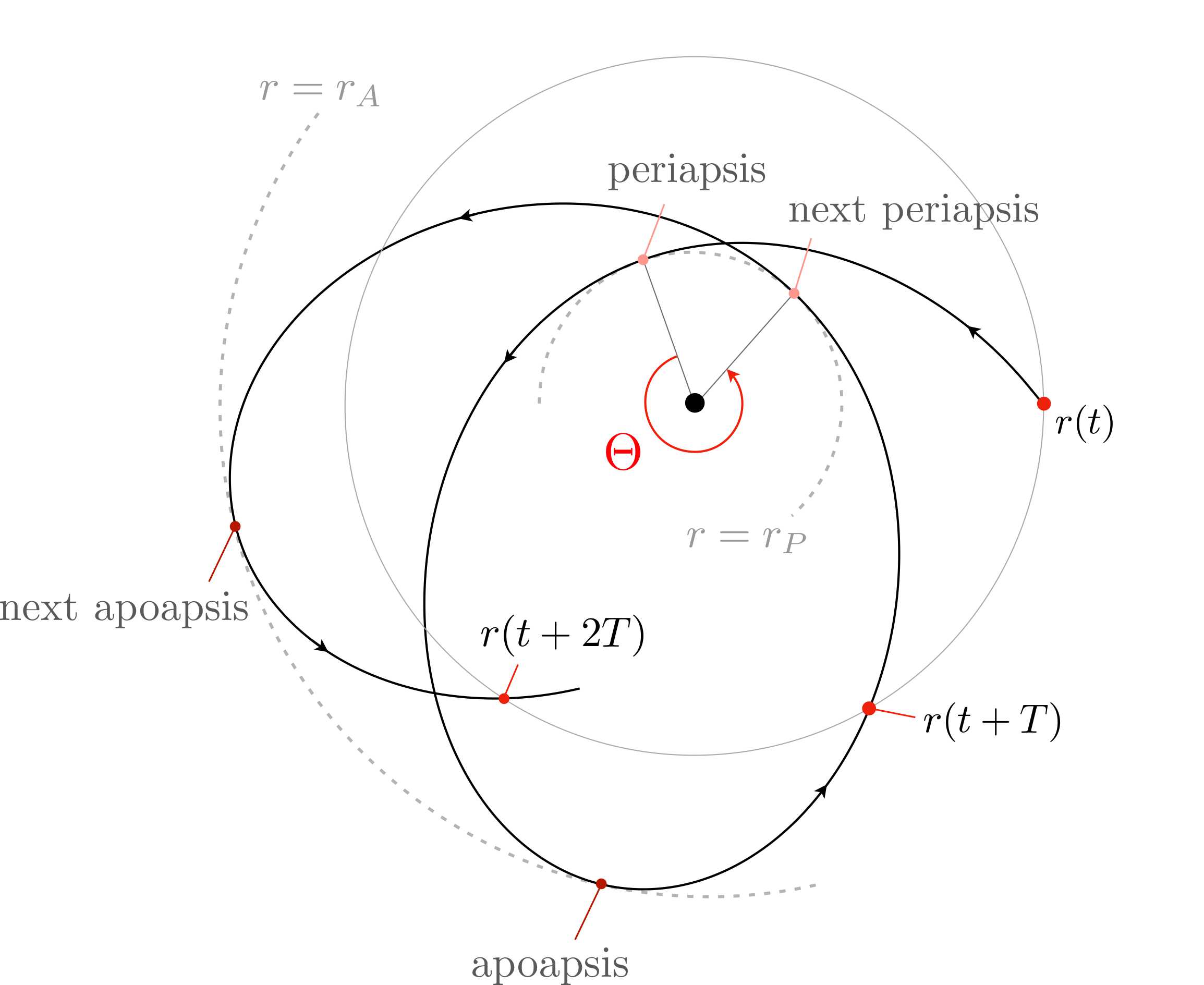}
	\caption{A typical orbit in a central potential (solid black), centered on the origin $O$, during $\sim 2$ periods $T$. At some initial time $t$, the particle is at a radius $r=r(t)$ (red dot on grey circle, right). At times $t+T$ and $t+2T$ it comes back to that same radius, crossing the grey circle with the same sign of $\dot{r}$. During the first period $[t,t+T]$, the particle reaches the periapsis (inner dashed circle) and then the apoapsis (outer dashed circle). During the second period $[t+T,t+2T]$, the process repeats. $\Theta$ is the angle between two successive periapsis, but also between any two successive positions a period $T$ appart. \label{fig:orbit}}
\end{figure}

Let a particle $(\xi,\Lambda)$ be at position $(r(t),\theta(t))$ on its orbit at some time $t$ (red point on the right of Fig.~\ref{fig:orbit}). The radial period $T$ corresponds to the time taken for the particle to go back to the radius $r(t)$ (with sign of $\dot{r}$). This does not mean, however, that the orbit itself is a closed curve in real space. It will be the case only if after a period $T$, the new angle $\theta(t+T)$ is equal to $\theta(t)+q\pi$, for some $q\in\mathbb{Q}$. The orbit then closes after a number of radial periods equal to the denominator of $q$. \\

To quantify this, let us define the quantity $\Theta:=\theta(t+T)-\theta(t)$. It is a constant angle along the orbit\footnote{Since $\Lambda=r^2\dot{\theta}$ we have $\dot{\Theta}=\Lambda/r(t+T)^2-\Lambda/r(t)^2$, which vanishes by $T$-periodicity of $r(t)$.} that corresponds physically to the angle difference between two positions, a period $T$ apart. In orbital mechanics it is customary to take the angle difference between two successive periapsis, as depicted in Fig.~\ref{fig:orbit}. Therefore, we shall call $\Theta>0$ the \textit{apsidal angle}. When $\Theta$ is a rational multiple of $\pi$, the orbit depicts a closed curve in real space. Otherwise, the orbit densely fills the shell region $r\in[r_P,r_A]$. \\

An integral formula can be obtained for $\Theta$, by using the conservation of angular momentum $\Lambda=r^2\dot{\theta}$. This equation gives $\ud \theta =\Lambda  \ud t / r^2$, which, when combined with Eq.~\eqref{dT} and integrated over one period, gives easily

\beq \label{defTheta}
\Theta := 2\Lambda \int_{r_P}^{r_A} \frac{\ud r}{r^2 \sqrt{2\xi - 2\psi(r) - \Lambda^2/{r^2}}} \, .
\eeq

Once again, we stress that Eq.~\eqref{defTheta} is valid for a generic, non-circular\footnote{As for the radial period $T$, any $\Theta\in\mathbb{R}$ is an apsidal angle for circular orbits, since $r(t)=r_C$ for all $t$ and, therefore, the particle is in some sense  always at periapsis.} orbit $[r_P,r_A]$, the convergence of the integral \eqref{defTheta} being justified by the same argument that was used for $T$ in Eq.~\eqref{defT}.
\subsubsection{Radial action}
The Hamiltonian formulation of a test particle orbiting in a central potential allows us to define the so-called \textit{radial action} $A(\xi,\Lambda) \propto \int_{r_P}^{r_A} \dot{r}(t) \ud r$, see e.g., \cite{Bitr}. Explicitly, using Eq.~\eqref{eomr}, this action is defined for any orbit $[r_P,r_A]$ and reads
\beq \label{radialaction}
A(\xi,\Lambda) := \frac{1}{\pi} \int_{r_P}^{r_A} \sqrt{2\xi - 2\psi(r) - \Lambda^2/{r^2}} \ud r \, .
\eeq
The radial action acts as a generating function for the radial period $T(\xi,\Lambda)$ and the apsidal angle $\Theta(\xi,\Lambda)$. Without going into too much detail, which can be found, e.g., in Sec.~6 of \cite{Bitr}, one can think of $T$ and $\Theta$ as the frequencies associated with the angle-action variables of the dynamics. In particular, we can set
\beq \label{defaction}
\frac{T}{2\pi} := \frac{\partial A}{\partial \xi} \quad \text{and} \quad \frac{\Theta}{2\pi} := - \frac{\partial A}{\partial \Lambda}  \, ,
\eeq
and this coincides with the definitions \eqref{defT} and \eqref{defTheta}, respectively. The definitions \eqref{defaction} give an important result that we shall keep in mind: \textit{$T$ depends only on $\xi$ if and only if $\Theta$ depends only on $\Lambda$}. This is immediate from Eq.~\eqref{defaction} since $\partial_{\Lambda}T \propto \partial_{\xi}\Theta$ by swapping the order of derivatives using Schwartz's theorem.
\subsection{Isochrony and H\'enon's variables} \label{sec:isopot}
\subsubsection{H\'enon's definition of isochrony }
For a generic central potential $\psi$, the radial period $T$ and the apsidal angle $\Theta$ are functions of both $\xi$ and $\Lambda$, as should be clear in view of Eqs.~\eqref{defT} and \eqref{defTheta}. In this paper, we are particularly interested in the class of \textit{isochrone} potentials. A central potential $\psi$ is called \textit{isochrone} if all periodic orbits it generates are such that $T$ is a function of the energy of the particle only, i.e., $T=T(\xi)$. In an isochrone potential, particles with the same energy share the same period, whence the name. As we have mentioned above (see Eq.~\eqref{defaction}), we thus have an alternative characterization of isochrony, namely that $\Theta$ depends on $\Lambda$ only (and not on $\xi$). In the end, one should keep in mind the following result
\beq
\psi\text{ is isochrone} \quad \Leftrightarrow \quad T=T(\xi,\Lambda\!\!\!/) \quad \Leftrightarrow \quad \Theta=\Theta(\xi\!\!\!/,\Lambda) \, .
\eeq

Isochrone potentials were introduced and studied first by Michel H\'enon in 1959. In a series of three papers \cite{HeI.59}, \cite{HeII.59} and \cite{HeIII.59}, he studied their physical properties, the orbits they generate and their application to astrophysics, respectively. Historically, the interest of H\'enon in the isochrone property was motivated during his study of globular clusters (a particular spherical collection of stars). He knew that these systems were isolated and had a homogeneous core. He also knew that a constant density profile is associated with a harmonic potential ($\psi\propto r^2$), and that outside any isolated spherical system, the potential is Keplerian ($\psi\propto-1/r$). Consequently, H\'enon wanted to find a potential that could interpolate these two. He noticed, quite remarkably, that one common feature of the Kepler and the harmonic potentials was isochrony, and this led him to try and find other isochrone potentials. After quite a remarkable analysis, he succeeded in finding a third potential with this property, nowadays commonly known as \textit{the}\footnote{In this paper as in \cite{SPD}, this one isochrone potential is reffered to as the H\'enon potential, \textit{isochrone} being a qualifier used here for the whole class of potentials such that $T=T(\xi,\Lambda\!\!\!/)$.} isochrone potential \cite{Bitr} that would help describe the density profiles of some stellar systems \cite{HeIII.59}. \\
\subsubsection{H\'enon's variables for central potentials}
The effective potential method, presented in Fig.~\ref{fig:newphie}, mixes the properties of the potential $\psi$ with that of the test particle $(\xi,\Lambda)$, as $\psi_e$ includes the centrifugal term $\Lambda^2/2r^2$. In particular, it is unpractical to draw and compare the orbits of two particles with different $(\xi,\Lambda)$ in a given potential $\psi$. In other words, a line $y=\xi$ crossing the curve $y=\psi_e(r)$ does not characterize a unique particle, as $\Lambda$ is encoded in $\psi_e$ and not in that line. We define in this section H\'enon's variables, which provide a way of working around this problem. \\

In his seminal paper on isochrony \cite{HeI.59}, Michel H\'enon introduced a change of variables in order to compute some complicated integrals. These variables have a much broader use that we shall exploit here. Instead of working with the physical radius $r$ and the physical potential $\psi(r)$, let us introduce the H\'enon variables $x$ and $Y(x)$ defined by
\beq \label{cov}
x := 2r^2 \quad \text{and} \quad Y:=2r^2\psi(r) \, .
\eeq

Since $Y(x)$ and $\psi(r)$ are in a one-to-one correspondence through Eq.~\eqref{cov}, the $x$ variable can still be thought of as a radius and $Y$ as a potential, and we shall sometimes abuse and speak of the radius $x$ and the potential $Y$, always referring to this duality. In terms of the H\'enon variables, the energy conservation \eqref{eomr} can be rewritten in the following evocative form
\beq \label{eomHenon}
\frac{1}{16} \biggl( \frac{\ud x}{\ud t} \biggr)^2 = (\xi x - \Lambda^2) - Y(x) \, .
\eeq
In view of the paragraph above Eq.~\eqref{rarp}, the apoapsis and periapsis $x_A:=2r_A^2$ and $x_P:=2r_P^2$ in H\'enon's variables are given by the intersection between the curve $\scC:y=Y(x)$ and the straight line $\scL:y=\xi x-\Lambda^2$, as can be read off the right-hand side of Eq.~\eqref{eomHenon}. Furthermore, it is clear that line $\scL$ should always lie \textit{above} curve $\scC$ since the left-hand side of Eq.~\eqref{eomHenon} is always positive. In what follows, $\scL$ will always denote a line of equation $y=\xi x-\Lambda^2$, associated with a particle $(\xi,\Lambda)$ ; and $\scC$ will always be the curve of equation $y=Y(x)$, associated with a potential $Y(x)=2r^2\psi(r)$. We note that, in terms of H\'enon's variables, the conservation of angular momentum reads

\beq \label{consmom}
\Lambda = \frac{x}{2} \, \frac{\ud \theta }{\ud t} \, .
\eeq

\begin{figure}[!htbp]
	\includegraphics[width=0.8\linewidth]{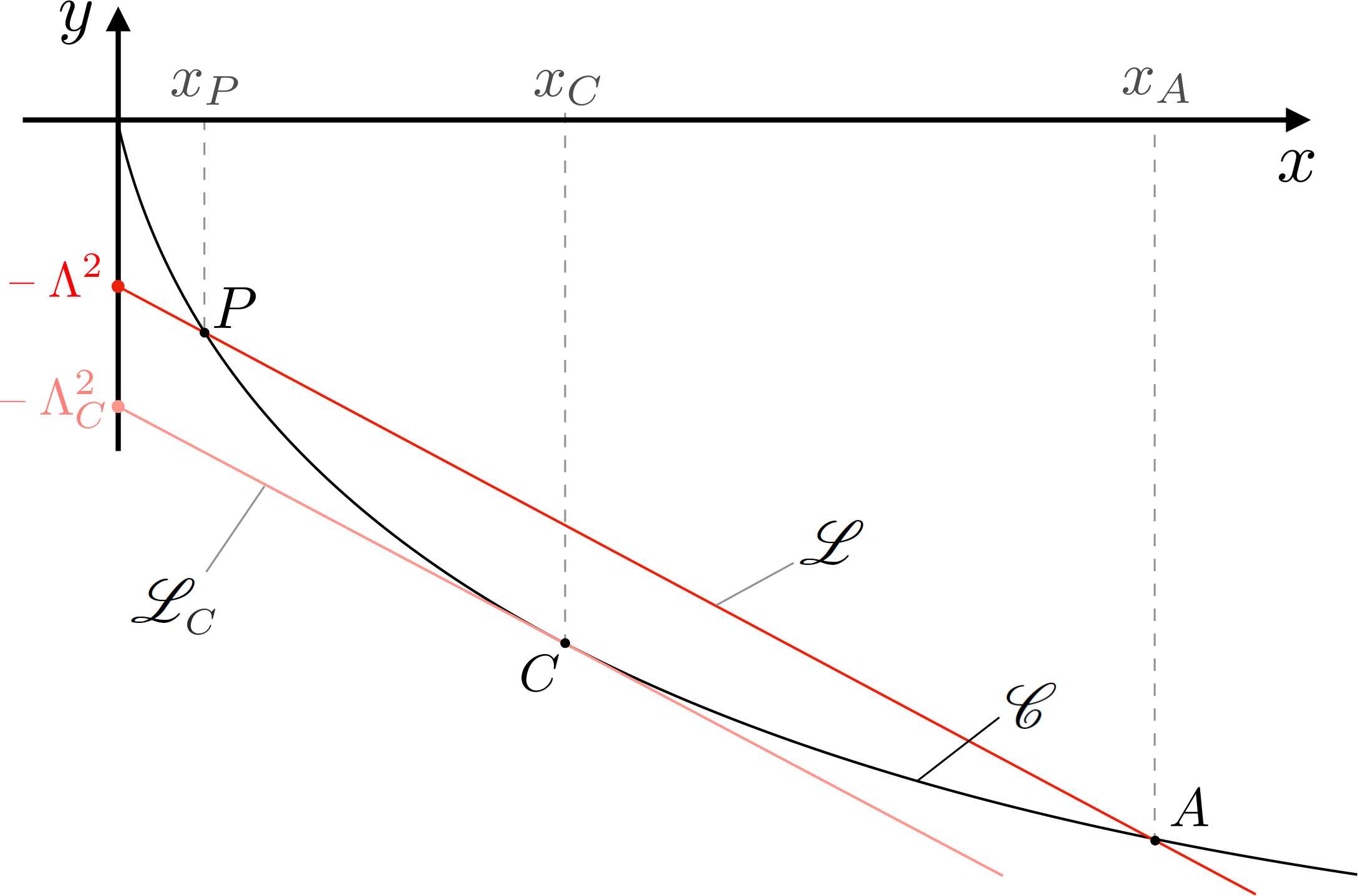}
	\caption{Same situation as in Fig.~\ref{fig:newphie} depicted here in the H\'enon plane, with H\'enon's variables. The curve $\scC$ of the graph $y=Y(x)$ corresponds to the potential $\psi$, in H\'enon's variables. Two particles are depicted as straight lines $\scL$ (red) and $\scL_C$ (light red). They have the same energy $\xi$ (same slope for both lines) but different angular momenta $\Lambda$ and $\Lambda_C>\Lambda$ (different $y$-intercept). Particle $(\xi,\Lambda)$ is on a generic orbit with periapsis $x_P$ and apoapsis $x_A$ given by the two intersections at $P$ and $A$. Particle $(\xi,\Lambda_C)$ is on a circular orbit of radius $x_C$ given the unique intersection at $C$. Notice that one can draw several different particles orbiting the potential without changing the curve $y=Y(x)$. \label{fig:newWhy}}
\end{figure}

H\'enon's variables $(x,Y(x))$ take advantage of the fact that a particle is entirely described by two numbers $(\xi,\Lambda)$, and is therefore in a one-to-one correspondence with a line, that has two degrees of freedom (e.g., the slope and the $y$-intercepts). The potential $Y(x)$ corresponds to a unique curve $\scC$, and a particle $(\xi,\Lambda)$ is associated with a unique straight line $\scL$. If $\scL$ intersects $\scC$ and lies above it, this particle orbits periodically the origin as detailed in Fig.~\ref{fig:newWhy}. The advantage is that for a given potential $Y(x)$, one can draw any particles and compare their orbital properties (which is not possible with the $(r,\psi_e)$ variables).
\clearpage
\section{A geometrical characterization of isochrony} \label{sec:geo}
This section aims at giving a geometrical characterization of isochrony. In Sec.~\ref{sec:Henonform} we derive the H\'enon formulae which give $T$ and $\Theta$ explicitly for isochrone potentials. In Sec.~\ref{sec:isopara}, we give a geometrical proof that the H\'enon formula for $T$ implies that the potential $Y$ must be an arc of parabola. We follow the notation introduced in the last section using H\'enon's variables: A particle $(\xi, \Lambda)$ is associated with a line $\scL:y=\xi x-\Lambda^2$, and $\scC:y=Y(x)$ is the curve of an arbitrary isochrone potential $Y(x)=2r^2\psi(r)$.
\subsection{H\'enon's formulae} \label{sec:Henonform}
This subsection is split into three parts. In the first, we integrate Eq.~\eqref{defT} explicitly, for any central potential, following a method of H\'enon \cite{HeI.59}. Assuming isochrony, we simplify in the second part this result to get the H\'enon formula for $T$. The third part presents more briefly this computation for $\Theta$.
\subsubsection{Computing the integral for $T$}
As we motivated below Eq.~\eqref{eomHenon}, we start by performing in Eq.~\eqref{defT} the change of variables $r\rightarrow x=2r^2$ and we introduce the potential $Y(x)=2r^2\psi(r)$. We readily obtain the following expression

\beq \label{defTHenon}
T = \frac{1}{2} \int_{x_P}^{x_A} \frac{\ud x}{\sqrt{D(x)}} \, , \quad \text{with} \quad D(x) := (\xi x - \Lambda^2) - Y(x) \, .
\eeq
The bounds of the integral are $x_P:=2r_P^2$ and $x_A:=2r_A^2\geq x_P$. In the $(x,y)$ plane, the quantity $D$ appearing in Eq.~\eqref{defTHenon} is the vertical distance between the curve $\scC$ and the line $\scL$. The fact that $D(x)\geq0$ is ensured by the very existence of the orbit, or equivalently by Eq.~\eqref{eomHenon}, as discussed in the last section. \\

Since the curve $\scC$ is smooth and lies below $\scL$ on $[x_P,x_A]$, there exists a line
$\scL_C$ that is both parallel to $\scL$ and tangent to $\scC$ at some point $C$ of abscissa $x_C \in [x_P,x_A]$.
This line intersects $\scC$ exactly once, and corresponds to a particle with a circular radius $r_C$, such that $x_C=2r_C^2$. Moreover, $\scL$ and $\scL_C$ are parallel and therefore associated with particles that share the same energy $\xi$. Consequently we may write $\scL_C:y=\xi x-\Lambda_C^2$, where $\Lambda_C$ is the angular momentum of the other particle, on the circular orbit. \\

\begin{figure}[!htbp]
	\includegraphics[width=0.6\linewidth]{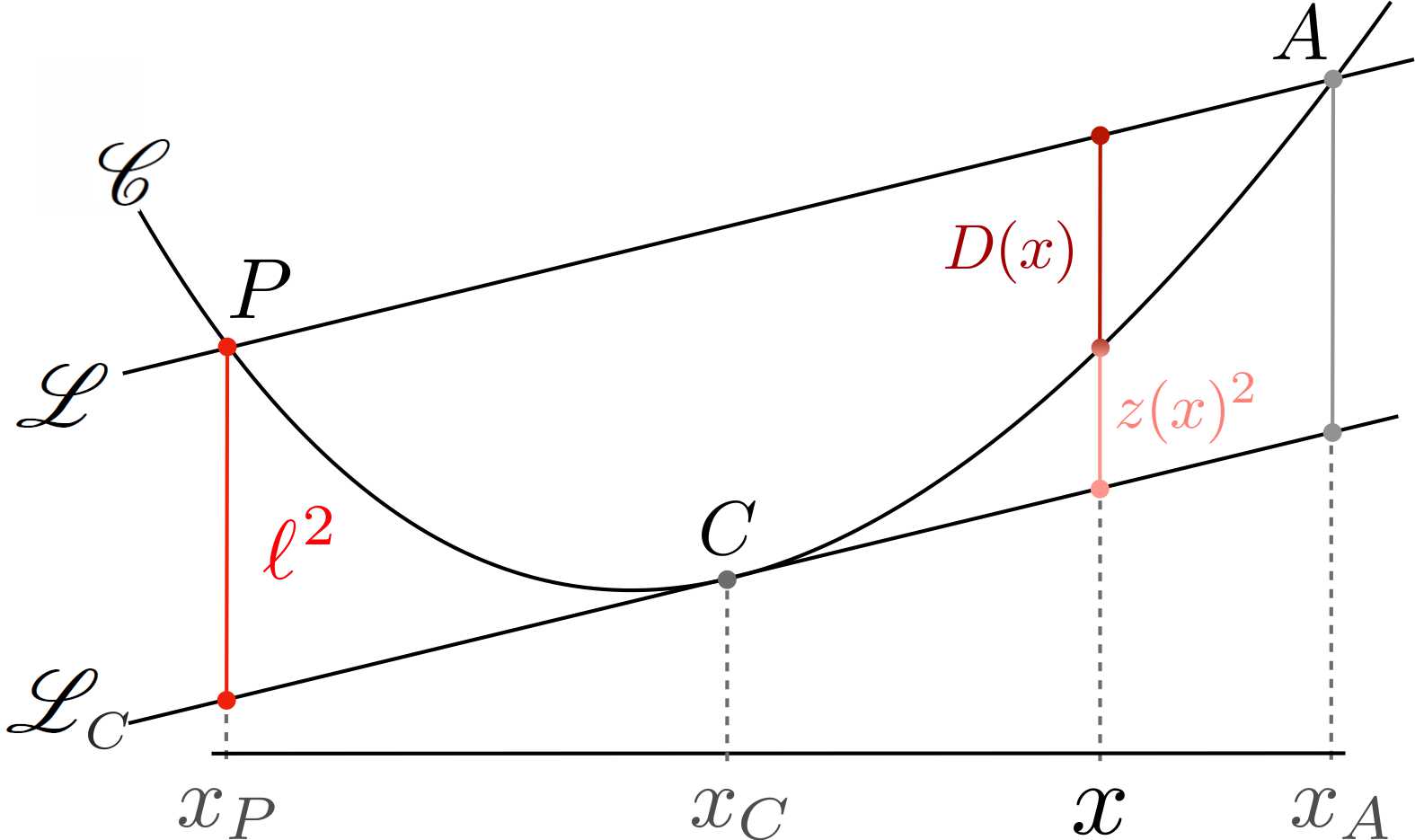}
	\caption{Summary of the geometrical quantities used to compute the integral for the period $T$. Both lines $\scL$ and $\scL_C$ define an orbit with the same period, and $\ell^2$ is the vertical distance between $\scL$ and $\scL_C$. The distance between $\scL_C$ and $\scC$ is $z(x)^2$, such that $D(x)+z(x)^2=\ell^2$.  \label{fig:geoiso_1}}
\end{figure}

With the help of this secondary line $\scL_C$, we may rewrite the distance $D$ of Eq.~\eqref{defTHenon} as the difference $\ell^2 - z(x)^2$, where $\ell^2:=\Lambda_C^2 - \Lambda^2 > 0$ is the vertical distance between $\scL$ and $\scL_C$, and $z(x)^2 := Y(x) - ( \xi x - \Lambda_C^2 ) >0 $ is the vertical distance between $\scC$ and $\scL_C$ (see Fig.~\ref{fig:geoiso_1}). We denote the latter by a squared quantity $z(x)^2$, so that we may conveniently choose $z(x)\leq0$ on $[x_P,x_C]$ and $z(x)\geq0$ on $[x_C,x_A]$. We stress that this is nothing but a convention: The positive distance is still $z(x)^2\geq0$, but the sign of $z(x)$ depends on where we are on $[x_P,x_A]$.
These new quantities are depicted in Fig.~\ref{fig:geoiso_1}, and upon insertion in Eq.~\eqref{defTHenon}, we obtain
\beq \label{Tsplit}
T = \frac{1}{2} \int_{x_P}^{x_A} \frac{\ud x}{\sqrt{\ell^2- z(x)^2}} \, ,
\eeq
Now, by construction, $z(x)$ varies monotonically on $[x_P,x_A]$: It is negative and increasing on $[x_P,x_C]$, it hits zero at $x_C$ and it is positive and increasing again on $[x_C,x_A]$. We can therefore perform the change of variables $x \rightarrow z(x)$ in Eq.~\eqref{Tsplit}. We readily obtain
\beq \label{TbefbefTaylor}
T = \frac{1}{2} \int_{-\ell}^{\ell} \frac{ f^{\prime}(z) \ud z}{\sqrt{\ell^2- z^2}} \, , \quad \text{with} \quad x := f(z) \, .
\eeq
It is now natural to perform in Eq.~\eqref{TbefbefTaylor} one last change of variables, namely $z \rightarrow \ell \sin \phi$, with $\phi$ varying between $-\pi/2$ and $\pi/2$, corresponding to $z=-\ell$ and $z=\ell$, respectively. We then get
\beq \label{TbefTaylor}
T = \frac{1}{2} \int_{-\pi/2}^{\pi/2} f'(\ell \sin \phi) \ud \phi  \, .
\eeq
We cannot, in general, compute explicitly the integral in Eq.~\eqref{TbefTaylor}, for $f'$ is but a generic, unspecified function that depends on the potential and the particle. However, assuming that the potential is regular enough, we can expand the function $f^{\prime}$ as a Taylor expansion at zero, i.e., write $f^{\prime}(z) = a_0 + \sum_{n\geq 1} a_n z^n$. Inserting this in Eq.~\eqref{TbefTaylor} and integrating term by term gives
\beq \label{Tsum}
T = \frac{\pi}{2}a_0 + \sum_{n \geq 1} a_{2n} W_{2n} \ell^{2n} \, , \quad \text{with} \quad W_n := \int_{0}^{\pi/2} \sin^{n} \! \phi \, \ud \phi \, .
\eeq
Note that only the even terms $a_{2n}$ remain since the integral of the odd function $\sin^{2n+1}$ vanishes over the symmetric interval $[-\pi/2,\pi/2]$. The integral $W_n$ is the celebrated Wallis integral, and can be given explicitly. Notice that Eq.~\eqref{Tsum} is valid for any potential and any particle orbiting within it. Both the coefficients $a_n$ and $\ell$ depend on $(\xi,\Lambda)$ and the properties of the potential. As such, it is not that useful. However, for isochrone potentials, it is of considerable interest.
\subsubsection{Explicit formula for $T$}
So far, what we have done does not take advantage of the isochrony property, and Eq.~\eqref{Tsum} is valid for any particle $(\xi,\Lambda)$ in any central potential $Y(x)$. In particular, we insist that the coefficients $a_n$ appearing in Eq.~\eqref{Tsum} are all function of $\xi$ and $\Lambda$, a priori. Now let us fix the energy $\xi$ of the particle. If the potential is isochrone, then by definition $T$ is independent of $\Lambda$, and so is the right-hand side of Eq.~\eqref{Tsum}. We may therefore choose $\Lambda=\Lambda_C$, i.e., $\ell=0$, so that the last term on the right-hand side of Eq.~\eqref{Tsum} vanishes, and we readily find that
\beq \label{T=pia0}
T(\xi)=\frac{\pi}{2} a_0 \, .
\eeq
Now Eq.~\eqref{T=pia0} is true for any $\Lambda$. (Actually, it is independent of $\Lambda$.) The combination of Eqs.~\eqref{Tsum} and \eqref{T=pia0} implies that the sum on the right-hand side of Eq.~\eqref{Tsum} is a power series in $\ell$ that vanishes for any $\ell$. By a classical result on power series, this is true if and only if all the coefficients of the power series vanish, i.e., $a_{2n}W_{2n}=0$ for all $n\geq1$. Since the Wallis integrals $W_n$ are all nonzero, we conclude that, if the potential is isochrone, $a_{2n}=0$ for all $n\geq1$. In particular, the Taylor expansion of $f'$ now reads
\beq \label{fprime}
f'(z) = a_0 + \sum_{n \geq 1} a_{2n+1} z^{2n+1} \, .
\eeq
The last step consists in finding explicitly the coefficient $a_0$ appearing in Eq.~\eqref{T=pia0}. To this end, we integrate Eq.~\eqref{fprime} over $[z(x_P),z(x_A)]=[-\ell,\ell]$. On the left-hand side, we use $f(z_A)=x_A$ and $f(z_P)=x_P$ (which follows from the definition $x=f(z)$). On the right-hand side, the first term is a mere constant, and the second term is an odd function of $z$: Its integral over $[-\ell,\ell]$ will vanish. Consequently, the integrated result is simply $x_A-x_P=2\ell a_0$. With the help of $\ell^2=\Lambda_C^2-\Lambda^2$ and Eq.~\eqref{T=pia0}, we obtain the following explicit formula for $T(\xi)$
\beq \label{HenonT}
T(\xi) = \frac{\pi}{4} \frac{x_A - x_P}{\sqrt{\Lambda_C^2-\Lambda^2}} \, .
\eeq \\

Let us make a few remarks on Eq.~\eqref{HenonT}. First, and quite remarkably, we stress that altough both $x_A-x_P$ and $\sqrt{\Lambda_C^2-\Lambda^2}$ depend explicitly on $\Lambda$, their ratio does not, since $T$ is independent of $\Lambda$ by assumption. \\

Second, if we square both sides of the equation, we observe that the horizontal distance $x_P-x_A$ squared is proportional to the vertical one $\Lambda_C^2-\Lambda^2$, and that the constant of proportionality, namely $16T^2/\pi^2$, is independent of $\Lambda$. We shall use this geometrical result to prove that the curve $\scC$ must be a parabola in the $(x,y)$ plane. \\

Third, we insist that this relation is valid for all isochrone potentials, even though their explicit form is unknown at this stage. In particular, given an isochrone potential, the radial period of any orbit can be read simply by drawing the line $\scL$ intersecting the curve $\scC$, and then finding the secondary line $\scL_C$ that is both parallel to $\scL$ and tangent to $\scC$. \\

Lastly, let us mention that formula \eqref{HenonT} for the period $T$ in not entirely new: It can be found as an intermediate equation in the seminal paper of H\'enon \cite{HeI.59}, (with a missing factor of 1/2 there). In fact, our method here is similar to his, altough more detailed. H\'enon did not seem to be interested in this particular equation, perhaps because his main goal was not to obtain a formula for the period. Yet, we shall see that this equation is rather central in the context of isochrony.
\subsubsection{Explicit formula of $\Theta$}
In the last paragraphs, we were able to obtain the explicit formula \eqref{HenonT} for $T(\xi)$. The recipe for the computation went in five steps that can be summarized as follows:
\begin{itemize}
	\item fix $\xi$ and rewrite the integrand in Eq.~\eqref{defT} as $1/\sqrt{D(x)}$ using H\'enon's variables,
	\item rewrite $D(x)$ as $\ell^2-z(x)$ using the line $\scL_C$ associated with the circular orbit of same energy (and thus same period),
	\item introduce $x=f(z)$, perform the change of variables $x\rightarrow z$ and then $z\rightarrow\phi$,
	\item perform a Taylor expansion of $f\prime$ around $0$ and integrate explicitly,
	\item assume that $\psi$ is isochrone and thus use $T=T(\xi,\Lambda\!\!\!/)$ to constrain $f$ and conclude.
\end{itemize}

Ultimately, the effectiveness of this recipe can be traced back to the H\'enon variable $x=2r^2$ which \textit{isolates} $\Lambda$ from the denominator of the integrand in Eq.~\eqref{defT}, as can be seen on Eq.~\eqref{defTHenon}. Knowing this, it is possible to try and adapt the recipe to find an explicit formula for $\Theta$, starting from its integral definition in Eq.~\eqref{defTheta}. As we argued earlier, examining the radial action \eqref{radialaction} shows that $T=T(\xi,\Lambda\!\!\!/)$ is equivalent to $\Theta=\Theta(\xi\!\!\!/,\Lambda)$. Therefore, one can apply the same recipe provided that one uses a variable that \textit{isolates} $\xi$ in the denominator in Eq.~\eqref{defTheta}. The Binet variable $u:=1/r$ turns out to be the appropriate variable this time. \\

More precisely, with the Binet variables $u=1/r$ and the Binet effective potential $\Psi_e(u):=\psi_e(1/u)$, it is possible to make a one-to-one dictionary between what was used for $T$ and what can be used for $\Theta$. The latter is presented in Table~\ref{TableIso}. The detailed computation is given in App.~\ref{app:Theta}. At the end of the computation, for any given $\Lambda$ we obtain the following formula in the case of isochrone potentials
\beq \label{HenonTheta}
\Theta(\Lambda) = \frac{\pi\Lambda}{\sqrt{2}} \frac{u_P-u_A}{\sqrt{\xi-\xi_C}} \, ,
\eeq
with $u_{P,A}:=1/r_{P,A}$. The value $\xi_C$ depends only on $\Lambda$ and is the energy to be given to a particle of angular momentum $\Lambda$ to obtain a circular orbit. Moreover, as we argued earlier for $T$ in Eq.~\eqref{HenonT} despite appearances the right-hand side of Eq.~\eqref{HenonTheta} is independent of $\xi$. \\

Just as Eq.~\eqref{HenonT} will be used in Sec.~\ref{sec:Kep} in order to write a generalized Kepler's third law for all isochrone orbits, Eq.~\eqref{HenonTheta} will be used to find a similar law for the apsidal angle of any isochrone orbit. We shall not use it directly and present a more astute computation, but it is possible to derive, without any trick, this periapsis law directly from Eq.~\eqref{HenonTheta}. \\

This apsidal angle law can, in turn, be used to give a proof of Bertrand's theorem, a well-known result of classical mechanics that states that the only two potentials in which all periodic orbits are closed are the Kepler and the harmonic potentials. In fact, since these two are also isochrone potentials, it should come as no surprise that Bertrand's theorem is closely related to isochrony. As demonstrated in \cite{SPD}, the theorem actually follows from the examination of Eq.~\eqref{HenonTheta}, once the latter is expressed in terms of $\Lambda$. Let us mention that the equivalent of Eq.~\eqref{Tsum} for $\Theta$ (Eq.~\eqref{Thetasum}) can be used to give a proof a Bertrand's theorem with brute force as in \cite{Sa.al.09} [compare Eq.~(20) of \cite{Sa.al.09} to Eq.~\eqref{Thetasum}].
\begin{table}[!htbp]
	\centering
			\begin{tabular}{|c|c|c|c|c|c|}
			\hline
			\, Isochrony $\Leftrightarrow$		\,			& \, Variable\,	& Potential  					  	 & \, Curve $\scC$	\,		& Line $\scL$   	      & Distance $D$	 	 	    	      \\ \hline
			$T=T(\xi,\Lambda\!\!\!/)$  					& $x=2r^2$	& \, $Y(x)=2r^2\psi(r)$  \,	 	 & $y = Y(x)$ 			& \, $y=\xi x -\Lambda^2$ \,	& $D(x)=\xi x -\Lambda^2 -Y(x)$ 	\\ \hline
			$\Theta=\Theta(\xi\!\!\!/,\Lambda)$ & $u=1/r$ 	& $\Psi_e(u)=\psi_e(r)$  & $y = \Psi_e(u)$	& $y=\xi$     				  & $D(u) = \xi - \Psi_e(u)$	\\ \hline
			\end{tabular}
		\caption{ Dictionary between the geometrical quantities involved in the derivation of the H\'enon formulae: \eqref{HenonT} for $T(\xi)$ and \eqref{HenonTheta} for $\Theta(\Lambda)$. \label{TableIso}}

\end{table}
\subsection{Geometry of parabolae} \label{sec:isopara}
In this section, we provide a geometrical proof that the curve $\scC:y=Y(x)$ must be a parabola\footnote{Strictly speaking, we will show that $\scC$ is \textit{an arc of parabola}, as it is the graph of a function. However any given arc of parabola defines a unique parabola, so there will be no possible confusion here.} 
in order for the associated potential $\psi$ to be isochrone. This result was first established by H\'enon in \cite{HeI.59}, using a very technical argument. It is also found in \cite{SPD} using techniques from complex analysis. In this paper, sticking to the geometrical approach, we present a new, geometrical proof of this remarkable fact, based only on Eq.~\eqref{HenonT} and a characterization of parabolae that can be traced back to Archimedes.
\subsubsection{Archimedean characterization} \label{sec:archicara}
Archimedes, in his treatise \textit{Quadrature of the Parabola}\footnote{We refer the interested reader to pp.233-252 of Heath \cite{Heath} for a modern English translation of this work, and to pp.51-62 of Stein \cite{Stein}, for a pedagogical version.}, proved in a series of 24 propositions the following remarkable property shared by all parabolae. On a given parabola $\scP$, take two points $A$ and $B$ defining a chord $AB$ and a third point $C$ where the tangent to $\scP$ is parallel to the chord $AB$. Then the area enclosed by $\scP$ and $AB$ is four thirds that of the triangle $ABC$. Although it was not known to Archimedes, it turns out that this property uniquely characterizes parabolae \cite{Ben.al.03}.
In other words, we have the following theorem : \\ 

\begin{theorem*} \label{thm:archi}
	Let $\scC$ be an arbitrary smooth curve in the plane, and $\scL$ any line that intersects $\scC$ exactly twice, say at points $P\!$ and $A$. Let $C$ be the point where the tangent to $\scC$ is parallel to $\scL$. Then, if $\scT$ denotes the triangle $P\!AC$ and $\scS$ the region enclosed by $\scL$ and $\scC$, the following equivalence holds
	\beq
	\text{Area} \, (\scT)=\frac{3}{4} \text{Area} \, (\scS) \quad \Leftrightarrow \quad \scC \text{ is an arc of parabola} \, .
	\eeq
\end{theorem*}

Notice that in order for $\scC$ to be a parabola, the area ratio should be $4/3$ for \textit{all of its chords}. As stated above, the $\Leftarrow$ result was the aim of Archimedes' work.
\subsubsection{Rewriting of H\'enon's formula}
Consider the curve $\scC:y=Y(x)$ associated with an isochrone potential $Y(x)=2r^2\psi(r)$. Following the notation used so far, let us take a line $\scL:y=\xi x - \Lambda^2$, such that $P\!$ and $A$ correspond to the periapsis and apoapsis of the orbit of particle $(\xi,\Lambda)$. Accordingly, the parallel line that passes through $C$ is $\scL_C:y=\xi x-\Lambda^2_C$, and defines a circular orbit with the same energy $\xi$, and thus the same period $T(\xi)$, given by Eq.~\eqref{HenonT}. If for any line $\scL$ the areas involved in the theorem are in proportion $4/3$, we will have shown that $\scC$ is a parabola. Therefore, the goal is to find an expression for these areas, using Eq.~\eqref{HenonT}. \\

\begin{figure}[!htbp]
	\includegraphics[width=0.7\linewidth]{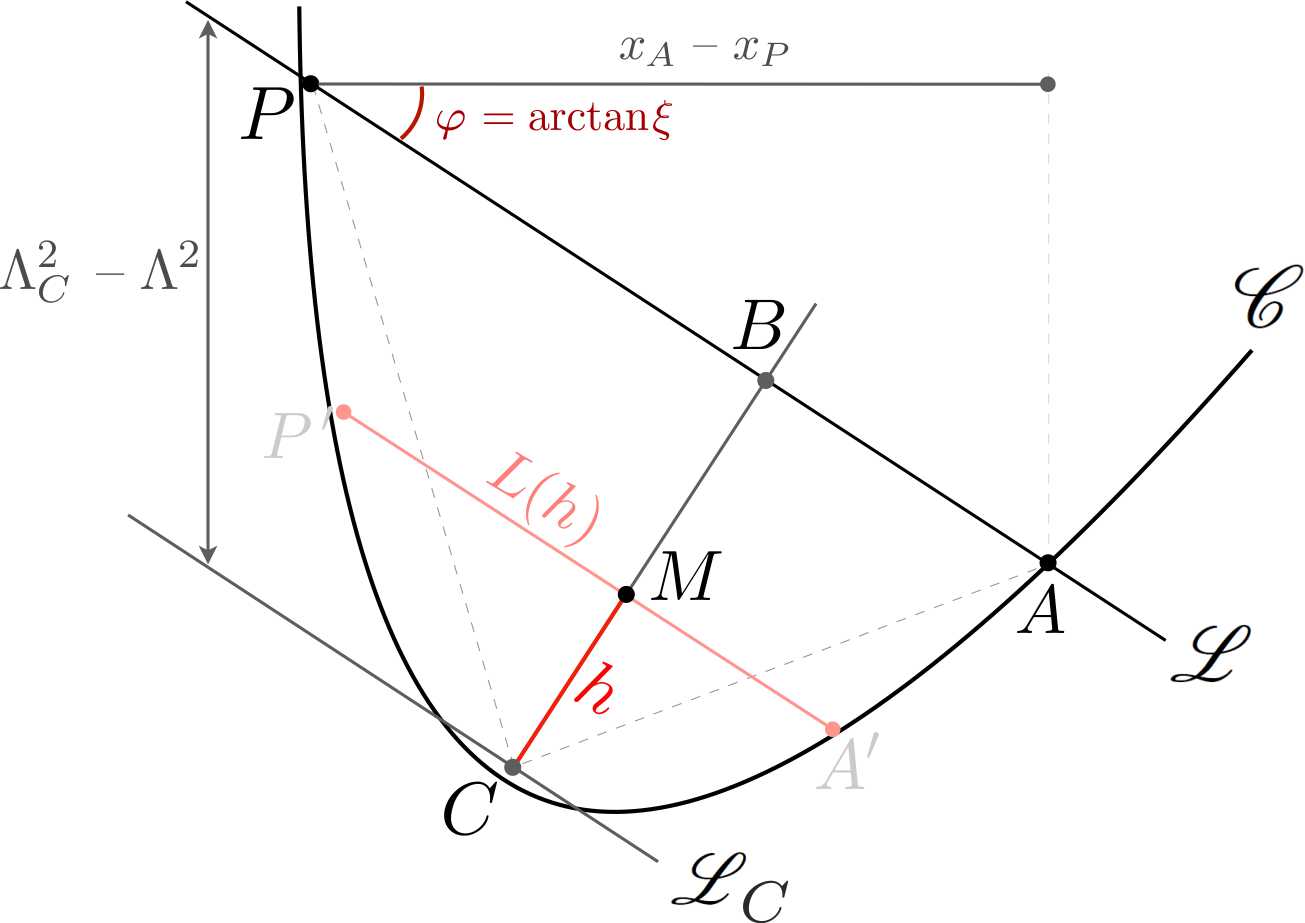}
	\caption{Initially, an arbitrary curve $\scC$ and an intersecting line $\scL$ of slope $\xi$ are drawn. They define the intersection points $P$ and $A$. The horizontal distance between $P$ and $A$ is $x_A-x_P$ (top). The line $\scL_C$, parallel to $\scL$ and tangent to $\scC$ at $C$, defines a circular orbit with energy $\xi$. The vertical distance between $\scL$ and $\scL_C$ is $\Lambda_C^2-\Lambda^2$ (left). The intermediary chord $P^\prime A^\prime$ defined in the text is parallel to $\scL$ and defines yet another orbit with energy $\xi$. \label{fig:paraarchi}}
\end{figure}

First let us do a bit of geometry. We define $B$ to be the orthogonal projection of $C$ on $\scL$, and take $M$ to be an arbitrary point on $CB$. We parameterize the length $CM$ by $h\geq0$, with the convention $h=0$ when $M=C$, and $h=CB$ when $M=B$. Next we define a chord $P^{\prime}A^{\prime}$ that is parallel to $\scL$ and passes through $M$. We denote by $L(h)$ the length of that chord $A^{\prime}$ and $P^{\prime}$. Note that $L(h)$ varies between $0$ (when $h=0$) and $PA$ (when $h=CB$). All these quantities are depicted in Fig.~\ref{fig:paraarchi}. \\

Now let us rewrite Eq.~\eqref{HenonT} in terms of these geometrical quantities. For the numerator, $x_A-x_P$ is but the horizontal projection of $P\!A$, and thus $x_A-x_P=P\!A\cos\varphi$, where $\varphi$ is the angle that $\scL$ makes with the horizontal axis, i.e.,. $\varphi=\arctan\xi$. Similarly, for the denominator, $\Lambda^2_C-\Lambda^2$ is simply the vertical projection of $CB$; consequently, we also have $\Lambda^2_C-\Lambda^2=CB/\cos\varphi$. Inserting these two identities in Eq.~\eqref{HenonT} gives its geometrical variant
\beq \label{newT}
T(\xi)=\frac{\pi}{4(1+\xi^2)^{3/4}}\frac{P\!A}{\sqrt{CB}}\, ,
\eeq
where we used the trigonometric identity $\cos(\arctan\xi)=(1+\xi^2)^{-1/2}$. Now, formula \eqref{newT} has been obtained for any chord $P\!A$ of the curve $\scC$, corresponding to a particle of energy $\xi$. However, by construction, for any $h$ the chord $P^{\prime}A^{\prime}$ is parallel to $P\!A$ and thus corresponds to an orbit with the same energy $\xi$. Therefore, the potential being isochrone, all parallel chords $P^{\prime}A^{\prime}$ generated by varying $h$ correspond to orbits with the same energy $\xi$, and therefore the same period $T(\xi)$. The conclusion is that Eq.~\eqref{newT}, which corresponds to the case $h=CB$, is also verified for any value of $h$ when the potential is isochrone. In other words, for any $h\in]0,CB]$, we have
\beq \label{newTh}
T(\xi)=\frac{\pi}{4(1+\xi^2)^{3/4}}\frac{L(h)}{\sqrt{h}}\, .
\eeq
Of course, Eq.~\eqref{newT} is just a particular case of Eq.~\eqref{newTh}, when $h=CB$ and $L(h)=P\!A$.
\subsubsection{Computing the areas}
With Eq.~\eqref{newTh} at hand, we can now turn to the compution of the areas involved in the theorem. For the triangle $P\!AC$, we have the basis $P\!A$ and the height $CB$. For the area between $\scC$ and $\scL$, we can simply integrate \textit{à la Lebesgue} the infinitesimal area $L(h) \ud h$ while $h$ varies between $0$ and $CB$. We thus have, respectively
\beq \label{areas}
\text{Area} \, (\scT)= \frac{1}{2} P\!A \times CB \, \, \quad \text{and} \quad \, \, \text{Area} \, (\scS)= \int_0^{CB} \, L(h) \ud h \, .
\eeq
Now we compute these areas and we show that they are in proportion $4/3$. For the area of the triangle $\scT$, we use Eq.~\eqref{newT} to express $P\!A$ in terms of $CB$ and plug the result in Eq.~\eqref{areas}. We obtain the following expression
\beq \label{area(T)}
\text{Area} \, (\scT)= \frac{2T}{\pi} (1+\xi^2)^{3/4} CB^{3/2} \, .
\eeq
In a similar manner, the area of the region $\scS$ can be found by isolating $L(h)$ from Eq.~\eqref{newTh} and expressing it in terms of $h$. Plugging the result in the area formula for $\scS$ in Eq.~\eqref{areas} and computing the integral explicitly give easily
\beq \label{area(S)}
\text{Area} \, (\scS)= \frac{8T}{3\pi} (1+\xi^2)^{3/4} CB^{3/2} \, .
\eeq
Comparing Eqs.~\eqref{area(T)} and \eqref{area(S)} shows that, indeed, $\text{Area} \, (\scS)/\text{Area} \, (\scT)=4/3$. By virtue of the theorem, the claimed result follows: If a potential is isochrone then the curve $\scC:y=Y(x)$ in H\'enon's variables is a parabola.
\subsubsection{Final remarks} \label{sec:finrem}
We end this section by answering a question: Why is it that $\scC$ should be a parabola, and not any other type of curve, when the potential is isochrone ? What is so special about parabolae ? To understand this, let us focus our attention on a point $M$ of a generic curve $\scC$ (i.e., non-necessarily a parabola). \\

Close enough to $M$, $\scC$ always looks like a parabola, as can be seen by writing its Taylor expansion\footnote{Qualitatively speaking, at least. The particular case when the second derivative vanishes at $M$ is not a problem, it corresponds to $\alpha=0$ in Eq.~\eqref{parabolicness_1}.} around $M$. To see this, consider the particular frame $(x,y)$ centered on $M$ where the tangent to $\scC$ at $M$ is horizontal. These two conditions indicate that $y(0)=0$ and $y'(0)=0$ respectively. Therefore the curve has an implicit equation of the type
\beq \label{parabolicness_1}
y=\alpha x^2+o(x^2) \quad \text{for some constant }\alpha \, .
\eeq
For a generic curve, the $o(x^2)$ in Eq.~\eqref{parabolicness_1} corrects the local \textit{parabolicness} of the curve as one moves away from $M$. However, Eq.~\eqref{newTh} shows that the $o(x^2)$ terms vanishes identically in the case of isochrony. Indeed, in Fig.~\ref{fig:paraarchi}, this particular frame $(x,y)$ we are considering is precisely the one centered on $C$ equipped with coordinates $(x,y)=(L,h)$. Now Eq.~\eqref{newTh} may be rewritten as
\beq \label{parabolicness_2}
h=\alpha L^2 \, , \quad \text{with } \alpha=\pi^2/16T^2(1+\xi^2)^{3/2} \, .
\eeq
Since this should be true for all $h$, or equivalently any $x$, comparing Eqs.~\eqref{parabolicness_1} and \eqref{parabolicness_2} shows that in the case of isochrony the $o(x^2)$ vanishes identically as claimed. \\

To summarize, in H\'enon's variables, any potential always looks, locally, like a parabola, a universal mathematical property encoded in its Taylor expansion. However, isochrony, through Eq.~\eqref{newTh}, propagates this local property to the global level, constraining the curve to \textit{be} a parabola, in addition to locally \textit{look like} one. \\

Following the same logic, one could ask whether Eq.~\eqref{HenonTheta} for $\Theta$ could not be used to reach the same result. The central difference is that Eq.~\eqref{HenonTheta} is to be read in the $(u,y)$ plane, where orbits correspond to horizontal lines $\scL:y=\xi$, whereas Eq.~\eqref{HenonT} is to be read in the $(x,y)$ plane, where orbits correspond to straight lines $\scL:y=\xi x-\Lambda^2$. Yet, the Archimedean characterization of parabola requires the areas ratio to be $4/3$ for \textit{any} chord, not just horizontal ones. Therefore, Eq.~\eqref{HenonTheta} cannot be used to conclude that $\scC$ should be a parabola, at least not with the Archimedean characterization.
\clearpage
\section{Isochrone parabolae} \label{sec:sec3}
The result of the last section implies that the curve $\scC$ of an isochrone potential $Y$ corresponds to (at least an arc of) a parabola. However, not all parabolae will \textit{contain} the potential of a physically realistic system. The aim of this section is to classify, based on their geometrical properties, the \textit{isochrone parabolae}, i.e., these that contain the curve $\scC$ of a well-defined, isochrone potential $Y$. From now on we shall always use the notation $\scP$ for an arbitrary parabola in the plane.
\subsection{Generalities on parabolae} \label{sec:genpara}
We start by a potpourri of algebraic and geometrical properties of parabolae, and derive some general results that shall be used throughout the next sections. The implicit equation of a parabola $\scP$ in the $(x,y)$ plane is

\beq \label{parabolaimplicit}
\scP : (ax + by)^2 + cx + dy + e=0 \quad \text{with} \quad \delta:=ad-bc\neq0 \, ,
\eeq
and where $(a,b,c,d,e)$ are five real numbers. The quantity $\delta$ is the discriminant of $\scP$ and is taken to be nonzero; otherwise, Eq.~\eqref{parabolaimplicit} degenerates into a pair of parallel lines. Without loss of generality, we will assume from now on that $\delta>0$.\footnote{If $\delta<0$ we can always replace $(a,b)$ by $(-a,-b)$. This leaves Eq.~\eqref{parabolaimplicit} unchanged and thus corresponds to the same parabola, but changes the sign of $\delta$.} \\

As an algebraic curve, a parabola is not, in general, the graph of a function. It is, however, always the union of such graphs. To see this,
let us take a point $(x,y)$ on the parabola $\scP$ given by Eq.~\eqref{parabolaimplicit}. Its ordinate $y$ can be given as a function of its abscissa $x$ by solving Eq.~\eqref{parabolaimplicit} for $y$. There are two cases depending on the parameter $b$: \\

\textbullet \, when $b=0$, the whole parabola $\scP$ is the graph of a function. Its equation is given by
\beq \label{harmo}
\scP : y = -\frac{e}{d} -\frac{c}{d}x -\frac{a^2}{d}x^2 \, ,
\eeq
where $d\neq0$ since the discriminant $\delta=ad\neq0$. For $d<0$ this parabola opens upwards and we shall say that it is \textit{top-oriented}. When $d>0$, it opens downwards and we will say \textit{bottom-oriented}. Any top- or bottom-oriented parabola crosses the $y$-axis once, and the ordinate of this point is
\beq
\lambda := -\frac{e}{d} \, .
\eeq

\begin{figure}[!htbp]
	\includegraphics[width=0.8\linewidth]{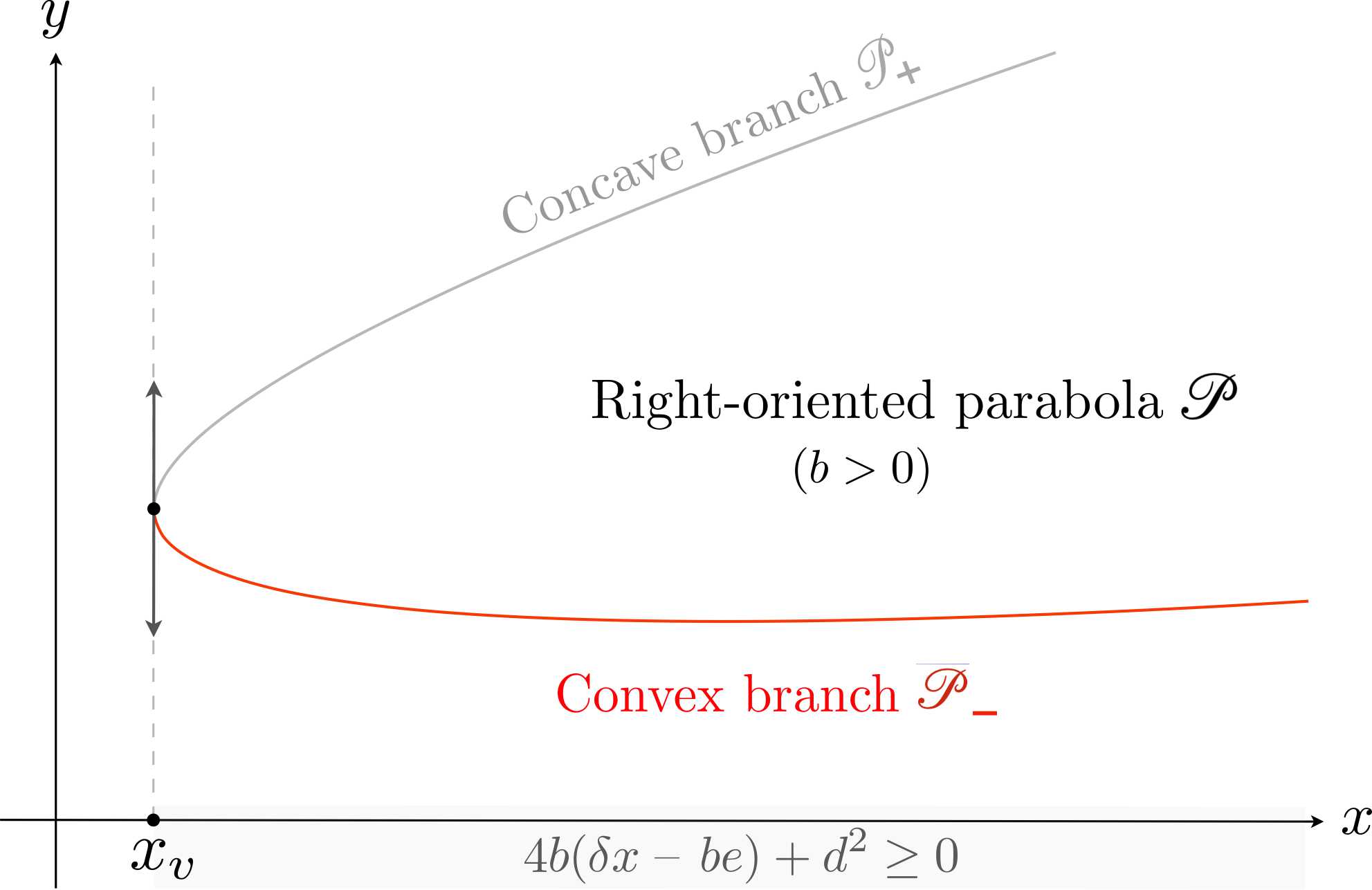}
	\caption{geometrical properties of a right-oriented parabola $\scP$, with its two branches $\scP_\pm$ that are actual graphs of functions. The parabola itself is $\scP=\scP_- \cup \scP_+$. On the left, the point of abscissa $x_v$ given by Eq.~\eqref{ixev} belongs to both $\scP_+$ and $\scP_-$, and the domain of the parabola is $[x_v,+\infty[$ (highlighted in grey on the $x$-axis). \label{fig:branches}}
\end{figure}

\textbullet \, when $b\neq0$, the curve $\scP$ is the union of two branches, which are actual graphs of a function. Indeed, for a fixed $x$, Eq.~\eqref{parabolaimplicit} is a quadratic-in-$y$ equation. Its solutions are easily found to be
\beq \label{branches}
\scP_\pm : y = -\frac{ax}{b} - \frac{d}{2b^2} \pm \frac{1}{2b^2} \sqrt{4b(\delta x - be) + d^2} \, ,
\eeq
where the condition $4b(\delta x - be) + d^2\geq0$ ensures the positivity inside the square root. The \textit{support} of the parabola is the set of $x$ such that $4b(\delta x - be) + d^2\geq0$. In particular, there is a unique value
\beq \label{ixev}
x_v := \frac{4b^2e - d^2}{4b\delta} \,
\eeq
that makes the square root in Eq.~\eqref{branches} vanish. The quantity $x_v$ is the abscissa of the common point between $\scP_+$ and $\scP_-$, where the branches meet and the tangent to $\scP$ is vertical (cf. Fig.~\ref{fig:branches}). Note that $\scP_-$ is always convex and always below $\scP_+$ which is concave. \\

The sign of $b$ controls the orientation of the parabola. If $b>0$, we will say that the parabola is \textit{right-oriented}. Its support is $[x_v,+\infty]$, and the parabola crosses the $y$-axis if and only if $x_v\leq0$. If $b<0$, we say that it is \textit{left-oriented}. Its support is $[-\infty,x_v]$ and it crosses the $y$-axis if and only if $x_v\geq0$. \\

A left or right-oriented parabola may not always cross the $y$-axis. When it does, the ordinates of the intersection points are obtained by setting $x=0$ in Eq.~\eqref{branches}. In particular, the convex branch $\scP_-$ crosses the $y$-axis at ordinate
\beq \label{lambda}
\lambda := -\frac{d + \sqrt{d^2 -4 b^2e}}{2b^2} \, .
\eeq
This implies that $d^2 -4 e b^2$ should always be positive for parabolae crossing the $y$-axis. In particular, we have $d^2 -4 b^2 e>0$ when there are two intersections. The case $d^2 -4 e b^2=0 \Leftrightarrow d^2 =4 e b^2$ happens when the two intersections degenerate into one, and its ordinate is simply $-d/2b^2$.

To summarize, the graph $\scC$ of an isochrone potential $Y(x)$ must be contained within a parabola $\scP$ in the plane. From the preceding generalities on parabolae, it follows that $\scC$ is to be looked for in any of the following families:
\begin{itemize} \label{families}
	\item top- and bottom-oriented parabolae, whose whole curve $\scP$ is that of a function defined on $\mathbb{R}$, cf. Eq.~\eqref{harmo}.
	\item left-oriented parabolae, whose curve $\scP$ is the union of a convex branch $\scP_-$ and a concave branch, $\scP_+$, each of the two being the graph of a function defined on $]-\infty,x_v]$, cf. Eq.~\eqref{branches} with $b<0$.
	\item right-oriented parabolae, whose curve $\scP$ is the union of a convex branch $\scP_-$ and a concave branch, $\scP_+$, each of the two being the graph of a function defined on $[x_v,+\infty[$, cf. Eq.~\eqref{branches} with $b>0$.
\end{itemize}
\subsection{Physical portion of the graph} \label{sec:portion}
In this section we examine under which conditions the curve $\scC$ of the (so far arbitrary) potential $Y(x)=2r^2\psi(r)$ is a physically and mathematically well-posed, isochrone potential. This will be done in four steps, each consisting on imposing a geometrical hypothesis $H_i$, $i=0,...,3$, on the curve. After each step, some curves will be discarded. At the end of the reduction process, we obtain the complete set of isochrone potentials.

In light of the results of Sec.~\ref{sec:isopara}, the first hypothesis is \\

\textbullet $\, \,$ \textit{$H_0$: $\scC$ must be an arc of parabola}. This geometrical requirement ensures that the potential $\psi$ is isochrone, regardless of its mathematical and physical properties. The next hypotheses are therefore concerned with the parabola $\scP$ that contains the curve $\scC$.
\subsubsection{Existence of orbits and well-posedness around origin } \label{sec:welpos}
\textbullet $\, \,$ \textit{$H_1$: $\scC$ must lie on the right half plane}. From a purely mathematical perspective, an isochrone potential is a function $\psi(r)$ defined on some subset of $\mathbb{R}_+$ (since $r$ is a positive radius). Since $x=2r^2>0$, \textit{we only keep parabolae that exhibit a portion on the right half plane} $x>0$. The only parabolae that do not are the left-oriented ones not crossing the $y$-axis (Eq.~\eqref{branches} with $b<0$ and $x_v\leq0$). \\

\textbullet $\, \,$ \textit{$H_2$: $\scC$ must be convex}. Indeed, a particle orbits periodically when the line $\scL$ intersects $\scC$ twice and \textit{$\scC$ is below $\scL$}. The geometrical  equivalent of this is that $\scC$ should lie under its chords, and therefore be convex. Since this is not possible on the concave branch of a parabola, \textit{we only keep parabolae that exhibit a convex branch}. these that do not are the bottom-oriented ones\footnote{Strictly speaking, these bottom-oriented parabolae will contain a unique, circular, unstable orbit.} (Eq.~\eqref{harmo} with $d<0$.). Therefore, we discard the bottom-oriented parabolae, and stress that on the right- and left-oriented ones, the curve of the potential will be an arc $\scA$ of the convex branch $\scP_-$. \\

At this stage, with only the three hypotheses $H_0,H_1$ and $H_2$, it turns out that all remaining parabolae are isochrone, in the following sense : \textit{Any parabola with a convex portion on the right half plane defines an isochrone potential $\psi(r)$}, i.e., such that Eq.~\eqref{eomr} has a periodic solution for some $\xi,\Lambda$ with $T(\xi)$. The following sections will be dedicated to the detailed analysis of these periodic orbits, and serve as a proof of this result. Nonetheless, we shall focus on parabolae that verify one more hypothesis. \\

\textbullet $\, \,$ \textit{$H_3$: $\scC$ should cross the $y$-axis}. This hypothesis discards \textit{right-oriented parabolae that do not cross the $y$-axis}. Indeed, these define a potential $\psi$ on an interval $[r_v,+\infty[$, where $x_v=2r_v^2>0$. The associated potential $\psi$ is therefore undefined in the region $r\in[0,r_v[$ surrounding the physical origin, as is the force $\vec{F}\propto\vec{\nabla}{\psi}$ and the mass density $\rho\propto\Delta\psi$. We shall coin such potentials ``hollow potentials" and leave them aside, stressing, however, that all our isochrone results apply to this type of potentials. \\

\begin{figure}[!htbp]
	\includegraphics[width=0.9\linewidth]{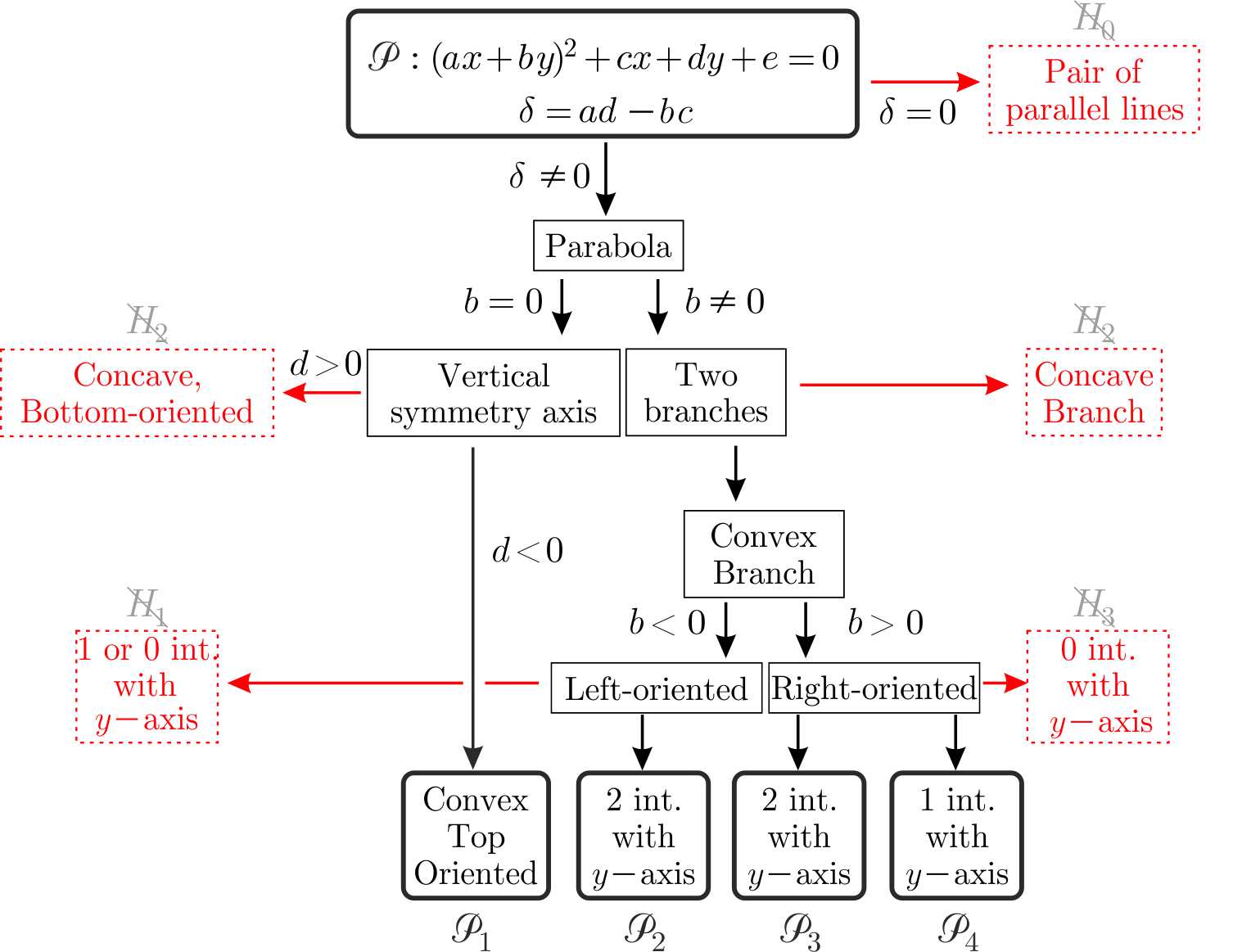}
	\caption{Tree showing the reduction process of isochrone parabolae. Starting at the top with the implicit equation \eqref{parabolaimplicit}, the reduction consists in exploring the properties of parabolae associated with the sign of $\delta,b,d$. Red horizontal lines correspond to a discarded parabola (that does not satisfy one of the hypotheses $H_i$). Black, downwards arrows lead naturally to the four families $\scP_i$, remain at the bottom. They are all associated with isochrone potentials. \label{fig:sumarpara}}
\end{figure}

After this reduction process, summarized in Fig.~\ref{fig:sumarpara}, all the remaining parabolae verify the four hypotheses and define the set of \textit{isochrone parabolae}. We recover, in particular, the algebraic classification of isochrone parabolae found in \cite{SPD}.
In this work, we shall group the isochrone parabolae into four families, according to their orientation in the $(x,y)$ plane and their number of intersections with the $y$-axis:
\begin{itemize}
	\item $\scP_1$: \textit{top}-oriented (Eq.\eqref{harmo}, with $d<0$),
	\item $\scP_2$: \textit{left}-oriented, crossing the $y$-axis \textit{twice} (Eq.~\eqref{branches} with $b < 0, 4b^2e < d^2$),
	\item $\scP_3$: \textit{right}-oriented, crossing the $y$-axis \textit{twice} (Eq.~\eqref{branches} with $b > 0, 4b^2e > d^2$).
	\item $\scP_4$: \textit{right}-oriented, crossing the $y$-axis \textit{once} (Eq.~\eqref{branches} with $b > 0, 4b^2e = d^2$).
\end{itemize}
\subsubsection{Finite mass and attractive nature} \label{sec:finitemass}
Before going further, we would like to discuss some physical properties of the isochrone potentials associated with the four families $(\scP_i)_{i=1,2,3,4}$. We start with a (non-necessarily isochrone) central potential $\psi$ and the Poisson equation \eqref{Poisson}, from which we can easily infer the mass contained within a (spherical) shell surrounding the origin. We choose the units so that $G=1$ in order to simplify the equations. \\

Let $\epsilon>0$ be the inner radius of such a shell, and $R>\epsilon$ be its outer radius, so that $\psi(R)$ is well-defined, and we let $M_\epsilon(R)$ be the mass contained within this shell $[\epsilon,R]$. By definition, $M_\epsilon(R)$ is given by $\int_\varepsilon^R \rho(r)4\pi r^2 \ud r$. Multiplying Eq.~\eqref{Poisson} by $r^2$ and integrating over the shell $[\epsilon,R]$ readily give
\beq \label{shellmass}
M_\epsilon(R) = R^2 \psi^{\prime}(R)-\epsilon^2 \psi^{\prime}(\epsilon) \, .
\eeq
From this equation, it is clear that the total mass $M(R)$ contained within the sphere or radius $r=R$ is simply given by the $\epsilon\rightarrow0$ limit of $M_\epsilon(R)$. Therefore, for any radius $R$, $M(R)$ is finite if and only if the rightmost term $\epsilon^2 \psi^{\prime}(\epsilon)$ in Eq.~\eqref{shellmass} remains bounded as $\epsilon\rightarrow0$. If this limit is infinite, the potential is sourced by an infinite amount of mass at the physical origin.  \\

With the H\'enon variables, it is very simple to see geometrically if $M(R)$ is infinite or not. Indeed, if we differentiate $Y(x)=2r^2\psi(r)$ with respect to $r$ we obtain $r^2\psi'(r)=(xY^\prime(x)-Y(x))/r$. Evaluating this at $r=\epsilon$ and Taylor-expanding around $\epsilon=0$ give easily
\beq \label{taylor}
\epsilon^2\psi^\prime(\epsilon)=-\frac{Y(0)}{\epsilon} + o(\epsilon) \, .
\eeq
It is clear from Eqs.~\eqref{shellmass} and \eqref{taylor} that the mass $M(R)$ is finite if and only if $Y(0)=0$; a result true for any central potential $\psi$. In other words, we have the following geometrical result: \textit{A potential $\psi$ is sourced by a finite mass at the origin if and only if its curve $\scC$ in H\'enon's variables passes through the origin}. In the isochrone context, this means that any isochrone parabola whose convex branch does not cross the origin is associated with an infinite mass at the origin. Moreover, we see on Eq.~\eqref{taylor} that if $Y(0)>0$, i.e., the convex branch crosses the $y$-axis above the origin, then the central mass is infinite and positive.\footnote{We shall see this feature at play in the classification of isochrone orbits, later in Sec.~\ref{sec:classorbits}.}. \\

\begin{figure}[!htbp]
	\includegraphics[width=0.6\linewidth]{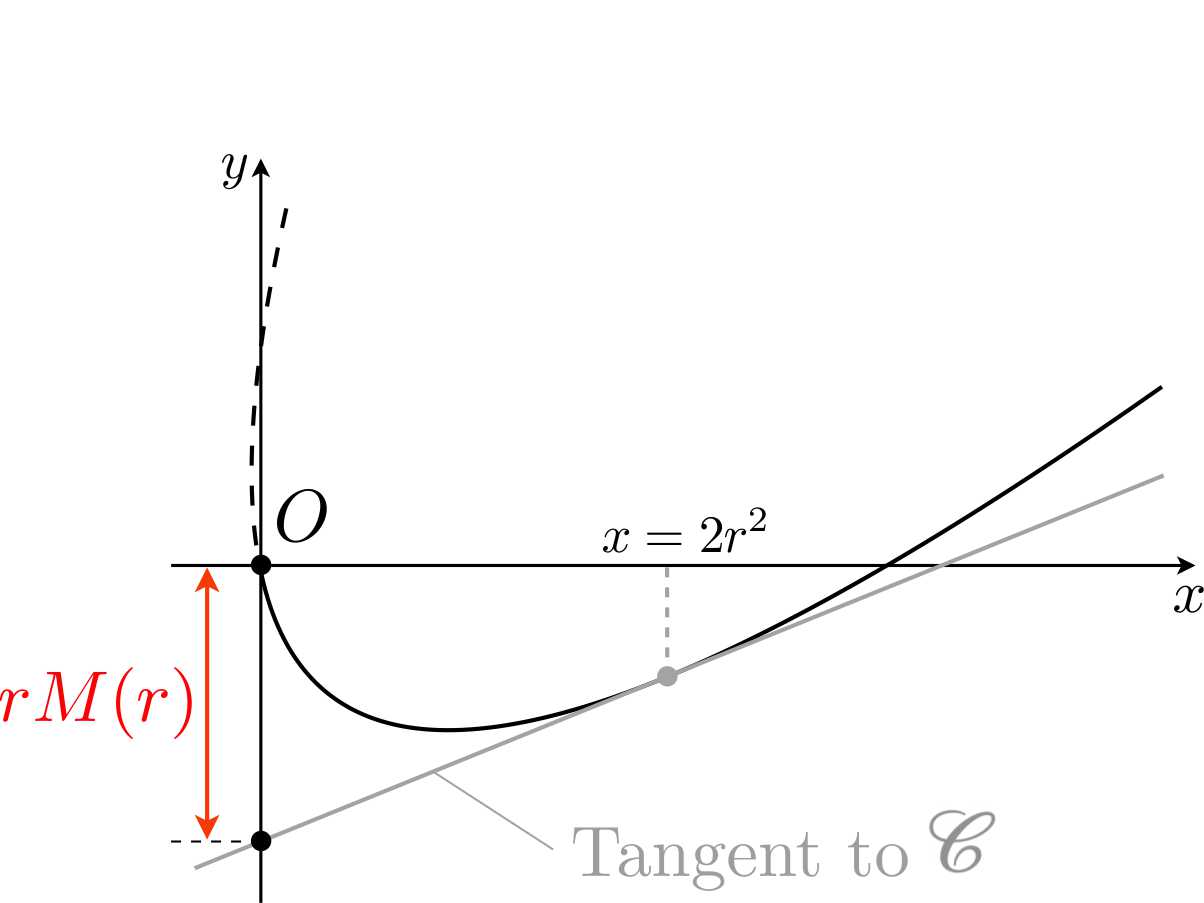}
	\caption{The curve $\scC$ is in solid black, and the rest of the parabola in dashed black. The curve passes through the origin; therefore, the mass $M(r)$ inside any sphere of radius $r$ is finite. It can be read off as the $y$-intercept of the tangent at the point of abscissa $x$ (in red). Note that this construction for the mass holds for any central potential.\label{fig:mass}}
\end{figure}

Now if we focus on a potential satisfying $Y(0)=0$, the mass $M(r)$ is finite within any sphere, and it can be read off of the curve $\scC$ as follows. Plugging $r^2\psi'(r)=(xY^\prime(x)-Y(x))/r$ into Eq.~\eqref{shellmass} allows us to write $M(r)$ in the following evocative form

\beq \label{geomass}
M(r) = -\frac{Y'(x)(0-x)+Y(x)}{r} \, .
\eeq
Notice that the numerator in Eq.~\eqref{geomass} is nothing but the $y$-intercept of the tangent of $\scC$ at the point of abscissa $x$. Therefore, given a central potential, the mass contained within a sphere of radius $r$ can be measured simply by reading this $y$-intercept. In particular, for the isochrone potentials the mass within a given sphere is also something that can be geometrically read off the parabola, as depicted in Fig.~\ref{fig:mass}.
\subsubsection{New parameters} \label{sec:newpara}
When working in the H\'enon plane, with geometry and parabolae, the Latin parameters $(a,b,c,d,e)$ are useful. In order to work with simpler expressions when dealing with the potentials, and prepare for the next steps, we rewrite the equations of the isochrone parabola $(\scP_i)$ with more adapted, Greek parameters $(\varepsilon,\lambda,\omega,\mu,\beta)$. We will follow the notations and definitions introduced in \cite{SPD}. \\

We first consider the top-oriented parabolae given by Eq.~\eqref{harmo} with $d<0$. We may combine the constants in the first two terms and define $\varepsilon:=-c/d\in\mathbb{R}$, $\lambda:=-e/d\in\mathbb{R}$. Moreover, since $d<0$ in this case, we may always write

\beq
\omega^2:=-\frac{16a^2}{d}>0 \, .
\eeq
Inserting the new parameters $(\varepsilon,\lambda,\omega)$ in Eq.~\eqref{harmo} and using $\psi(r)=Y(2r^2)/2r^2$, we obtain for the first family of potentials $\psi_1$ and/or parabolae $\scP_1$, associated with top-oriented parabolae, \\

\textbullet \, \textit{the Harmonic family}:
\beq \label{pot1}
\psi_1(r) = \varepsilon + \frac{\lambda}{2r^2} + \frac{1}{8}\omega^2 r^2 \quad \leftrightarrow \quad \scP_1:y=\varepsilon x + \lambda + \frac{1}{16}\omega^2 x^2 \, .
\eeq

The name \textit{Harmonic} comes from the fact that $\psi_1$ is a harmonic potential, up to a constant $\varepsilon$ and a centrifugal-like term $\lambda/2r^2$. The normalizing factor $1/16$ is chosen such that the radial period $T$ coincides exactly with the angular frequency $\omega$; i.e.,. $T=2\pi/\omega$, as we shall find later on. \\

Now consider the $b\neq0$ case, i.e., the convex branch $\scP_-$ of Eq.~\eqref{branches}. Once again, we may define $\varepsilon:=-a/b\in\mathbb{R}$. With a bit of rewriting, we can also introduce the $\lambda$ parameter of Eq.~\eqref{lambda} and move the square root of the rightmost term down, using the usual conjugate trick. Furthermore, independently of the sign of $b\neq0$, we may always define
\beq
\mu:=\sqrt{\frac{\delta|b|}{2b^4}}>0 \, \quad \text{and} \quad \beta:=\sqrt{\frac{d^2-4b^2e}{8\delta|b|}}\geq0 \, .
\eeq

When $\beta\neq0$, the parabola crosses the $y$-axis twice. Then either $b>0$ (right-oriented parabola) or $b<0$ (left-oriented parabola), as depicted in Fig.~\ref{fig:sumarpara}. Inserting the new parameters $(\varepsilon,\lambda,\mu,\beta)$ in Eq.~\eqref{branches} and using $\psi(r)=Y(2r^2)/2r^2 $, we obtain the second family of potentials $\psi_2$ and/or parabolae $\scP_2$, associated with left-oriented parabolae, \\

\textbullet \, \textit{the Bounded family}:
%
%
\beq \label{pot2}
	\psi_2(r) = \varepsilon + \frac{\lambda}{2r^2}  + \frac{\mu}{\beta + \sqrt{\beta^2 - r^2}} \quad \leftrightarrow \quad \scP_2 : y = \varepsilon x + \lambda + 2\mu(\beta-\sqrt{\beta^2 - x/2}) \, .
\eeq
The name \textit{Bounded} comes from the fact that $\psi_2$ is defined only on the bounded interval $[0,\beta]$. Finally, the same parameters can be used to define the third family of potentials $\psi_3$ and/or parabolae $\scP_3$, associated with right-oriented parabolae crossing the $y$-axis twice, namely \\

\textbullet \, \textit{the H\'enon family}:
\beq \label{pot3}
	\psi_3(r) = \varepsilon + \frac{\lambda}{2r^2}  - \frac{\mu}{\beta + \sqrt{\beta^2 + r^2}} \quad \leftrightarrow \quad \scP_3 : y = \varepsilon x + \lambda +2\mu(\beta-\sqrt{\beta^2 + x/2}) \, .
\eeq

The name \textit{H\'enon} comes from the fact that $\psi_3$ is, up to a constant and a centrifugal-like term, the potential found by Michel H\'enon \cite{HeI.59}. Finally, the case $\beta=0$ makes up for the fourth and last family of potentials $\psi_4$ and/or parabolae $\scP_4$, associated with right-oriented parabolae crossing the $y$-axis once, namely \\

\textbullet \, \textit{the Kepler family}:
\beq \label{pot4}
	\psi_4(r) = \varepsilon + \frac{\lambda}{2r^2} - \frac{\mu}{r} \quad \leftrightarrow \quad \scP_4 : y = \varepsilon x + \lambda - \mu\sqrt{2x} \, .
\eeq

The name \textit{Kepler} comes from the fact that $\psi_4$ is, up to a constant and a centrifugal-like term $\lambda/2r^2$, the usual Kepler potential. \\

With the new parameters, the result of this section can be summarized easily: \textit{If a potential $\psi$ is isochrone, then it must be equal to one of the $\psi_i$ with $i=1,2,3$ or $4$}. In other words, if $\psi$ is isochrone, then there exists some constants $(\varepsilon,\lambda)$ and $(\mu,\beta,\omega)$ such that $\psi(r)=\psi_i(r)$ for some $i=1,2,3,4$. We stress that by definition, $\omega\neq0$ , $\mu>0$ and $\beta\geq0$. However, $\varepsilon$ and $\lambda$ are defined in such a way that they can take any real value, \textit{a priori}. \\

Lastly, let us make contact with \cite{SPD} once again and introduce some qualifiers for the different potentials. Notice that the parameter $\lambda$ in Eqs.~\eqref{pot1}-\eqref{pot4} is precisely the $y$-intercept of the parabola, i.e., $Y(0)=\lambda$. According to our findings in Sec.~\ref{sec:finitemass}, any potential with $\lambda\neq0$ will be associated with an infinite mass at the origin. Such potentials are coined \textit{gauged potentials} in \cite{SPD}, as the term $\lambda/2r^2$ that makes them special looks like a gauged angular momentum $\Lambda^2\rightarrow\Lambda^2+\lambda$ in the effective potential formalism. Gauged potential are opposed to so-called \textit{physical potentials} that have $\lambda=0$, and that are associated with a finite mass at the center.
\subsection{Complete set of isochrone parabolae} \label{sec:complete}
In the last section, we have isolated four families of potentials $(\psi_i)$ and their associated parabolae $(\scP_i)$, and discussed some of their properties. We have shown that if a potential is isochrone, then its curve $\scC:y=Y(x)$ is the portion of a parabola $\scP_i$ that is convex and lies on the right half plane $x>0$. \\

What remains to be shown is the reciprocal of this statement, i.e., that any such $\scC$ is the curve of an isochrone potential. To this end, we just need to show that $\scC$ can always be intersected twice by some line $\scL:y=\xi x-\Lambda^2$. We shall do this by finding explicitly which lines $\scL$ can intersect $\scC$. In so doing, we will find two important results:
\begin{itemize}
	\item the set $(\psi_i)$ is complete, i.e., it contains all and only the isochrone potentials, and
	\item necessary and sufficient conditions on $(\xi,\Lambda)$ such that the particle's orbit is bounded.
\end{itemize}
In what follows, we consider a curve $\scC$ and a line $\scL:y=\xi x-\Lambda^2$. By assumption, $\scC$ is on the right half plane and on the convex portion of a parabola $\scP_i$ given by Eqs.~\eqref{pot1}-\eqref{pot4}. When they exist, we denote by $x_P$ and $x_A>x_P$ the abscissae of $P$ and $A$, the two intersections of $\scL$ with $\scC$.
\subsubsection{Top-oriented parabolae $\scP_1$ }
Let us start with the family $\scP_1$ given in Eq.~\eqref{pot1}. We fix the parameters $(\varepsilon,\lambda,\omega)$ and look for the conditions on $(\xi,\Lambda)$ under which the line $\scL:y=\xi x-\Lambda^2$ intersects $\scP_1$ twice. By definition, $P,A$ belong to both $\scL$ and $\scP_1$; therefore, $x_P,x_A$ are solutions to $\xi x -\Lambda^2 = \varepsilon x + \lambda + \omega^2 x^2 / 16$. We may equivalently write this equation in the following evocative form
\beq \label{harmocross}
x^2 - s x + p = 0 \, , \quad \quad \text{where} \quad \,
	\begin{cases}
		s := 16(\xi-\varepsilon)/\omega^2 \, , \\
		p := 16(\Lambda^2+\lambda)/\omega^2 \, ,
	\end{cases}
\eeq
with the sum $s=x_P+x_A$ and product $p=x_Px_A$ of the two roots. Note that $x_P,x_A$ and therefore $s$ and $p$ are all functions of $\xi$ and $\Lambda^2$. For a generic quadratic equation such as Eq.~\eqref{harmocross}, two solutions exist if and only if $\Delta:=s^2-4p>0$, and are given by $(s\pm\sqrt{\Delta})/2$, with a minus sign for $x_P$ and a plus sign for $x_A$. We want these solutions to lie on the convex branch of the parabola, in order for an orbit to actually exist. This is always satisfied since the parabola $\scP_1$ is everywhere convex. Furthermore, we want them to be strictly positive, in order for $P$ and $A$ to be in the right half plane $x\geq0$. It is sufficient to require $x_P>0$, because then $x_A>x_P>0$. \\

\begin{figure}[!htbp]
	\includegraphics[width=0.9\linewidth]{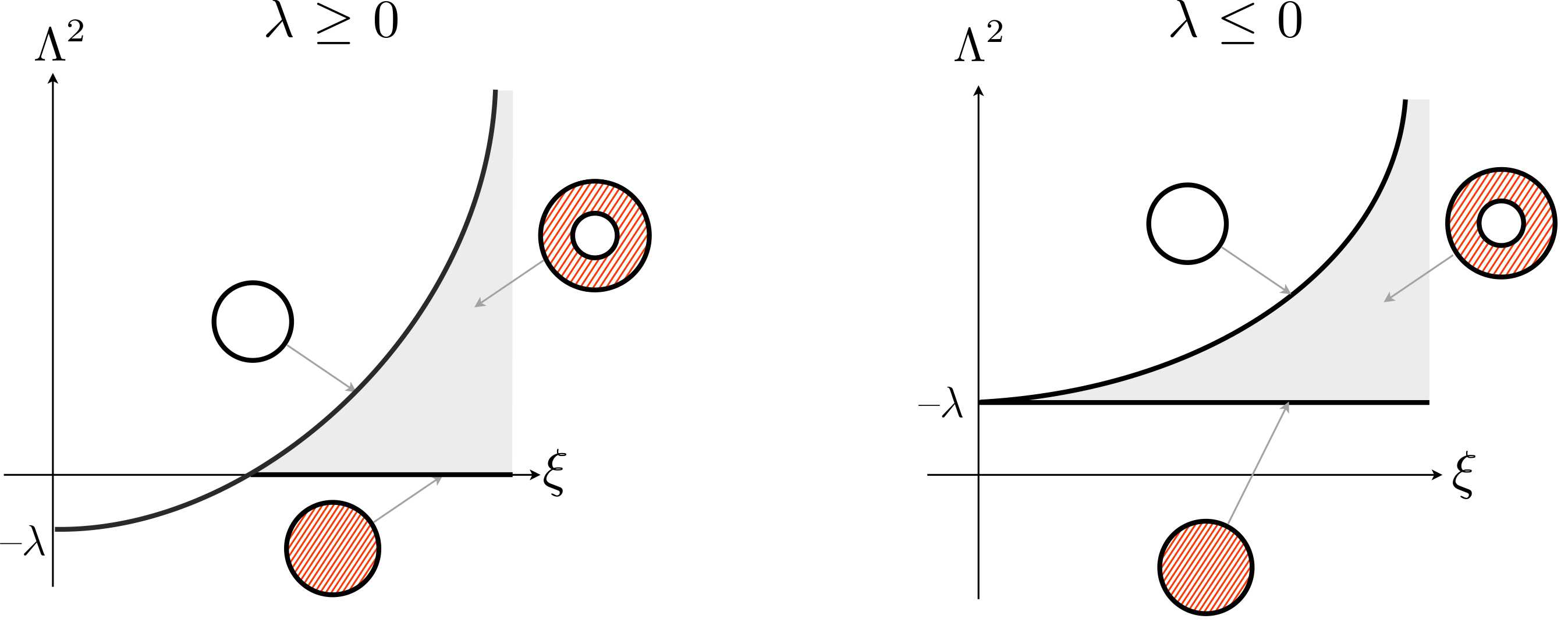}
	\caption{Bifurcation diagram for a harmonic potential $\psi_1$, with $\lambda\geq0$ (left) and $\lambda\leq0$ (right). The axes are $\xi=\varepsilon$ and $\Lambda^2=0$. The light grey region defined by the inequalities \eqref{condiharmo} contains the $(\xi,\Lambda^2)$ associated with bounded motion. Following the convention of Arnold (Fig.~(2.3) of \protect\cite{Arn}), the region of possible motion in the physical space is depicted as a light-red region in the orbital plane. For a generic orbit, $0<r_P<r_A$ and the motion takes place in an annulus. The black boundaries correspond to degeneracies: The top one to circular motion $(r_P\rightarrow r_A)$ and the bottom one to trajectories spiraling toward the center $(r_P\rightarrow 0)$. \label{fig:existP1}}
\end{figure}

The parameters of the potential are fixed; therefore, the condition $x_P(\xi,\Lambda^2)>0$, along with $\Lambda^2\geq0$, defines a region in the $(\xi,\Lambda^2)$ plane that contains every pair $(\xi,\Lambda^2)$ such that the orbit is periodic. Using the formula for $x_P$ in terms of $s$ and $p$, this domain is explicitly delimitated by the two inequalities
\beq \label{condiharmo}
\Lambda^2\geq0  \quad \text{and} \quad x_P(\xi,\Lambda^2)>0 \, ,
\eeq
with $x_P=(s-\sqrt{s^2-4p})/2$ and $s,p$ given by Eq.~\eqref{harmocross}. Outside this region, there may be collision orbits (the particle avoids the origin and goes to infinity), or no orbit at all (for instance in the $\Lambda^2<0$ region). This is depicted in Fig.~\ref{fig:existP1}.
\subsubsection{Left-oriented parabolae $\scP_2$}
We proceed similarly for the left-oriented parabolae $\scP_2$, associated with a Bounded potential. In particular, we fix the parameters $(\varepsilon,\lambda,\mu,\beta)$ and look for a domain in the $(\xi,\Lambda^2)$ plane that contains all and only the periodic orbits. The condition $P,A\in\scL\cap\scP_2$ translates algebraically into
\beq \label{pain}
\xi x-\Lambda^2 = \varepsilon x + \lambda + 2\mu\beta - 2\mu \sqrt{\beta^2 - x/2} \, .
\eeq

As for $\scP_1$, with a bit of algebra we may write Eq.~\eqref{pain} as $x^2 - s x + p = 0$ where $s:=x_P+x_A$ and $p:=x_Px_A$. In terms of $(\xi,\Lambda^2)$, we have explicitly

\beq \label{cross}
s := 2 \, \frac{(\Lambda^2+\lambda + 2\mu\beta)(\xi-\varepsilon)-\mu^2}{(\xi-\varepsilon)^2} \quad \text{and} \quad p := \frac{ (\Lambda^2+\lambda + 4\mu\beta)(\Lambda^2+\lambda)}{(\xi-\varepsilon)^2} \, .
\eeq

As before, the solutions $x_P,x_A$ must be strictly positive and this is ensured by the condition $x_P>0$. At this point, choosing $(\xi,\Lambda^2)$ in the region $x_P(\xi,\Lambda^2)>0$ ensures that there are two intersections between $\scP_2$ and $\scL$. Adding the condition $\Lambda^2\geq0$ ensures that $x_P$ is on the convex branch. However, a third condition is needed, namely that $A$ belongs to the convex branch. To this end, notice that in Eq.~\eqref{pain}, the minus sign in front of $2\mu$ came from selecting the convex branch of the parabola. Therefore, to ensure that $A$ is on this branch, it is sufficient to impose that $\xi x_A  - \Lambda^2 \leq \varepsilon x_A + \lambda+2\mu\beta$. All in all, three conditions are sufficient to draw the bifurcation diagram. They read explicitly
\beq \label{condiP2}
\Lambda^2\geq0 \, , \quad x_P(\xi,\Lambda^2)>0 \quad \text{and} \quad (\xi-\varepsilon ) \, x_A(\xi,\Lambda^2) \leq \Lambda^2+\lambda+2\mu\beta \, .
\eeq

\begin{figure}[!htbp]
\includegraphics[width=0.9\linewidth]{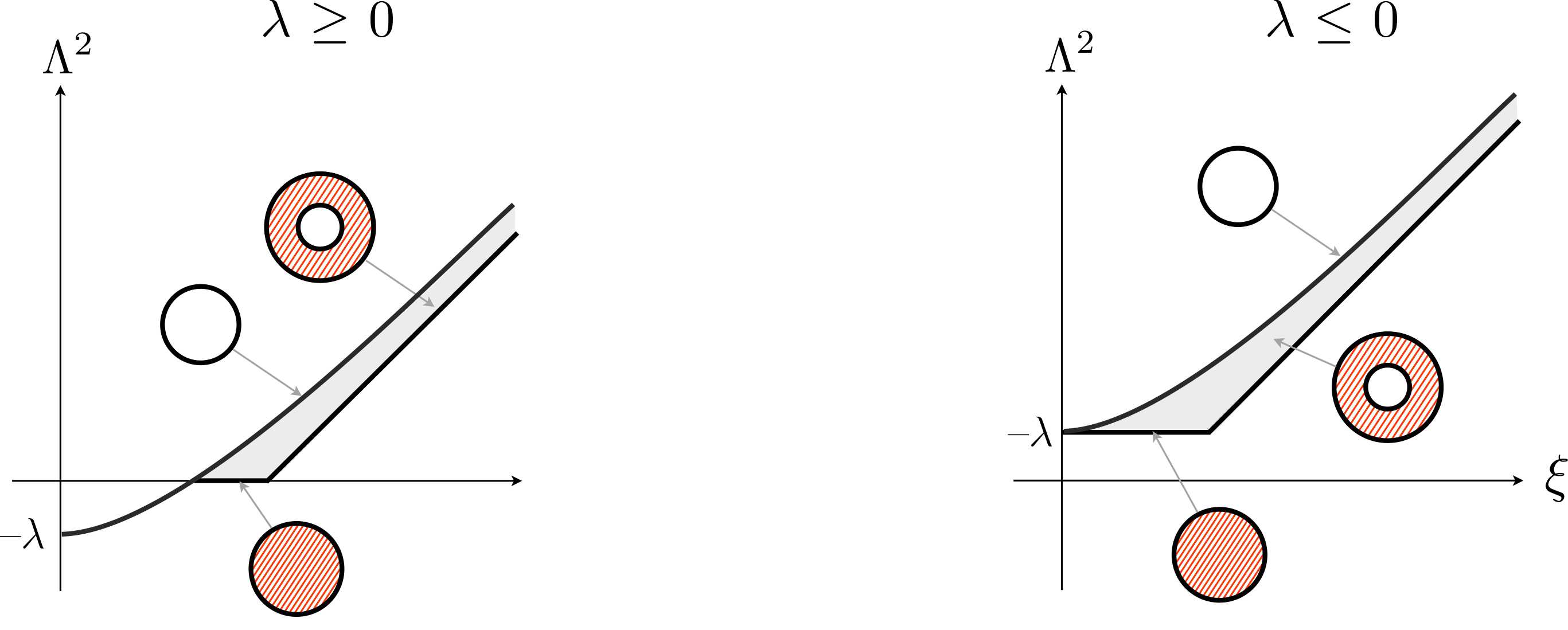}
	\caption{Bifurcation diagram for a bounded potential $\psi_2$, with $\lambda\geq0$ (left) and $\lambda\leq0$ (right). The axes are $\xi=\varepsilon+\mu/2\beta$ and $\Lambda^2=0$. All information is encoded the same way as in Fig.~\ref{fig:existP1}. \label{fig:existP2}}
\end{figure}

Once we express $x_P$ and $x_A$ in terms of $s,p$, and thus in terms of $(\xi,\Lambda^2)$ via Eq.~\eqref{cross}, the three inequalities \eqref{condiP2} delimit a region with all and only the periodic orbits. This region is depicted for a typical bounded potential $\psi_2$ in Fig.~\ref{fig:existP2}.
\subsubsection{Right-oriented parabolae $\scP_3$ and $\scP_4$}
For the right-oriented parabolae, we proceed the same way. Noticing that a parabola $\scP_3$ can be obtained as the $\beta\rightarrow0$ limit of a parabola $\scP_4$, we may focus on the latter. As before, the condition $P,A\in\scP_4$ translates into $x^2-sx+p=0$, where $s$ and $p$ are have the same expression as in Eq.~\eqref{cross}, albeit with a plus sign in front of $\mu^2$ for the former. \\

\begin{figure}[!htbp]
\includegraphics[width=0.9\linewidth]{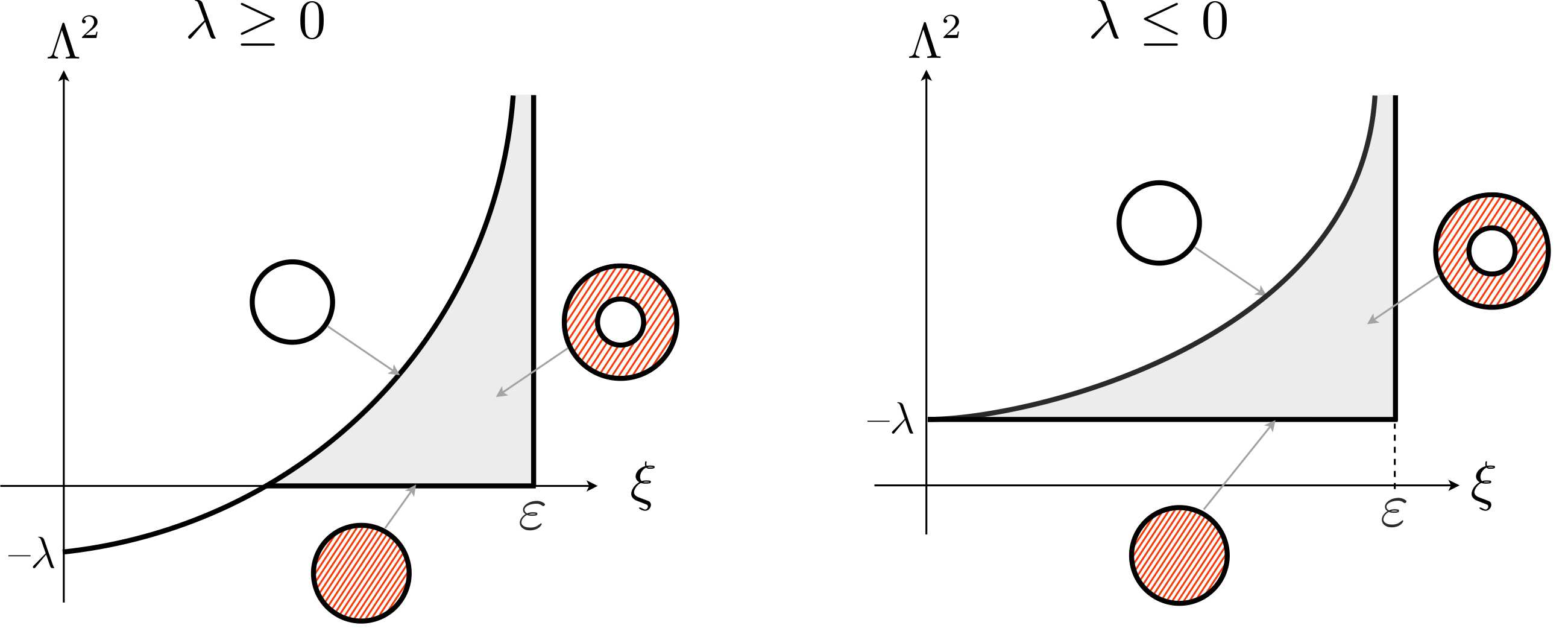}
\caption{Bifurcation diagram for a H\'enon potential $\psi_3$, with $\lambda\geq0$ (left) and $\lambda\leq0$ (right). The axes are $\xi=\varepsilon-\mu/2\beta$ and $\Lambda^2=0$. All information is encoded the same way as in Fig.~\ref{fig:existP1}. \label{fig:existP3}}
\end{figure}

The conditions to be imposed to have a well-defined orbit are as before, $x_P>0$, $\Lambda^2\geq0$. These two ensure that $P$ is in the right half plane and on the convex branch. However, this does not imply that $A$ is on the convex branch, so we must, again, add a third condition. Therefore, all three requirements are the same as in the $\scP_2$ case, and the bifurcation diagram can be depicted using the inequalities Eq.~\eqref{condiP2} (again, with a plus sign in front of $\mu^2$). For the H\'enon family, $\beta\neq0$ and the bifurcation diagram is depicted in Fig.~\ref{fig:existP3}. The bifurcation diagram for the Kepler family $\scP_4$ is the $\beta\rightarrow0$ limit of Fig.~\ref{fig:existP3} and coincides precisely with Fig.~(2.3) of \cite{Arn}.
\clearpage
\section{Dynamics in isochrone potentials} \label{sec:Kep}
Johannes Kepler published in his \textit{Astronomia Nova} a set of three laws that nowadays rightfully bear his name. The first law states that planets follow elliptical orbits around the Sun. The second law states that an orbiting planet always sweeps equal areas in equal times, and this holds for any central potential, hence for isochrone ones. Last but not least, the third law is arguably one of the most celebrated and useful equations in astronomy and astrophysics. In modern notation, this law reads
\beq \label{Keporiginal}
T^2 = \frac{\pi^2}{2} \frac{\mu^2}{(-\xi)^3}
\eeq
with $T$ the period of motion\footnote{Around a point mass, a test particle orbits on a closed ellipse whose semimajor axis $a$ is such that $a\propto \xi^{-1}$ \cite{Arn}. Consequently, Kepler's third law is commonly written as $T^2 \propto a^3$. Note that here the radial period coincides with that of the motion, i.e., $\vec{r}(t+T)=\vec{r}(t)$.} of a test particle of energy $\xi<0$ orbiting a point mass $\mu=GM$. \\

This fourth and last section is mainly dedicated to a generalization of Kepler's first and third laws. Regarding the third law, we will show that it is actually inherent to isochrony, in the sense that any periodic orbit in any isochrone potential satisfies a strikingly similar law. We shall interpret this law in various geometrical contexts and will also point out a similar and unified law for the apsidal angle $\Theta$. For the first law, we will provide an explicit formula for isochrone orbits in polar coordinates, and show that all isochrone orbits can be parameterized by a Keplerian ellipse. Finally, we shall use these results to exhibit and classify \textit{isochrone orbits}, i.e., orbits of test particles in an isochrone potential.
\subsection{Isochrone Kepler's laws for $T$ and $\Theta$} \label{sec:T&Theta}
This subsection is divided into three parts. In Sec.~\ref{sec:lawsT} we derive the generalized Kepler's third law for the period $T$ in any isochrone potential, by solving quadratic equations. Then in Sec.~\ref{sec:lawsTheta} we use the \textit{circular orbit trick} to get a similar law for $\Theta$. Lastly in Sec.~\ref{sec:georead} we provide an alternative formulation of these laws, in terms of purely geometrical quantities that can all be inferred solely from the parabola $\scP$ and the line $\scL$.
\subsubsection{Laws for the radial period $T$} \label{sec:lawsT}
Let us consider a generic isochrone potential $\psi$ and a particle $(\xi,\Lambda)$ that orbits periodically within it. As we did many times above, in the H\'enon plane $\psi$ is associated with a parabola $\scP$, and the particle is associated with a line $\scL$ that intersects $\scP$. The equations for $\scP$ and $\scL$ are of the form
\beq \label{PetL}
\scP : (ax + by)^2 + cx + dy + e = 0 \quad \text{and} \quad \scL : y = \xi x- \Lambda^2 \, .
\eeq
Keeping the parabola $\scP$ fixed, we take a line $(\xi,\Lambda)$ with two intersections $P$ and $A$, both functions of $(\xi,\Lambda)$. Now since $A$ and $P$ belong to both $\scP$ and $\scL$, we can eliminate $y$ from the two equations in \eqref{PetL} and get an equation on $x$ whose solutions are $x_P$ and $x_A$, the abscissa of $P$ and $A$. Re-arranging the result gives the following quadratic equation:
\beq  \label{Quadx}
(a+b\xi)^2x^2 + [(c+d\xi) - 2b\Lambda^2(a+b\xi)]x + [b^2\Lambda^4+e-d\Lambda^2] = 0 \, ,
\eeq
whose discriminant $\Delta$ is given by
\beq  \label{Discx}
\Delta(\xi,\Lambda) := (c+d\xi)^2 - 4b\Lambda^2(c+d\xi)(a+b\xi) -4(a+b\xi)^2(e-d\Lambda^2) \, .
\eeq
We will now compute the period $T$ with the help of H\'enon's formula for $T$ given in Eq.~\eqref{HenonT} proved in Sec.~\ref{sec:T&Theta}. First, we need the difference between $x_A$ and $x_P$. This is simply a matter of writing the solutions to a quadratic equation. We obtain easily
\beq  \label{xa-xp}
(x_A-x_P)^2= \frac{\Delta(\xi,\Lambda)}{(a+b\xi)^4} \, .
\eeq
Second we need a formula for $\Lambda_C$, the angular momentum of the circular orbit with energy $\xi$. To find it, we notice that when we keep $\xi$ fixed, the discriminant $\Delta(\xi,\Lambda)$ is strictly positive when there are two intersections, and by definition vanishes for some $\Lambda_C$, when there is only one intersection. This corresponds to a circular orbit, obtained by translating $\scL$ downward until $P$ and $A$ degenerate into a single point $C$. Therefore, we have $\Delta(\xi,\Lambda_C)=0$, which may be solved for $\Lambda_C$. With the help of Eq.~\eqref{Quadx} and some elementary algebra, we obtain
\beq  \label{Lambdac}
\Lambda_C^2 = \frac{4e(a+b\xi)^2-(c+d\xi)^2}{4\delta(a+b\xi)} \, ,
\eeq
where $\delta$ is the discriminant of the parabola, cf. \eqref{parabolaimplicit}. Now we can combine H\'enon's formula \eqref{HenonT} with Eqs.~\eqref{xa-xp} and \eqref{Lambdac}. Once again, after some easy algebra we obtain the following generalization of Kepler's third law
\beq \label{GenKep1}
T^2 = -\frac{\pi^2}{4} \frac{\delta}{(a+\xi b)^3}.
\eeq
A few remarks are in order here. First of all, we stress that Eq.~\eqref{GenKep1} is valid for any particle $(\xi,\Lambda)$ orbiting periodically in \textit{any} isochrone potential. In particular, this law is valid even for the potentials that were discarded earlier, i.e., these that are decreasing around the origin, these that contain infinite mass at the center and even the hollow ones, undefined around the origin. As long as there is a periodic orbit in a isochrone potential, physical or not, there is an associated parabola $\scP$ given by Eq.~\eqref{PetL} and its radial period $T$ verifies Eq.~\eqref{GenKep1}. \\

Second, we see that it involves in the numerator $\delta$ which is strictly positive. Therefore, Eq.~\eqref{GenKep1} implies that the denominator is strictly negative and thus that $a+b\xi<0$. This is a general property that can be traced back to the very existence of solutions to the quadratic equation \eqref{Quadx}. We shall use this result later in Sec.~\ref{sec:param} to find a parameterization of isochrone orbits. Moreover, speaking of the parameters, we recover the two well-known cases: When $a=0$ the parabola has horizontal symmetry and we have $T^2\propto\xi^{-3}$, as in the Kepler potential. Similarly, when $b=0$, the parabola has vertical symmetry and we have $T=\text{cst}$, i.e., $T$ is independent of the properties of the particle, as for the harmonic potential.   \\
\subsubsection{Laws for the apsidal angle $\Theta$} \label{sec:lawsTheta}
All the results presented in the last paragraphs regarding $T$ are also true for the apsidal angle $\Theta$. In particular, we can use the H\'enon formula \eqref{HenonTheta} in order to write the apsidal angle for any orbit solely in terms of $\Lambda$ and the parameters $(a,b,c,d,e)$. To this end, we start, as usual, with some geometry. \\

Consider a line $\scL$ intersecting a generic isochrone parabola $\scP$, both given by Eq.~\eqref{PetL}. Since $\Theta$ is independent of $\xi$, we may choose a value of $\xi$ such that the orbit is circular. This can be done as follows. Keeping $\Lambda$ fixed, decreasing $\xi$ defines other lines with the same $\Lambda$ and thus the same apsidal angle $\Theta(\Lambda)$ for the associated orbits. In particular, $\xi$ can reach a critical value $\xi_C$ such that the line $\scL$ becomes tangent to $\scP$, at some point of abscissa $x_C$. It is important to notice that $\xi_C$ and $x_C$ are function of $\Lambda$ only. \\

Let us focus on this very line $\scL_C:y=\xi_C x - \Lambda^2$ and the associated circular orbit. Its orbital radius is $r_C$, such that $2r_C^2=x_C$. The period $T(\xi_C)$ of this orbit is given by Eq.~\eqref{GenKep1}. Now by definition of the angular momentum, we have, for this circular orbit $\Lambda = r_C^2 \dot{\theta}$. Since $x_C=2r_C^2$, this can be turned into a differential equality $2\Lambda \ud t = x_C \ud \theta$. Now, by definition of $\Theta$, integrating the latter over a period $T(\xi_C)$ readily gives
\beq \label{GrosTheta}
\Theta = 2\Lambda \frac{T(\xi_C)}{x_C}\, .
\eeq
Let us stress again that both $\xi_C$ and $x_C$ are functions of $\Lambda$, thus we just need a formula for these in terms of $(a,b,c,d,e)$ and $\Lambda$. Now we can apply the results of the last section, regarding the intersections of $\scP$ and $\scL$, below Eq.~\eqref{PetL}. In particular, in the present context $\xi_C$ is such that $\Delta(\xi_C,\Lambda)=0$, with $\Delta$ given by Eq.~\eqref{Discx}. After some algebra, the solution for $\xi_C$ is easily found to be
\beq \label{xic}
\xi_C(\Lambda) = - \frac{a \Xi_C-c}{b\Xi_C-d} \quad \text{where} \quad \Xi_C := 2b\Lambda \pm 2 \sqrt{b^2\Lambda^4-d\Lambda^2+e} \, ,
\eeq
and the $+/-$ sign should be used for the left-/right-oriented parabolae, respectively. Regarding the quantity $x_C$, it can be found by writing the solution to Eq.~\eqref{Quadx} when $\Delta=0$. We obtain easily
\beq \label{ixc}
x_C(\Lambda) = \frac{b \Lambda^2}{a+b\xi_C} - \frac {c+d \xi_C} {2(a+b \xi_C)^2} \, ,
\eeq
where $\xi_C$ is given in terms of $\Lambda$ by Eq.~\eqref{xic}. \\

Now it is just a matter of inserting Eqs.~\eqref{ixc} and \eqref{xic} into Eq.~\eqref{GrosTheta} and do some algebra to obtain a formula for $\Theta$. After a rather lengthy but simple computation, we obtain the following law, valid for any particle orbiting in any isochrone potential
\beq \label{KepTheta1}
\frac{\Theta^2}{\pi^2 \Lambda^2 } = \frac{2b^2\Lambda^2-d}{b^2\Lambda^4-d\Lambda^2+e} + \frac{2b}{\sqrt{b^2\Lambda^4-d\Lambda^2+e}} \, .
\eeq
As for Eq.\eqref{GenKep1}, we stress that Eq.~\eqref{KepTheta1} is valid for any orbit in any isochrone potential, even the gauged and hollow ones (discarded at the end of Sec.~\ref{sec:portion} due to unusual physical properties. As a corollary of this general formula, one may insert the Greek parameters $(\varepsilon,\lambda,\omega,\mu,\beta)$ introduced earlier, and find agreement with the results of \cite{SPD}.
\subsubsection{geometrical reading of the third laws} \label{sec:georead}
\begin{figure}[!htbp]
	\includegraphics[width=0.8\linewidth]{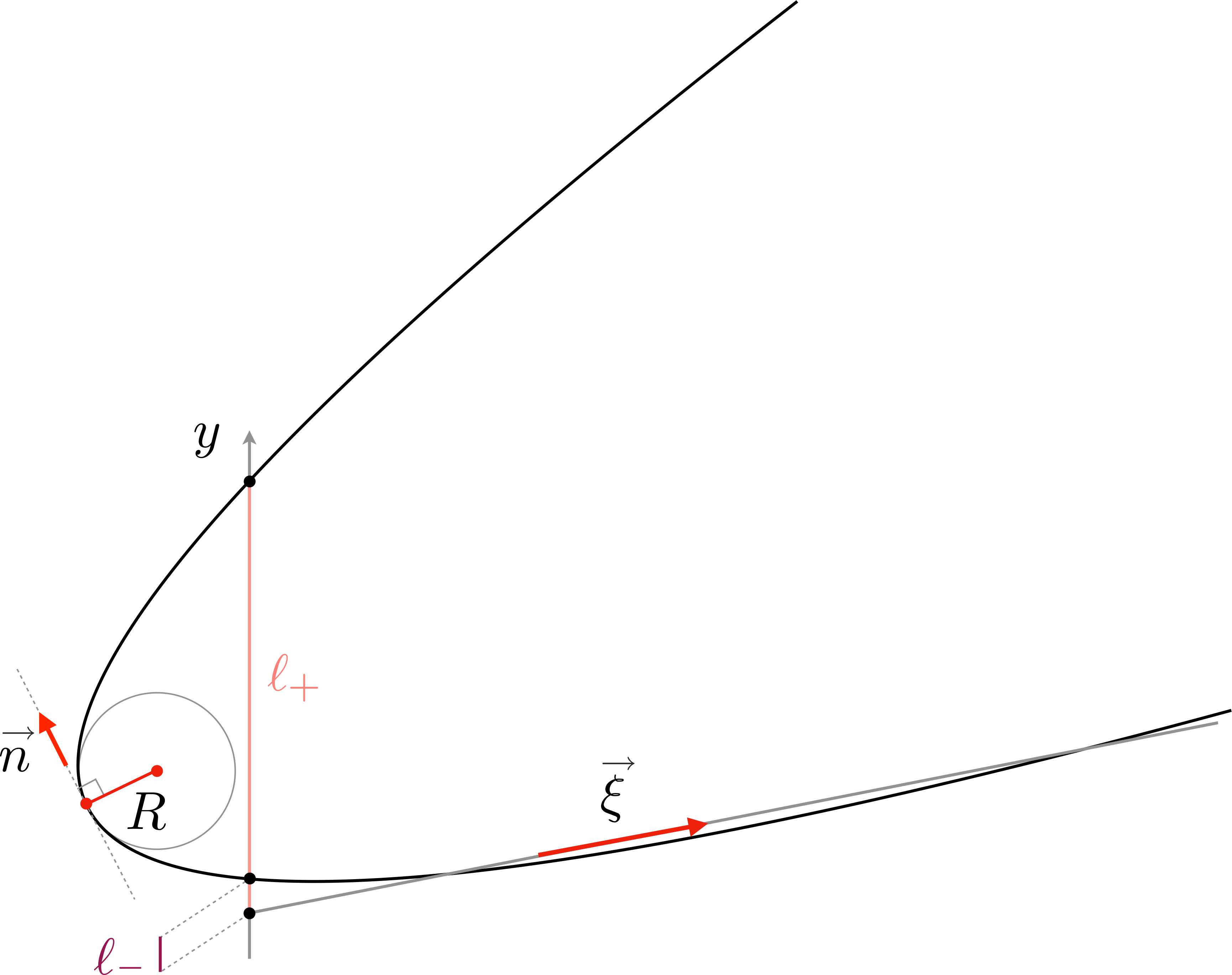}
	\caption{geometrical quantities involved in the geometrical laws for $T$ (Eq.~\eqref{GenKep2}) and for $\Theta$ (Eq.~\eqref{ThetaGeo}). In this figure, $\vec{n}$ is a unit vector and gives the asymptotic direction of the parabola, $R$ is the curvature radius at the apex, and $\ell_{\pm}$ are the vertical distances between the $y$-intercept of $\scL$ and the $y$-intercept of $\scP_{\pm}$. \label{fig:georead}}
\end{figure}

The computation of the period $T$ and apsidal angle $\Theta$ via Eqs.~\eqref{GenKep1} and \eqref{KepTheta1} involves the parameters $a,b,c,d,e$, and can thus be made only if we know the algebraic equation of the parabola. Here, we show that it is also possible to express $T$ and $\Theta$ entirely in terms of geometrical quantities, i.e., compute them solely with Euclidean geometry, once a parabola $\scP$ and a line $\scL$ is drawn in the H\'enon plane. For the period $T$, we need to define three geometrical quantities
\begin{itemize}
	\item $\vec{\xi}:=(1,\xi)$, the natural tangent vector to the line $\scL$,
	\item $\vec{N}=(a,b)$, the natural tangent vector to the symmetry axis of $\scP$ (which controls its orientation), and
	\item and $R=\delta/2|\vec{N}|^3$, the radius of curvature of $\scP$ at its apex (which controls its aperture).
\end{itemize}
The expressions for these quantities can be easily derived with the help of Sec.~\ref{sec:genpara}. We can now rewrite Eq.~\eqref{GenKep1} without the $(a,b,c,d,e)$ parameters, using the unit vector $\vec{n}=\vec{N}/|\vec{N}|$, simply as
\beq \label{GenKep2}
T^2 = \frac{\pi^2}{2} \frac{R}{|\vec{n}\cdot{\vec{\xi}} \, | ^3} \, .
\eeq
This formula should be compared to Kepler's third law for the radial period as given by Eq.~\eqref{Keporiginal}. In a similar fashion, we can make a geometrical construction for the law of the apsidal angle $\Theta$. In particular, let $\ell_\pm$ be the ordinate of the intersection points between the branch $\scP_\pm$ and the $y$-axis. As we have seen already in Sec.~\ref{sec:genpara}, we have
\beq \label{lpm}
\ell_\pm = -\frac{d\pm\sqrt{d^2-4b^2e}}{2b^2} \,
\eeq
In terms of the Greek parameters, we can show easily that the quantity $\ell_+$ is nothing but $\lambda$, and that $\ell_-$ is $\lambda+4\mu\beta$. Now we can easily turn Eq.~\eqref{lpm} into $b^2\Lambda^4-d\Lambda^2+e=b^2(\ell_++\Lambda^2)(\ell_-+\Lambda^2)$
and notice that the left-hand side appears precisely in the denominator in Eq.~\eqref{KepTheta1}. Therefore, we insert this result in Eq.~\eqref{KepTheta1}, make a partial fraction decomposition for the first term and obtain
\beq \label{ThetaGeo}
\Theta = \frac{\pi}{\sqrt{1+\ell_+/\Lambda^2}} + \frac{\iota \pi}{\sqrt{1+\ell_-/\Lambda^2}} \, ,
\eeq
where $\iota\in\{-1,0,1\}$ is simply the sign of $b$ and determines the orientation (resp. left, top, bottom) of the parabola. As for the period $T$, we see that the apsidal angle $\Theta$ can be found with only geometrical quantities that can be read off the parabola. In particular, $\ell_\pm/\Lambda^2$ is simply the ratio of the vertical distances between the $y$-intercept of $\scL$ and the branches\footnote{For the Kepler family, the two intersections degenerate into one and $\ell_+=\ell_-$. For the Harmonic family, $\ell_+$ goes to $+\infty$ (think of a $\pi/2$-rotation turning $y=-\sqrt{x}$ into $y=x^2$).} $\scP_\pm$, as depicted in Fig.~\ref{fig:georead}.

Finally, at the end of App.~\ref{app:alt} we provide yet another form for the third law $T(\xi)$ and $\Theta(\Lambda)$. For $T$, this alternative form involves the slope and curvature of the parabola at the circular point $C$ uniquely associated with the orbit. For $\Theta$, this alternative involves the curvature of the effective potential, when looked at in the Binet variable $u=1/r$. The interested reader may find these results useful in order to go further into the geometrical properties of isochrony, e.g., for academic purposes. Along these lines, we stress that what led to the mathematical equalities \eqref{GenKep1}, \eqref{KepTheta1} and other third laws in App.~\ref{app:alt} can be seen as a geometrical method to compute the rather complex-looking integrals \eqref{defTHenon} and \eqref{HenonTheta}, with $Y(x)$ given by Eq.~\eqref{harmo} or Eq.~\eqref{branches}. In particular, the fact that these rather complex-looking integrals do not depend on $\Lambda$ for $T$, and on $\xi$ for $\Theta$, is quite remarkable.
\subsection{Isochrone orbits transformations} \label{sec:param}
In this section, we provide a geometrical analysis that leads naturally to an explicit and analytic parameterization of any isochrone orbit in polar coordinates. The essential idea is the following: An isochrone orbit is associated with an arc of parabola in the H\'enon plane. There is one isochrone orbit for which we know an analytic expression: The Keplerian ellipse. Using linear transformations in the H\'enon plane, we show how to map any arc of parabola to a Keplerian one, and therefore establish a one-to-one correspondence between any isochrone orbit and a Keplerian ellipse, the latter being used to parameterize the former. \\
\subsubsection{Reduced orbit}
As we have seen many times before, an arc of parabola $\scA$ in the H\'enon plane $(x,y)$ is associated with an isochrone orbit in the physical space that will be denoted by $\scO$. By conservation of angular momentum, the particle orbits within a plane, equipped with the usual polar coordinates $(r,\theta)$. We shall always choose the angle $\theta$ such that $\theta=0$ at periapsis $r=r_P$. \\

\begin{figure}[!htbp]
	\includegraphics[width=1.0\linewidth]{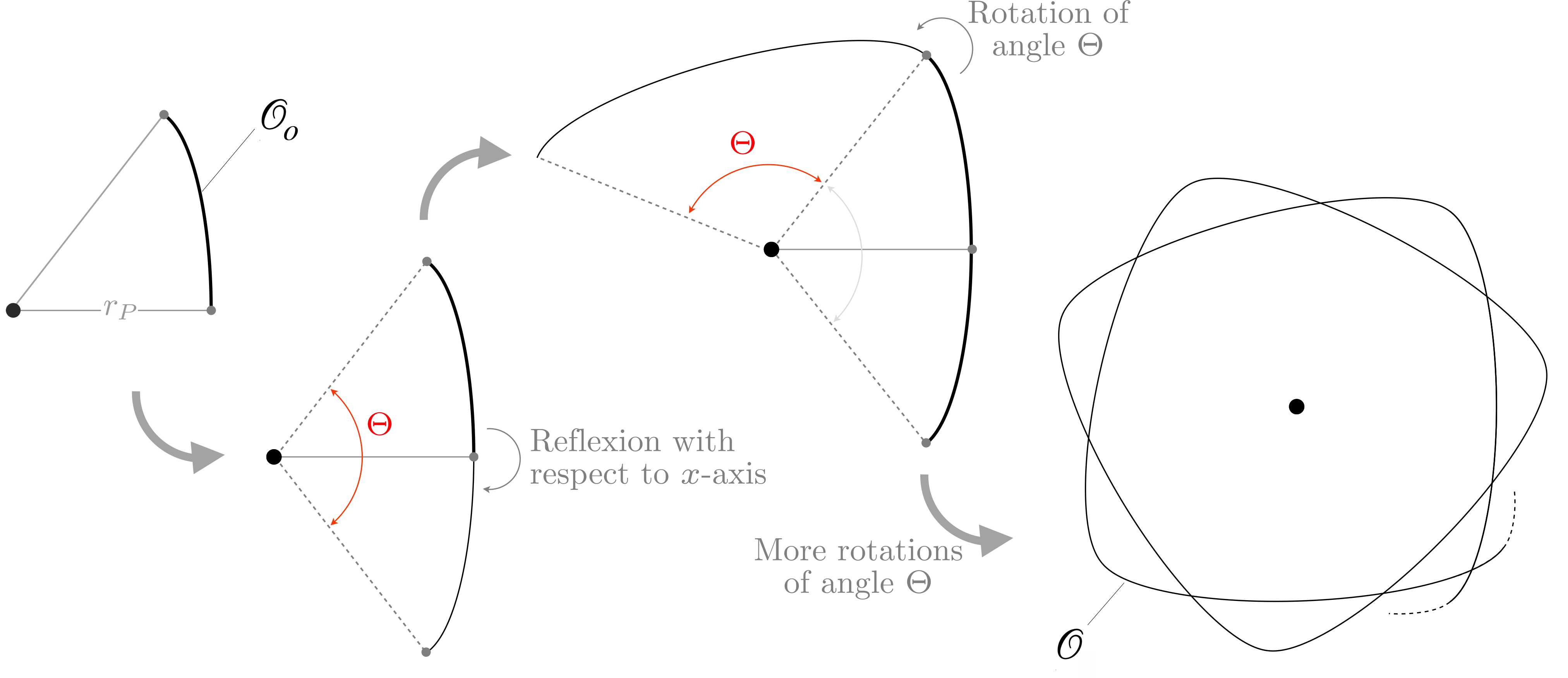}
	\caption{The construction of a full orbit $\scO$ (right) from a a reduced orbit $\scO_o$ (left). All points of $\scO_o$ have a different radius $r$ varying increasingly between periapsis $r=r_P$ and apoapsis $r=r_A$. By reflexion with respect to $\theta=0$, one obtains the portion travelled in one radial period $T$, and the opening angle is therefore $\Theta$. By successive rotations of angle $\Theta$, the full orbit $\scO$ can be constructed. \label{fig:redorbit}}
\end{figure}

When the particle moves on an isochrone orbit $\scO$, its radius $r$ changes periodically and can be mapped to a point $M$ that travels back and forth on the arc $\scA$. However, the converse is not true: A point $M\in\scA$ of abscissa $x=2r^2$ corresponds to an infinite number of points on $\scO$, namely the points $(r,\theta+k\Theta)_{k\in\mathbb{Z}}$, precisely because of the radial periodicity. To get a one-to-one correspondence, we can quotient the full orbit $\scO$ by reflexions and rotations, to get the \textit{reduced orbit $\scO_o$}, as depicted in Fig.~\ref{fig:redorbit}. The full orbit $\scO$ can be constructed from $\scO_o$, which acts as a generator of the orbit and which, contrary to the \textit{full} orbit $\scO$, is in a one-to-one correspondence with the arc $\scA$: a point $(r,\theta)\in\scO_o$ is uniquely linked to a point $(x,y)\in\scA$ via $x=2r^2$.
\subsubsection{Kepler parabola} \label{sec:keplerorbit}
We will need in this section a few results about the Kepler parabola $y^2=2\mu^2x$. It is associated with the usual Kepler potential $\psi(r)=-\mu/r$. The bifurcation diagram in terms of $(\xi,\Lambda)$ for the Kepler potential is given in Fig.~\ref{fig:existP3} with $\beta=\varepsilon=\lambda=0$, or equivalently in Fig.~(2.3) of \cite{Arn}. Requiring that the values of $(\xi,\Lambda)$ generate a periodic orbit (i.e., that they are in the grey region of Fig.~\ref{fig:existP3}) is equivalent to the following algebraic inequalities:
\beq \label{Keplercond}
\Lambda^2\geq0 \quad \text{and} \quad 0>\xi\geq -\frac{\mu^2}{2\Lambda^2} \, .
\eeq

In the Kepler potential, any bounded orbit is an ellipse whose focus is at the center of polar coordinates. An elliptic orbit $\scO$ is made of one periapsis and one apoapsis, joined by two symmetrical portions, as can be seen in the bottom left of Fig.~\ref{fig:transfoiso}. We will take the reduced orbit $\scO_o$ to be the upper portion. From the classical solution to the Kepler problem (see, e.g., \cite{Arn}), the reduced orbit $\scO_o$ can be given the following parametric representation\footnote{Note that if the periapsis is at $(r,\theta)=(r_P,0)$, the reduced orbit $\scO_o$ is simply the upper half of $\scO$.}
\beq \label{paraellipse}
\begin{cases}
	r(s) = p \, (1+\varepsilon \cos s)^{-1} \\
	\theta(s) = s
\end{cases}
\!\!\!\!\!\!\!, \quad s\in [0,\pi] \, .
\eeq
In these equations, $\varepsilon$ is the eccentricity of the orbit, and $p$ is its semi-latus rectum \cite{Arn}. They depend explicitly on the energy $\xi$ and angular momentum $\Lambda$ of the particle, as well as the central mass $\mu:=GM$ of the Keplerian potential. They are given by
\beq \label{elementsellipse}
\varepsilon := \sqrt{1+ \frac{2\Lambda^2\xi}{\mu^2}} \quad \text{and} \quad p:=\frac{\Lambda^2}{\mu} \, ,
\eeq
and one can see that the conditions \eqref{Keplercond} are actually equivalent to $\varepsilon\in[0,1[$ and $p\geq0$.
\subsubsection{Linear transformation}
We consider a reduced isochrone orbit $\scO_o$ in a central isochrone potential with finite mass at the origin. Recall that $\scO_o$ is in a one-to-one correspondence with an arc of parabola $\scA$ in the H\'enon plane that passes through the origin (as explained in Sec.~\ref{sec:finitemass}). This arc $\scA$ is defined as the portion of a parabola $\scP$ that lies below a given line $\scL$, both given by Eq.~\eqref{PetL} with $e=0$
\beq \label{PetL_transfo}
\scP : (ax + by)^2 + cx + dy  = 0 \quad \text{and} \quad \scL : y = \xi x- \Lambda^2 \, .
\eeq
Now let us apply the following linear transformation $F$ to the H\'enon plane

\beq \label{lineartr}
F:(x,y) \mapsto (-cx-dy,ax+by) =: (\bar{x},\bar{y}) \, .
\eeq

\begin{figure}[!htbp]
	\includegraphics[width=1.0\linewidth]{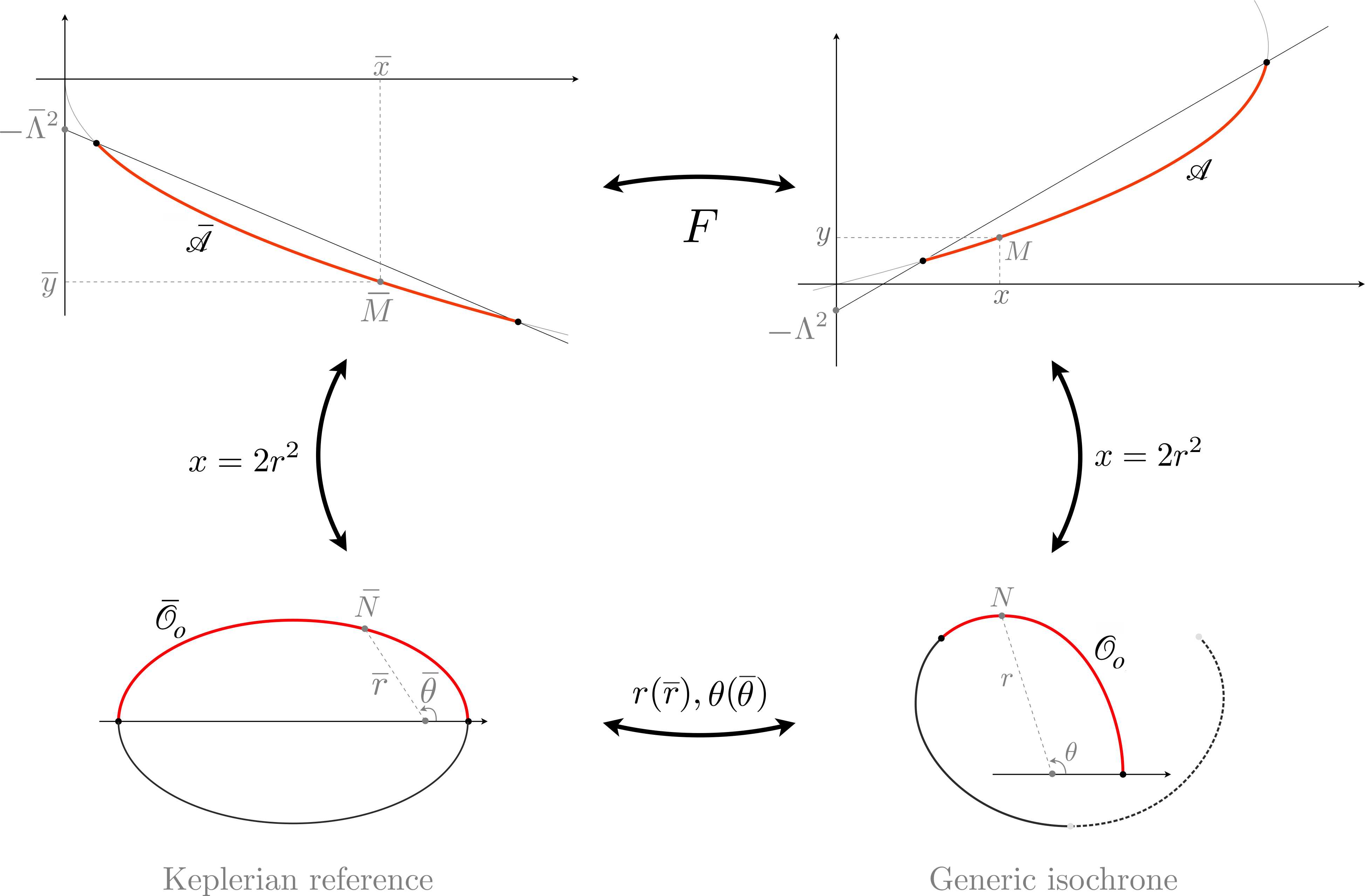}
	\caption{Any reduced isochrone orbit $\scO_o=(r,\theta)$ (red, bottom right) can be put in correspondence with a reduced Keplerian ellipse $\bar{\scO}_o=(\bar{r},\bar{\theta})$ (red, bottom left). The mapping $r(\bar{r}),\theta(\bar{\theta})$ can be obtained by going into the H\'enon plane and making a linear transformation $F$ to relate the arcs of parabolae associated with $\scO_o$ (red, top right) and $\bar{\scO}_o$ (red, top left) (see details in the text). \label{fig:transfoiso}}
\end{figure}

Following Eq.~\eqref{lineartr}, any quantity $X$ that has been mapped by $F$ will be denoted with a bar as $\bar{X}:=F(X)$. For instance, a point $M$ of coordinates $(x,y)$ on $\scA$ will be mapped to the point $\bar{M}$ with coordinates $(\bar{x},{\bar{y}})=F(x,y)$ given by Eq.~\eqref{lineartr}. Since the set of parabolae and the set of lines are stable under affine transformations (and thus linear ones), $\bar{\scP}=F(\scP)$ is still a parabola and $\bar{\scL}=F(\scL)$ still a line. The parameters $(a,b,c,d)$ in Eq.~\eqref{lineartr} are precisely these of the parabola $\scP$ in Eq.~\eqref{PetL_transfo} and are not chosen randomly. It is straightforward to find its implicit equation which reads
\beq \label{paraimage}
\bar{\scP} : \bar{y}^2 = \bar{x}  \, .
\eeq
In view of the previous subsection, it is clear from Eq.~\eqref{paraimage} that $F$ maps $\scP$ to a Keplerian parabola with mass parameter $\mu=1/\sqrt{2}$. Regarding the image $\bar{\scL}$ of the line $\scL$, a quick computation gives the following equation
\beq \label{lineimage}
\bar{\scL} : \bar{y} = \bar{\xi}\bar{x} -\bar{\Lambda}^2 \, , \quad \text{with} \quad \bar{\xi} = -\frac{a+b\xi}{c+d\xi} \quad \text{and} \quad \bar{\Lambda}^2 = -\frac{\delta\Lambda^2}{c+d\xi} \, .
\eeq
The image arc $\bar{\scA}$ is a portion of $\bar{\scP}$, although we do not know yet if it lies below $\bar{\scL}$. If it does, then $\bar{\scA}$ is a Keplerian arc and the associated orbit is an ellipse. Let us first ensure that $\bar{\scA}$ indeed corresponds to a well-defined elliptic orbit. According to the inequalities \eqref{Keplercond} with $\mu=1/\sqrt{2}$, a Keplerian orbit is periodic provided that $\bar{\Lambda}^2\geq0$, $\bar{\xi}<0$ and $\bar{\xi}>-1/4\bar{\Lambda}^2$. We now argue that these three conditions are always satisfied, in the three following steps:
\begin{itemize}
	\item \, From Kepler's generalized third law \eqref{GenKep1} and $\delta>0$, we have $a+b\xi<0$. Consequently, by Eq.~\eqref{lineimage}, the two conditions $\bar{\xi}<0$ and $\bar{\Lambda}^2\geq0$ hold if and only if $c+d\xi<0$.
	\item \, The condition $c+d\xi<0$ is a geometrical consequence of all hypotheses $(H_i)$ that are, by assumption, verified since the initial orbit $\scO$ is isochrone. (The proof is easy but not central here; it can be found in App.~\ref{app:c+dxi}.) At this stage, we thus have $\bar{\xi}<0$ and $\bar{\Lambda}^2\geq0$.
	\item \, Linear transformations preserve the existence of intersection points; therefore, $\bar{\scL}$ intersects the Kepler parabola $\bar{\scP}$ twice. Along with $\bar{\xi}<0$ and $\bar{\Lambda}^2\geq0$, we can check easily that these intersections are necessarily on the convex branch. Consequently, the orbit is an ellipse with eccentricity $\bar{\varepsilon}=\sqrt{1+4\bar{\Lambda}^2\bar{\xi}}\in[0,1[$, and therefore, $\bar{\xi}>-1/4\bar{\Lambda}^2$ holds.
\end{itemize}
To summarize, we can map any isochrone arc $\scA$ to a Keplerian one $\bar{\scA}$ with mass $\mu=1/\sqrt{2}$ using an appropriate linear transformation $F$ given by Eq.~\eqref{lineartr}. According to Keplerian dynamics, the orbit $\bar{\scO}$ associated with $\bar{\scA}$ is an ellipse whose polar equation is
\beq \label{polarKepler}
\bar{r}(\bar{\theta}) = \frac{\bar{p}}{1+\bar{\varepsilon} \cos \bar{\theta}} \quad \text{with} \quad \bar{\varepsilon}=\sqrt{1+4\bar{\Lambda}^2\bar{\xi}} \quad \text{and} \quad \bar{p}=\sqrt{2}\bar{\Lambda}^2 \, .
\eeq
We thus have a mapping between the generic isochrone arc $\scA$ and the Keplerian one $\bar{\scA}$, i.e., we have established the upper part of Fig.~\ref{fig:transfoiso}. The next step is to extend this to the lower part of Fig.~\ref{fig:transfoiso}, i.e., link the polar coordinates of each orbit.
\subsubsection{General isochrone orbit}
Consider a point $N=(r,\theta)$ on a reduced isochrone orbit $\scO_o$. It is in a one-to-one correspondence with the point $M\in\scA$ of abscissa $x=2r^2$. The point $M=(x,y)$ is in turn associated with a unique point $\bar{M}=(\bar{x},\bar{y})$ on the Keplerian parabola \eqref{paraimage} and thus with a unique point $N=(\bar{r},\bar{\theta})$ on the reduced elliptic orbit $\bar{O}_o$. This is all depicted in Fig.~\ref{fig:transfoiso}. The goal now is to express $(r,\theta)$ of the generic isochrone orbit in terms of $(\bar{r},\bar{\theta})$. \\

We start with the radius. By inverting Eq.~\eqref{lineartr} we write $x=\delta^{-1}(b\bar{x}+d\bar{y})$. Now $\bar{M}$ is on the convex branch of the Keplerian parabola \eqref{paraimage}; therefore, $\bar{y}=-\sqrt{\bar{x}}$. Combining these two equations readily gives a relation between the abscissa of the two points $M$ and $\bar{M}$, namely
\beq \label{xxbar}
x=\delta^{-1}(b\bar{x}-d\sqrt{\bar{x}}) \quad \Rightarrow \quad r=\sqrt{\frac{\bar{r}}{\delta}   \biggl(   b\bar{r}-\frac{d}{\sqrt{2}} \biggr)  } \, ,
\eeq
where we simply used the definition of H\'enon variables for each orbit, i.e., $\bar{x}=2\bar{r}^2$ and $x=2r^2$ to get the second equation, and $\delta=ad-bc\neq0$ (see Sec.~\ref{sec:genpara}). Now if we take a point $(\bar{r},\bar{\theta})$ on the Keplerian reduced ellipse, then while $\bar{\theta}$ varies in $[0,\pi]$, $\bar{r}$ changes according to Eq.~\eqref{polarKepler}, and $r$ changes as well according to Eq.~\eqref{xxbar}. Therefore, we can use $\bar{\theta}$ as a parameter, denoted $s$, to track the radius $r$ on $\scO_o$. To this end, we insert Eq.~\eqref{polarKepler} into Eq.~\eqref{xxbar} and perform some algebraic manipulations to find that \textit{the radius $r$ of the particle on a generic reduced isochrone orbit $\scO_o$ can be parameterized by $r=\varrho(s)$, $s\in[0,\pi]$, where}
\beq \label{rrbar}
\varrho(s) := \frac{\sqrt{c_1+c_2 \cos s}}{1+c_3 \cos s} \, ,
\eeq
\textit{for some constants $c_1,c_2$ and $c_3$ that depend algebraically on $(a,b,c,d)$ and $(\xi,\Lambda)$ given by}
\beq \label{c1c2c3}
c_1 = \frac{2b\delta\Lambda^4}{(c+d\xi)^2} + \frac{c_2}{c_3}\, , \quad c_2 = \frac{\Lambda^2d}{c+d\xi}c_3  \quad \text{and} \quad c_3 = \sqrt{1+\frac{4\delta\Lambda^2(a+b\xi)}{(c+d\xi)^2}} \, .
\eeq
We note incidentaly that $c_3$ is actually the eccentricity $\bar{\varepsilon}$ of the parameterizing ellipse \eqref{polarKepler}. Now we seek to find a similar parameterization for the angle $\theta$. Since $\bar{\theta}=s$ is our parameter, all we need is an expression of $\theta$ in terms of $\bar{\theta}$. To this end, we start by writing the following chain rule
\beq \label{chainrule}
\frac{\ud \theta}{\ud \bar{\theta}} = \frac{\ud \theta}{\ud t}  \frac{\ud t}{\ud x}  \frac{\ud x}{\ud \bar{x}}  \frac{\ud \bar{x}}{\ud \bar{t}}  \frac{\ud \bar{t}}{\ud \bar{\theta}} \, .
\eeq
In this equation, the time $\bar{t}$ is the one associated with the dynamics of the Keplerian orbit $\bar{\scO}$, i.e., such that energy and angular momentum conservation read
\beq \label{consbar}
\frac{1}{16} \biggl( \frac{\ud \bar{x}}{\ud \bar{t}} \biggr)^2 = \bar{\xi} \bar{x} - \bar{\Lambda}^2 - \bar{y} \quad \, \, \text{and} \quad \, \, \bar{\Lambda} = \frac{\bar{x}}{2} \frac{\ud \bar{\theta}}{\ud \bar{t}} \, .
\eeq

We now express each factor on the right-hand side of Eq.~\eqref{chainrule} in terms of $\bar{\theta}$, in order to integrate a first order ODE. For the first and last terms, we use angular momentum conservation in each orbit: Eq.~\eqref{consmom} for $\scO$ and Eq.~\eqref{consbar} for $\bar{\scO}$. Similarly, in the second and second-to-last terms we use energy conservation: Eq.~\eqref{eomHenon} for $\scO$ and Eq.~\eqref{consbar} for $\bar{\scO}$. Inserting these results in Eq.~\eqref{chainrule} readily gives
\beq \label{chainrule_2}
\frac{\ud \theta}{\ud \bar{\theta}} = \frac{\bar{x}}{x} \frac{\ud x}{\ud \bar{x}} \frac{\Lambda}{\bar{\Lambda}} \biggl( \frac{\bar{\xi} \bar{x} - \bar{\Lambda}^2 - \bar{y}}{\xi x - \Lambda^2 - y} \biggr)^{1/2} \, .
\eeq
To simplify this equation, we express $(\bar{x},\bar{y})$ in terms of $(x,y)$ using Eq.~\eqref{lineartr} and $(\bar{\xi},\bar{\Lambda})$ in terms of $(\xi,\Lambda)$ using Eq.~\eqref{lineimage}. When doing so, the last two terms on the right-hand side compensate each other exactly. The only terms contributing on the right-hand side of Eq.~\eqref{chainrule_2} are the first two, and they can be simplified with the help of Eq.~\eqref{xxbar}. In the end, we find
\beq \label{EDOthetabar}
\frac{\ud \theta}{\ud \bar{\theta}} = \frac{1}{2} +  \frac{1}{2}\frac{b}{b-d / \sqrt{2} \bar{r}} \, .
\eeq
The final step is to insert Eq.~\eqref{polarKepler} into Eq.~\eqref{EDOthetabar}. We then obtain a first-order ODE that can then be integrated using the usual change of variables $u = \tan \bar{\theta}/2$. Once this integration is done and the initial condition is chosen,\footnote{We choose $\theta$ such that $\theta=0$ at initial $r=r_P$. Since $\bar{r}_P$ is sent to $r_P$, we require $\theta=0$ when $\bar{\theta}=0.$}, we find that \textit{the angle $\theta$ of the particle on a generic reduced isochrone orbit $\scO_o$ can be parameterized by $\theta = \vartheta(s)$, with $s\in[0,\pi]$, where}
\beq \label{thetathetabar}
\vartheta(s) := \frac{s}{2} + c_4 \arctan \biggl( c_5 \tan \frac{s}{2} \biggr) \, ,
\eeq
\textit{for some constants $c_4$ and $c_5$ that depend algebraically on $(a,b,c,d)$ and $(\xi,\Lambda)$; given by}
\beq \label{c4c5}
c_4 = \frac{b\Lambda}{\sqrt{b^2\Lambda^2-d}} \quad \text{and} \quad c_5 = \sqrt{1 - \frac{2d(c+d\xi)c_3}{2b\delta\Lambda^2+d(c+d\xi)(1+c_3)}} \, .
\eeq
\subsubsection{Summary and remarks}
To summarize, we have found a parameterization for any reduced orbit $\scO_o$ in an isochrone potential $\psi$ with finite central mass. To get the full orbit $\scO$ from $\scO_o$, we follow Fig.~\ref{fig:redorbit}. In particular, we add to $\scO_o$ its symmetric with respect to the $x$-axis, by extending the range of the parameter $s$ from $[0,\pi]$ to $[-\pi,\pi].$\footnote{Indeed, from Eqs.~\eqref{rrbar} and \eqref{thetathetabar}, two points $(\varrho(s),\vartheta(s))$ and $(\varrho(s),\vartheta(-s))$ are symmetric with respect to $\theta=0$ for $s\in[0,\pi]$ simply because $\cos$ is even and $\text{id}$, $\tan$ and $\arctan$ are odd.}. We then obtain a piece $\scO_T$ of the orbit that spans a full radial period $T$, or equivalently a full apsidal angle $\Theta$. The full orbit $\scO$ is then obtained by copying and pasting the piece $\scO_T$, albeit rotated anti-clockwise by an angle $n\Theta$, for all $n\in\mathbb{Z}$. In particular, any orbit $\scO$ in a potential with finite central mass can be parameterized by
\beq \label{parameterizationfinal}
\scO = \bigcup_{n\in\mathbb{Z}} \scO_n \quad \text{where} \quad \scO_n := \{ (r,\theta)=(\varrho(s),\vartheta(s)+n\Theta) \,, s\in[-\pi,\pi]\} \, ,
\eeq
with $\varrho(s)$ and $\vartheta(s)$ given by Eqs.~\eqref{rrbar} and \eqref{thetathetabar}, respectively. By construction, when $s=0$, the particle is at periapsis $(r_P,0)$ and when $s=\pi$ it is at apoapsis $(r_A,\Theta/2)$. The latter implies that $\Theta=\pi(1+c_4)$, a result that can be checked by comparing Eqs.~\eqref{KepTheta1} and \eqref{c4c5}. The special case of the Keplerian ellipse of eccentricity $\varepsilon$ and semi-latus rectum $p$ corresponds to $(c_1,c_2,c_3,c_4,c_5)=(p^2,0,\varepsilon,1,1)$. \\

This parameterization covers \textit{any} isochrone orbit in a potential associated with a finite mass at the center. However, it can be extended easily to orbits in gauged potentials (with $\lambda\neq 0$, i.e., infinite central mass) and to hollow potentials (with $x_v>0$, i.e., undefined around the origin), by considering affine transformations instead of only linear ones. Indeed, starting from the appropriate parabola crossing the origin, with a vertical (resp horizontal) translation, one can reach any orbit in a gauged (resp hollow) potential. In particular, one can follow the previous method and send any parabola $\scP:(ax+by)^2+cx+dy+e=0$ to the Keplerian parabola $\bar{\scP}:\bar{y}^2=\bar{x}$ by applying to $\scP$ the affine transformation $F\circ G$, composed of the linear map $F$ given by Eq.~\eqref{lineartr} and the translation $G:x\mapsto x-e$. The computation can be done to find an analytic parameterization, with a little more work in the integration of the ODE expressing $\theta$ in terms of $\bar{\theta}$, cf.. Eq.~\eqref{chainrule_2}. We leave this as an interseting exercise to the interested reader. \\

Speaking of Eq.~\eqref{chainrule_2}, we have seen that the last two terms on the right-hand side cancel each other. Notice that it would also have been the case if the following assumptions had been made: $\xi x - y=\bar{\xi}\bar{x} - \bar{y}$ and $\Lambda=\bar{\Lambda}$. In \cite{SPD}, the authors precisely make these assumptions and the consequence was twofold. On the one hand, not all orbits can be reached from the Keplerian one (only the ones associated with arcs that verify theses two geometrical constraints). Therefore, the so-called bolst transformations cannot bridge between any two isochrone orbits. On the other hand, these bolsts form a subgroup of the linear transformations, whose additive representation exhibits similarities with the Lorentz group (to some extent), allowing for some analogies with special relativity, for a particular subclass of bolsts (the so-called `$\boldsymbol{i}-$bolst). However, as we have seen, the integration of Eq.~\eqref{chainrule_2} is tractable without any additional assumption, and considering linear transformations is the only way to describe all isochrone orbits. To summarize, although we believe that the special relativistic analogies presented in \cite{SPD} may be of pedagogical interest, the fundamental group associated with isochrony is that of parabola arcs equipped with affine transformation. Any other subgroup will necessarily miss the description of some isochrone orbits.
\subsection{Classification of isochrone orbits} \label{sec:classorbits}
Now that we have an analytic expression for any isochrone orbit, we will classify each of them according to the isochrone potential in which it exists. In classical textbooks, the two academic orbits turn out to be ellipses: In the Kepler problem, these ellipses have the origin at one of their foci, and in the harmonic problem, the ellipse is centered on the origin. We will of course recover these results here, and try to exhibit the plethora of orbits arising from all four families of isochrone potentials, one by one. \\

\subsubsection{General properties of isochrone orbits}

It is well known that in gravitational mechanics, a periodic orbit in a generic radial potential consists in a \textit{rosette} \cite{Arn}. Although no clear definition of a rosette exists, all have in common a generally not-closed flower-shaped plane curve that may wrap numerous times around the origin while oscillating between an perapsis and apoapsis. Of course, we will recover all these results here. In Figs.~\ref{fig:orbitharmo} through \ref{fig:orbitHenon}, we depict the orbits of particles in each of the four families of isochrone potentials, such as defined in Sec.~\ref{sec:newpara}. However, all isochrone orbits exhibit similar properties, due to the following fact: Any isochrone potential $\psi(r)$ can be written as
\beq \label{f(r)}
\psi(r) = \varepsilon + \frac{\lambda}{2r^2} + f(r) \, ,
\eeq
where $f$ is an increasing function or $r$, as can be checked on the definitions of the potentials $\psi_i$ in Sec.~\ref{sec:newpara}. \\

In the case $\lambda\leq0$, it is immediate from Eq.~\eqref{f(r)} that $\psi$ will be increasing and therefore be associated with a \textit{gravitational} potential, i.e., with positive mass density $\rho$. Consequently, the orbiting particle will feel an ever-attracting force and its orbit will be some kind of rosette \cite{Arn}, as is well known in classical mechanics. In particular, the apsidal angle in such a case will always verify $\Theta\geq\pi$ since the particle, when approaching the origin, \textit{misses} it as its angular velocity $v_\theta=\Lambda/r$ increases, while $r\rightarrow0$. Given that the harmonic and Kepler potentials are isochrone, it is no surprise that all isochrone orbits with $\lambda\leq 0$ will be some kind of precessing ellipses, as we shall see in the following sections. \\

We mention that in the $\lambda\leq0$ case, any particle $(\xi,\Lambda)$ with $\Lambda^2=-\lambda$ will fall toward the center without stopping, and will take an infinite amount of time to reach it, as on the innermost orbit depicted in Fig.~\ref{fig:orbitharmo}. This is because the $\lambda/2r^2$ term will balance the centrifugal term $\Lambda^2/2r^2$ and the particle will not feel that centrifugal wall anymore. In terms of the parabola, this happens when the line $\scL$ intersects the parabola $\scP$ once, on the $y$-axis. This is a generalization of the well-known radial orbits \cite{Arn}, i.e., these with $\Lambda=0$. When $\lambda=0$, only $\Lambda=0$ travel on radial orbits. As their angular velocity $v_\theta=\Lambda/r$ vanishes they go in a straight line to the center. When $\lambda<0$, although the centrifugal wall is not here anymore, they still have $v_\theta\neq0$ and will thus inspiral toward the center. These orbits are all depicted as the innermost ones in Fig.~\ref{fig:orbitharmo} through Fig.~\ref{fig:orbitHenon}. \\

In the case $\lambda>0$, Eq.~\eqref{f(r)} shows that $\psi$ will be decreasing around the origin $r=0$, and therefore be associated with a repulsive force in this region, associated with a negative mass density. The particle will therefore be repelled at periapsis. Far from the origin, however, the potential is always decreasing, and the particle will be attracted at apoapsis. This situation is closer to electrostatics than to gravitational dynamics, and shows that isochrony is not unique to gravitational systems, and can be found in the motion of charged, test particles in central electrostatic potentials. This special property of $\lambda>0$ potentials will imply that $\Theta$ can take value in $[0,\pi]$, and isochrone orbits will be drastically different. \\

We now turn to the analysis of orbits in each of the four families of isochrone potentials. We stress that the general shape of the orbits can be classified only by the value of the $\lambda$ parameter. In particular, we will set $\varepsilon=0$ for each potential as it just amounts to re-scaling the potential energy, and with a good choice of units for time and space, we may always set $\omega=1$ for the harmonic family and $\mu=\beta=1$ in the Bounded, H\'enon and Kepler families. These choices do not change the general characteristics of the orbit. A dynamical system formulation of the problem (detailed in App.~\ref{app:DS}) has been integrated numerically and used to check (and found perfect agreement with) all the isochrone formulae: \eqref{GenKep1} for the radial period $T$, \eqref{KepTheta1} for the apsidal angle $\Theta$ and the parameterization \eqref{parameterizationfinal} for the shape of the orbit.

\subsubsection{Harmonic family} \label{sec:harmoclass}
According to Sec.~\ref{sec:newpara}, a potential $\psi$ in the Harmonic family is given by
\beq \label{resumepotharmo}
\psi(r) = \varepsilon + \frac{\lambda}{2r^2} + \frac{1}{8}\omega^2r^2 \, ,
\eeq
for some $\omega>0$ and $(\varepsilon,\lambda)\in\mathbb{R}^2$. Given the potential, the values of $(\xi,\Lambda)$ that yield periodic orbits are given by the inequalities \eqref{condiharmo}. With the help of Sec.~\ref{sec:newpara}, we may insert the Greek parameters in place of the Latin ones into Eq.~\eqref{GenKep1} and \eqref{KepTheta1} to find the period $T(\xi)$ and apsidal angle $\Theta(\Lambda)$ in terms of $\varepsilon,\lambda,\omega$
\beq \label{resumeharmo}
T = \frac{2\pi}{\omega} \, , \quad \text{and} \quad \Theta = \frac{\pi\Lambda}{\sqrt{\Lambda^2+\lambda}} \, .
\eeq

\begin{figure}[!htbp]
	\includegraphics[width=1.0\linewidth]{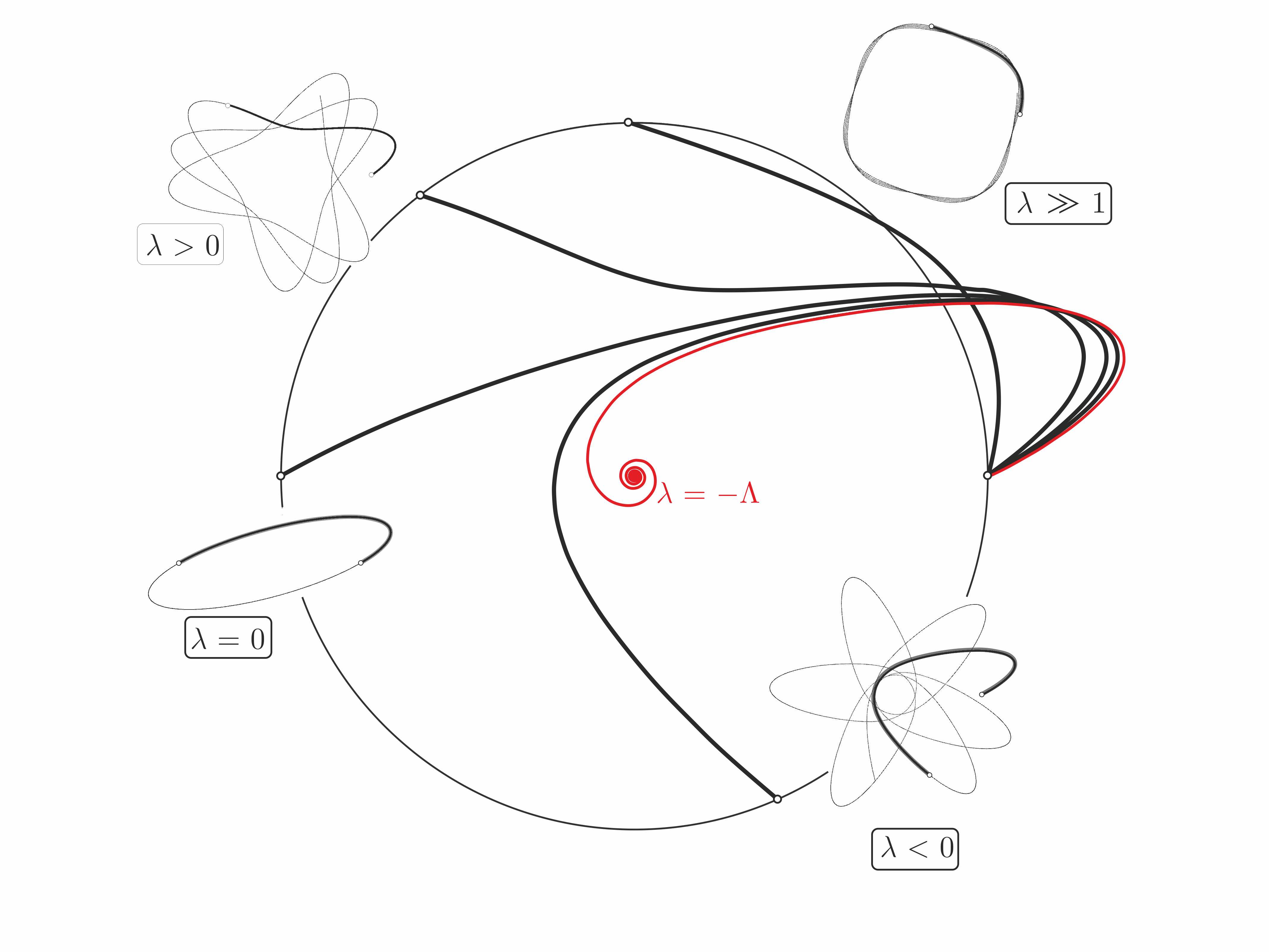}
	\caption{Five isochrone orbits, generated by the same particle $(\xi,\Lambda)$ orbiting in five different potentials of the harmonic family $\psi_1$, with varying $\lambda$. In the corners, the four orbits are depicted over several periods $T$. The curves in the middle, aside each orbit in the corner, is the highlighted, first period $[0,T]$ of each orbit, allowing for an easier comparison. Each of these orbits has the same initial position $(r,\theta)=(r_0,0)$ (black dot on the right) and same initial velocity. The innermost orbit (in red) spirals toward the origin, and the four outer ones are found at $r=r_0$ again (black dots on the dashed circle $r=r_0$) after one radial period $T$, by definition. The spiraling one corresponds to the $\lambda=-\Lambda^2$ case, discussed in the text. \label{fig:orbitharmo}}
\end{figure}

\textbullet \, In the case $\lambda=0$, we have $\Theta=\pi$ for all orbits. Up to the additive constant $\varepsilon$, $\psi_1$ is the well-known harmonic (or Hooke) potential. The dynamics can be solved analytically, and the shape of the orbit is an ellipse centered on the origin (bottom left of Fig.~\ref{fig:orbitharmo}). \\

\textbullet \, In the case $\lambda<0$, from Eq.~\eqref{resumeharmo} we have $\Theta>\pi$. When $\pi<\Theta<2\pi,$ the particle makes less than one turn in one radial period $T$. When $\Theta>2\pi$ the orbit winds up at least once around the origin, and the winding number can become arbitrarily large with $\lambda$. One such orbit is depicted on the bottom-right of Fig.~\ref{fig:orbitharmo}. \\

\textbullet \, In the case $\lambda>0$, we have $\Theta\in]0,\pi[$. In this peculiar case, the orbits need many periods in order to make a complete turn around the origin. This is because particles are repelled when reaching their periapsis. Such orbits are depicted at the top of Fig.~\ref{fig:orbitharmo}. When the orbit is close to a circular one this gives rather odd shapes, such as the top-right of Fig.~\ref{fig:orbitharmo}.
\subsubsection{Bounded family} \label{sec:Boundedclass}
According to Sec.~\ref{sec:newpara}, a potential $\psi$ in the Bounded family is given by
\beq \label{resumepotborne}
\psi(r) = \varepsilon + \frac{\lambda}{2r^2}  + \frac{\mu}{\beta + \sqrt{\beta^2 - r^2}} \, ,
\eeq
for some $\mu>0,\beta>0$ and $(\varepsilon,\lambda)\in\mathbb{R}^2$. Given the potential, the values of $(\xi,\Lambda)$ that yield periodic orbits are given by the inequalities \eqref{condiP2}. As established in Sec.~\ref{sec:T&Theta}, the period $T(\xi)$ and apsidal angle $\Theta(\Lambda)$ of the orbit are given by
\beq \label{resumeborne}
T = \frac{\pi}{\sqrt{2}}\frac{\mu}{(\varepsilon-\xi)^{3/2}} \quad \text{and} \quad \Theta = \frac{\pi\Lambda}{\sqrt{\Lambda^2+\lambda}} - \frac{\pi\Lambda}{\sqrt{\Lambda^2+\lambda+4\mu\beta}} \, .
\eeq

\begin{figure}[!htbp]
	\includegraphics[width=1.0\linewidth]{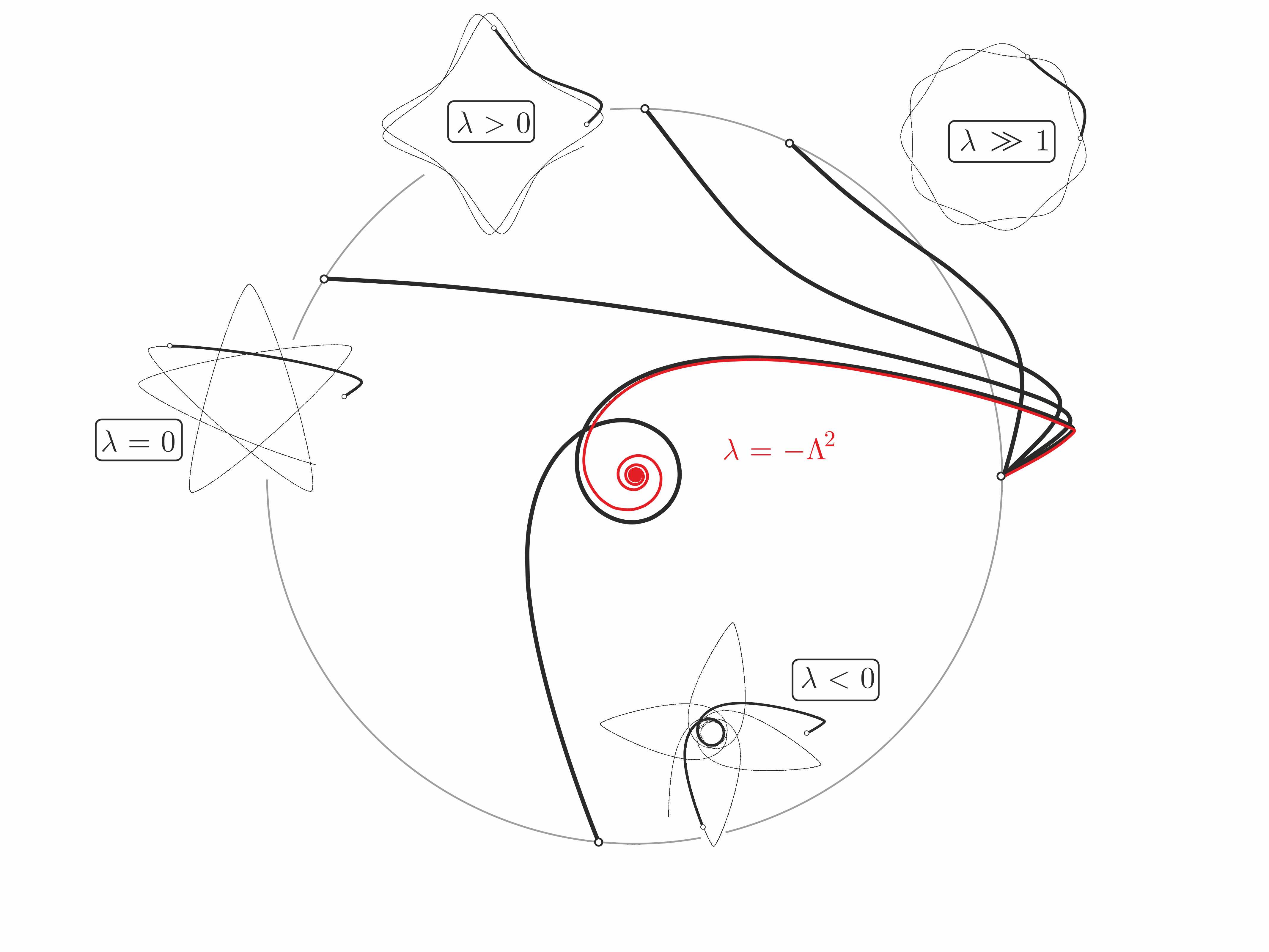}
	\caption{Some orbits in various potentials of the Bounded family $\psi_2$, with varying $\lambda$. See Fig.~\ref{fig:orbitharmo} for the explanation of the figure. \label{fig:orbitborne}}
\end{figure}

The most striking feature of orbits in Bounded potentials is the angular, almost non-differentiable, turning point at the apoapsis, as depicted in Fig.~\ref{fig:orbitborne}. In fact, these orbits are smooth and we provide some insight as to why they \textit{seem} pointy in App.~\ref{app:pointy}. Regarding the classification of orbits, it will be very similar to that of the harmonic family, by examining the function $\Theta(\Lambda)$ given by Eq.~\eqref{resumeborne}, the properties of which can be found in App.~\ref{app:ThetaofLambda}. We set $\varepsilon=0$ and $\mu=\beta=1$ in Eq.~\eqref{resumepotborne} by a good choice of units, and the shape of the orbits depends on the sign of $\lambda$. \\

\textbullet \, In the case $\lambda=0$, $\Theta$ decreases with $\Lambda$, but varies in $]0;\pi]$. The particle needs many periods to make a full rotation around the center. If the apoapsis are peaked, then this can lead to peculiar, star-shaped orbits, such as the bottom left one in Fig.~\ref{fig:orbitborne}. It is even possible to tune $\Lambda$ so that $\Theta$ is commensurable with $\pi$ in order to obtain any regular polygon whose vertices are the apsides of the orbit. \\

\textbullet \, In the case $\lambda<0$, $\Theta$ decreases with $\Lambda$ and can take arbitrary values in $]0,+\infty[$. As we said for the harmonic family, the orbit may wrap around the origin numerous times in one period, as depicted at the bottom of Fig.~\ref{fig:orbitborne}. \\

\textbullet \, In the case $\lambda>0$, $\Theta$ is not monotonous with respect to $\Lambda$. It is increasing from $0$ to some maximum value $\Theta_{\text{max}}<\pi$ when $\Lambda$ equals some critical value $\Lambda_o$, and then, it decreases to zero for $\Lambda_o$. In particular, all orbits have a maximum apsidal angle that is less than $\pi$. However, we are again in the case where the particle is repelled at periapsis, giving the orbits a different look than the $\lambda=0$ case. Two exemples are depicted at the top of Fig.~\ref{fig:orbitborne}.
\subsubsection{H\'enon family} \label{sec:Henonclass}
According to Sec.~\ref{sec:newpara}, a potential $\psi$ in the H\'enon family is given by
\beq \label{resumepotHenon}
\psi(r) = \varepsilon + \frac{\lambda}{2r^2}  - \frac{\mu}{\beta + \sqrt{\beta^2 + r^2}} \, ,
\eeq
for some $\mu>0,\beta\geq0$ and $(\varepsilon,\lambda)\in\mathbb{R}^2$. Given the potential, the values of $(\xi,\Lambda)$ that yield periodic orbits are given by the inequalities \eqref{condiP2} (with a minus sign in front of $\mu^2$). As established in Sec.~\ref{sec:T&Theta}, the period $T(\xi)$ and apsidal angle $\Theta(\Lambda)$ of the orbit are given by
\beq \label{resumeHenon}
T = \frac{\pi}{\sqrt{2}}\frac{\mu}{(\xi-\varepsilon)^{3/2}} \quad \text{and} \quad \Theta = \frac{\pi\Lambda}{\sqrt{\Lambda^2+\lambda}} + \frac{\pi\Lambda}{\sqrt{\Lambda^2+\lambda+4\mu\beta}} \, .
\eeq

Regarding the classification of orbits, we apply the same method that we used for the two other families. In particular, the variations of the function $\Theta(\Lambda)$ by Eq.~\eqref{resumeHenon} are given in App.~\ref{app:ThetaofLambda} and we set $\varepsilon=0$ and $\mu=\beta=1$ in Eq.~\eqref{resumepotborne} by a good choice of units. The shape of the orbits depends on the sign of $\lambda$. \\

\begin{figure}[!htbp]
	\includegraphics[width=1.0\linewidth]{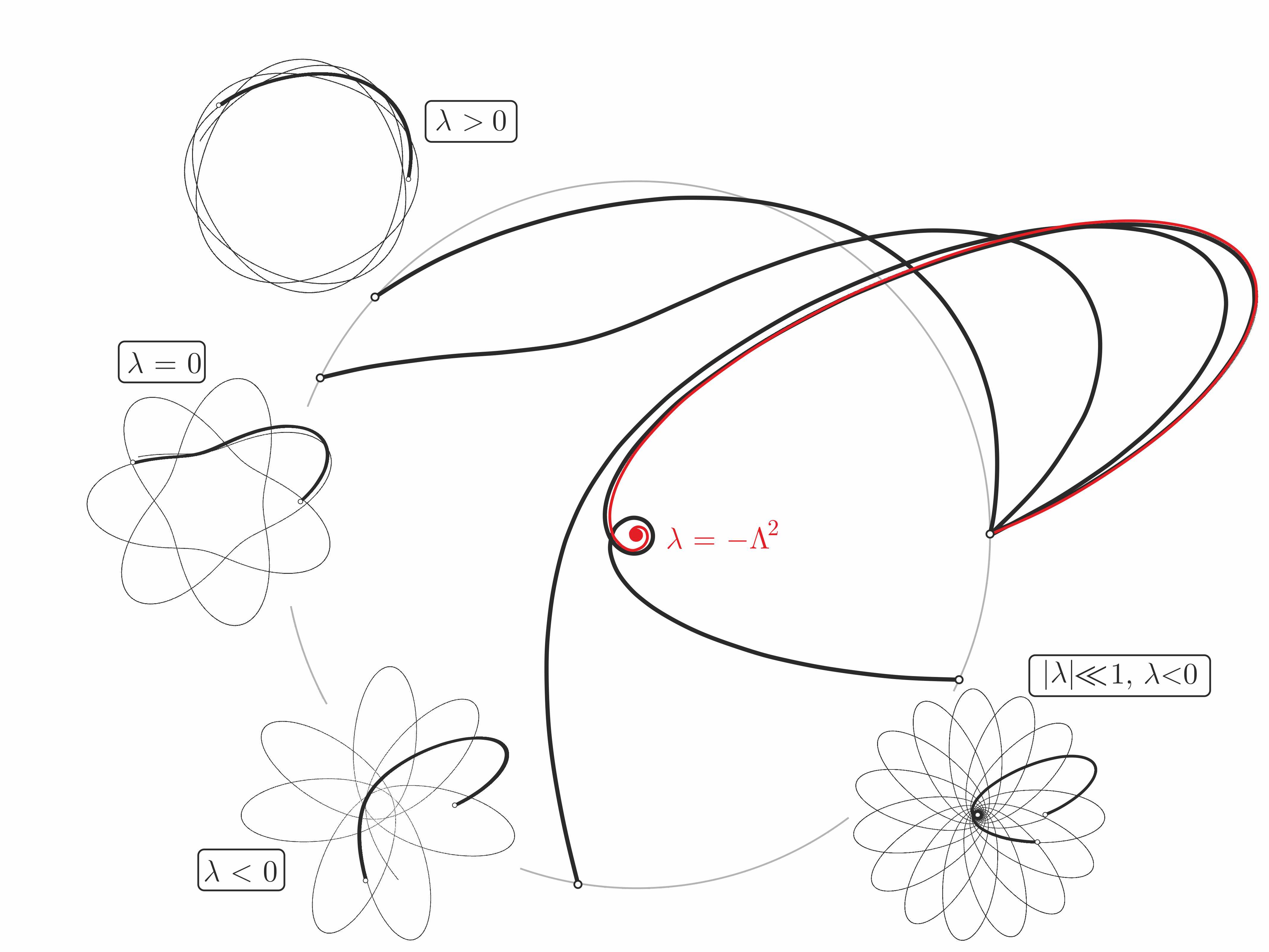}
	\caption{Some orbits in various potentials of the H\'enon family $\psi_3$, with varying $\lambda$. See Fig.~\ref{fig:orbitharmo} for the explanation of the figure. The H\'enon family contains the usual Kepler potential when $(\beta,\lambda)=(0,0)$, with closed ellipses. Consequently, for $(\beta,\lambda)\neq(0,0)$ the orbits are similar to precessing ellipses. \label{fig:orbitHenon}}
\end{figure}

\textbullet \, In the case $\lambda=0$, $\Theta$ increases with $\Lambda$ and varies in $[\pi;2\pi[$. The particle needs at least two periods to make a full rotation around the center. The Kepler potential belongs to the H\'enon family with $\lambda=\beta=0$, and has $\Theta=2\pi$, recovering the elliptic orbit. It is thus not a surprise that most orbits in the H\'enon family resemble precessing ellipses. One such orbit is depicted at the bottom of Fig.~\ref{fig:orbitHenon}. \\

\textbullet \, In the case $\lambda<0$, $\Theta$ is, in general, not monotonous with respect to $\Lambda$. The precise shape of the function $\Theta(\Lambda)$ can be found in App.~\ref{app:Theta}, but generally speaking, $\Theta$ is decreasing from $+\infty$ to some minimum value $\Theta_{\text{min}}<\pi$ when $\Lambda$ equals some critical value $\Lambda_o$, and it increases to reach $2\pi$ for $\Lambda>\Lambda_o$. For some values of $\lambda$, the critical angular momentum $\Lambda_o$ goes to $+\infty$, and $\Theta(\Lambda)$ is then strictly decreasing, varying between $2\pi$ and $+\infty$. In either case, $\Theta>\pi$ and the periapsis can be at an arbitrarily large radius, leading to an orbit with numerous windings around the center, as depicted on the bottom-right of Fig.~\ref{fig:orbitHenon}. \\

\textbullet \, In the case $\lambda>0$, the apsidal angle $\Theta$ is strictly increasing between $0$ and $2\pi$. This case is peculiar because the shape of the orbit will depend on the location of the periapsis. Indeed, note that since $\lambda>0$, the potential is always decreasing in some region surrounding the origin. If the periapsis is in this region, then the particle will be repelled, and we will have necessarily $0<\Theta<\pi$, as usual for repelled orbits. On the contrary, if the periapsis is outside the region where the potential decreases, the particle is always attracted and $\pi<\Theta<2\pi$. \\

\clearpage
\section*{Conclusion}

In this paper, we have tried to answer as thoroughly as possible the following mathematical physics question: What are the characteristics of orbits in isochrone potentials? To this end we have extended some results of \cite{SPD} and managed to fill in some gaps along the way, privileging a geometrical approach as much as possible, in the spirit of Michel H\'enon's work. We often made contact with the results of \cite{SPD}, and argued that this paper should be seen as the second part of a work toward the completion of Michel H\'enon's endeavor, namely the systematic study of isochrone potentials and orbits. \\

After some reminders about orbital dynamics in central potentials in the first section, in the second one we have presented and argued how the H\'enon variables $(x,Y)$ are well suited for the study of a test particle in an arbitrary central potential. The reason is that, in these variables, the particle is associated with a line $\scL$ whose two degrees of freedom (slope and $y$-intercept) are in a one-to-one correspondence with the properties of the particle (energy and angular momentum). We have used this H\'enon formalism to give a geometrical proof of the central result of isochrony: A potential $\psi$ is isochrone if and only if its curve $\scC$ in H\'enon's variables is an arc of parabola, using a beautiful result of Euclidean geometry tracing back to Archimedes. We have given insight as to why these special curves are \textit{parabolae} and not any other type of curve in Sec.~\ref{sec:finrem}. \\

To establish that the set of isochrone potentials is really in a one-to-one correspondence with parabolae in the H\'enon plane, we needed to find the explicit bifurcation diagrams for each potential. In particular, we built on the Keplerian picture exposed in \cite{Arn} and provided the set of particles $(\xi,\Lambda)$ that have bounded motion in any isochrone potential, as depicted in Fig.~\ref{fig:existP1} through \ref{fig:existP3}. By a careful analysis of the set of parabolae summarized in Fig.~\ref{fig:sumarpara}, we have derived a classification of isochrone potentials. We have found the one exposed in \cite{HeI.59} and \cite{SPD}, as well as the hollow potentials, with non-usual physical properties, but nonetheless bounded isochrone orbits, and therefore necessary to exhaustively complete the set of isochrone potentials. We have also provided a geometrical way of finding the mass in a given sphere, associated with a given isochrone potential. \\

After this analysis on isochrone potentials, we turned to the study of isochrone orbits per se in Sec.~\ref{sec:Kep}. We first derived a generic and analytic formula for the radial period $T$ and the apsidal angle $\Theta$ of any isochrone orbit in terms of the parabola parameters $(a,b,c,d)$ and the particles properties ($\xi$ for $T$ and $\Lambda$ for $\Theta$). We provided various geometrical interpretations of these laws and given a recipe for computing both $T$ and $\Theta$ in the H\'enon plane, as depicted in Fig.~\ref{fig:georead}. The final results involve the solution of the orbital differential equation itself. Based on all our previous findings, we devised and completed a geometrical program in the H\'enon plane that allowed to find an analytic parameterization for any isochrone orbit.\footnote{We note that this procedure could be repeated for any family of curves in the H\'enon plane that is stable under affine transformations, provided that one of them is already analytically known. This might be a fruitful and interesting academic exercise, perhaps by replacing parabolae by other conics, or algebraic curves of higher degree.} Last but not least, we have computed numerically and plotted various isochrone orbits in order to check the consistency of our analytic results in the orbital plane. We described and classified these orbits according to the isochrone potential $\psi_i$ to which they belong and to the parameter $\lambda$, that is essentially the only responsible for the different shapes of the orbits. In particular, we have seen that orbits in the Bounded potential have peculiar properties at apoapsis and understood these properties analytically. \\

Isochrony has much to offer when it comes to useful and fruitful problems of mathematical physics: It contains the two most fundamental potentials of gravitation, the Keplerian and the harmonic ones, and exhibits interesting geometrical and algebraic properties (generalized Kepler's third laws, affine transformations of parabolae in the H\'enon plane), and on top of that, everything can be found analytically, from the shape of the potentials (\eqref{pot1} through \eqref{pot4}), the radial period and apsidal angle (\eqref{GenKep2} and \eqref{KepTheta1}) to the parameterization of orbits therein (\eqref{rrbar}, \eqref{thetathetabar} and \eqref{parameterizationfinal}). Last but not least, it seems to be of physical interest in the evolution of cluster dynamics, as suggested in \cite{SPDnum} and in ongoing works. We encourage the interested reader to take inspiration out of the geometrical methods and results of this paper to devise exercises material that serve academic purposes. It appears that mathematical physics problems such as isochrony are rarely seen in the literature nowadays, and the authors are convinced that such problems are of great interest as both academic and research material.

\acknowledgments

PR is grateful to M. Langer and A. Le Tiec for helpful discussions, suggestions and comments; and to the Centro Brasileiro de Pesquisas Fis\`{ı}cas for its hospitality, where part of this work was done.
\clearpage
\appendix

\section{Dynamical system} \label{app:DS}
In order to draw the orbit, we write the equations of motion as a three-dimensional dynamical system. Although, in general, a generic three-dimensional motion in classical mechanics involves 6 degrees of freedom, namely the three coordinates and their associated momenta, the spherical symmetry here at play reduces this number to three. Moreover, the radial motion is decoupled from the polar one.
To see this, differentiate Eq.~\eqref{eomr} with respect to $r$ to obtain a second-order ODE for $r(t)$, or equivalently, a two-dimensional dynamical system for the radial motion in $(r,\dot{r})$.
To get the polar motion, and thus the full orbit $(r(t),\theta(t))$, one may simply use the definition of the angular momentum $\Lambda=r^2\dot{\theta}$, which gives $\theta(t)$ directly from $r(t)$. These three pieces together give the following three-dimensional dynamical system in $(r,\dot{r},\theta)$
\beq \label{dynsys}
\frac{\ud r}{\ud t} = \dot{r} \,, \quad \frac{\ud \dot{r}}{\ud t} = \frac{\Lambda^2}{r^3} - \psi^{\prime}(r) \quad \text{and} \quad \frac{\ud \theta}{\ud t} = \frac{\Lambda}{r^2} \, .
\eeq

The system \eqref{dynsys} is sufficient to compute the trajectory of any particle in any central potential $\psi(r)$. In particular, once $\psi(r)$ is plugged into Eqs.~\eqref{dynsys} and some initial conditions $(r(0),\dot{r}(0),\theta(0))$ are provided, the motion can be solved using, e.g., a classical Runge-Kutta numerical method. Since we are interested in periodic, bounded orbits, we must, however, choose the initial conditions carefully. In order to find these orbits more easily, we choose to express $(r(0),\dot{r}(0))$ in terms of the two constants of motion $\xi$ and $\Lambda$, and take $\theta(0)=0$, as the latter does not change the periodic nature of an orbit. Since the set of $(\xi,\Lambda)$ that produces periodic orbits is precisely the one we found in Sec.~\ref{sec:complete} depicted in Figs.~\ref{fig:existP1}, \ref{fig:existP2} and \ref{fig:existP3}, this procedure allows for an easy picking of initial conditions and allows us to draw any periodic orbit in any isochrone potential. This has been used to draw the orbits in Figs.~\ref{fig:orbitharmo} through \ref{fig:orbitHenon}, and to check the validity of all our analytic isochrone formulae.
\section{Alternative form of the third laws} \label{app:alt}
We have seen that the H\'enon's formula \eqref{HenonT} gives the period $T(\xi)$ of an orbit $(\xi,\Lambda)$ in any isochrone potential. For any value of $\xi$, there exists a unique value $\Lambda_C$ such that the orbit is circular, corresponding to the line $\scL_C:y=\xi x-\Lambda^2_C$ being tangent to the isochrone parabola. Using the notations $h$ and $L(h)$ introduced in Sec.~\ref{sec:isopara}, this circular limit corresponds to $h\rightarrow 0$. Taking this well-defined limit in Eq.~\eqref{newTh} gives
\beq \label{limTh}
T^2=\frac{\pi^2}{16(1+\xi^2)^{3/2}} \lim_{h\rightarrow 0} \frac{L(h)^2}{h}\, .
\eeq
As it can be intuited from the discussion of Sec.~\ref{sec:finrem}, it turns out that the limit on the right-hand side of Eq.~\eqref{limTh} is independent of the global aspect of the curve. In fact, this limit is simply eight times the radius of curvature $R_C$ at the point $C$ corresponding to the circular orbit.\footnote{The intuition comes from the following remark: The information on the period should be encoded somewhere on the curve, but be independent of $\Lambda$ and thus of the height of the line $\scL$. By varying $\Lambda$ we see that the only place that is not altered by this translation is the point $C$. In particular, the slope of the tangent encodes $\xi$, and the curvature at that point encodes $T$.} In other words, we have
\beq \label{KeplerCurvature}
T^2=\frac{\pi^2}{2}\frac{R_C}{(1+\xi^2)^{3/2}} \, .
\eeq
Equation \eqref{KeplerCurvature} provides a geometrical way to find the period of any given orbit in an isochrone potential, without any algebraic reference to the parabola itself. First take a line $\scL$ intersecting an isochrone parabola $\scP$ at $P$ and $A$, and then, perform a translation of this line to construct $\scL_C$, tangent to $\scP$ at $C$. The curvature radius $r_C$ of the parabola at the tangency point $C$ gives the period, via Eq.~\eqref{KeplerCurvature}. This is the local version of the result given in Eq.~\eqref{HenonT}. \\

In a similar fashion, the law for the apsidal angle can also be written in terms of curvature, albeit for the effective potential. If $\Psi_e(u)=\psi_e(r)$ with $u=1/r$, we have
\beq
\Theta^2 = \frac{4\pi^2\Lambda^2}{\Psi_e^{\prime\prime}{(u_C)}} \, .
\eeq
This law provides a way to compute the apsidal angle in the effective potential $\Psi_e$ in the Binet variable $u=1/r$, or in the real effective potential $\psi_e$, using $\Psi_e^{\prime\prime}(u_C)=r_C^4\psi_e^{\prime\prime}(r_C)$.
\section{H\'enon's formula for \texorpdfstring{$\Theta$}{Theta}} \label{app:Theta}
We detail the computation of the integral \eqref{defTheta} for $\Theta$, with the method used to derive the H\'enon formula \eqref{HenonT} for $T$. According to the dictionary in Table.~\ref{TableIso}, this time we use the Binet variable $u:=1/r$ and define a potential $\Psi_e(u)$ by $\psi_e(r)=\Psi_e(u)$. Inserting these notations in \eqref{defTheta} readily gives
\beq \label{defThetaBinet}
\Theta = \sqrt{2}\Lambda \int_{u_A}^{u_P} \frac{\ud u}{\sqrt{D_\Theta(u)}} \, , \quad \text{with} \quad D_\Theta(u):=\xi - \Psi_e(u) \, .
\eeq
This is the equivalent for $\Theta$, of Eq.~\eqref{defTHenon} for $T$. The bounds of the integral are $u_A:=1/r_A$ and $u_P=1/r_P \geq u_A$. In the $(u,y)$ plane, the quantity $D_\Theta(u):=\xi- \Psi_e(u)$ appearing in Eq.~\eqref{defThetaBinet} is the vertical distance between the curve $\scC:y=\Psi_e(u)$ and the line $\scL:y=\xi$.
Once again, the fact that $D_\Theta(u)\geq0$ follows from the requirement $\xi-\psi_e(r)\geq0$, cf. \eqref{positivity}.
\begin{figure}[!htbp]
	\includegraphics[width=0.5\linewidth]{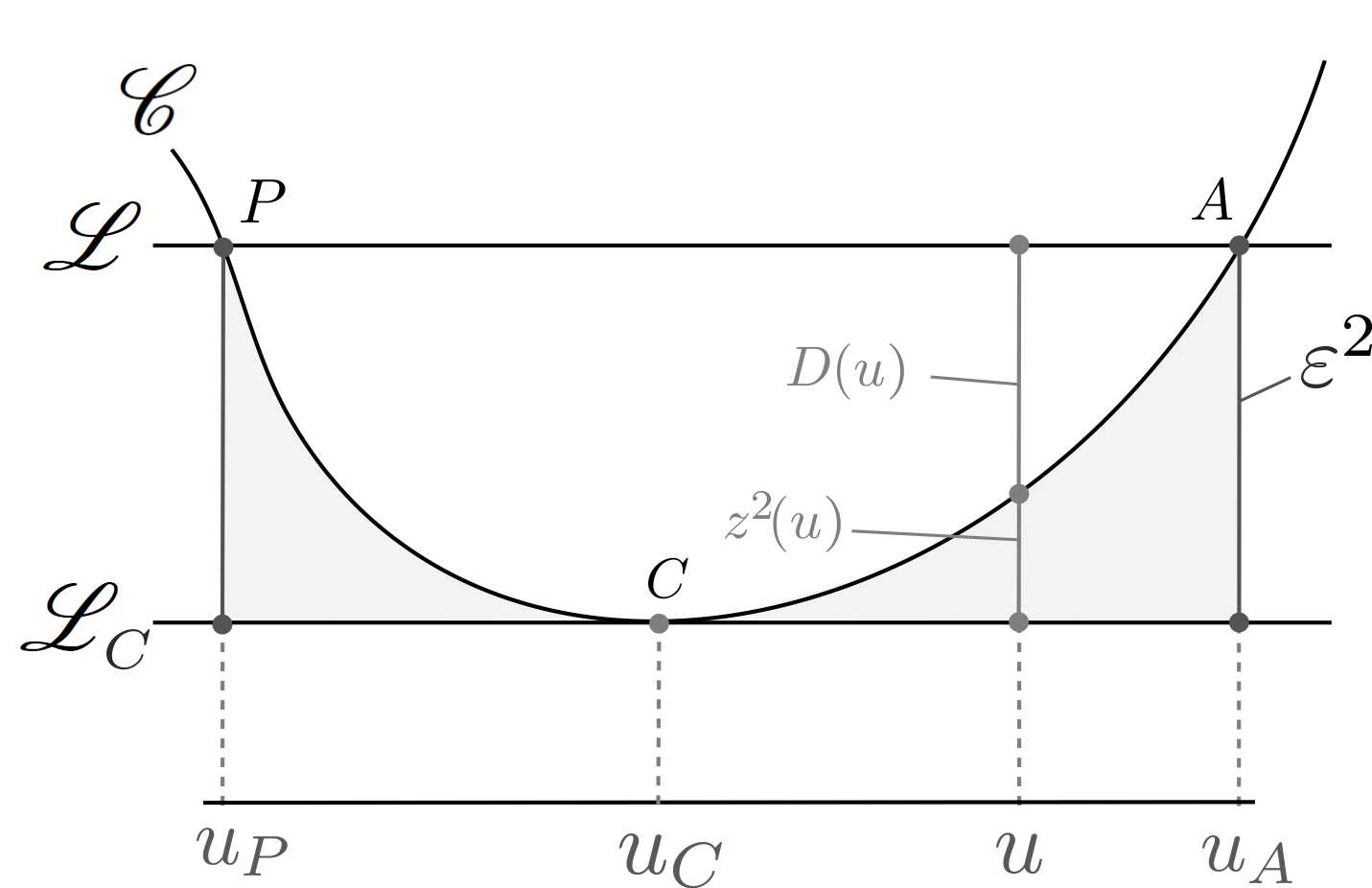}
	\caption{Illustration of the geometrical analysis involved in the computation of $\Theta$, according to the dictionary of Table.~\ref{TableIso}. $D_\Theta$ is the vertical distance between the straight line $\scL:y=\xi$ and a generic curve $\scC:y=\Psi_e(u)$. The line $\scL_C:y=\xi_C$ is the unique line both parallel to $\scL$ and tangent to $\scC$. \label{geoiso_0}}
\end{figure}
Next we rewrite the distance $D_\Theta$ as $D_\Theta(u)=\varepsilon^2 - z(u)^2$, with $\varepsilon^2 := \xi - \xi_C$ the vertical distance between the two lines $\scL$ and $\scL_C$, and $z(u)^2:=\Psi_e(u)-\xi_C$, as depicted in Fig.~\ref{geoiso_0}. As we did for $T$, we may conveniently choose $z(u)$ to be negative on $[u_A,u_C]$ and positive on $[u_C,u_P]$. The formula for $\Theta$ becomes
\beq
\Theta = \sqrt{2} \Lambda \int_{u_A}^{u_P} \frac{\ud u}{\sqrt{\varepsilon^2 - z(u)^2}} \, .
\eeq
As we argued for $T$, the function $u\mapsto z(u)$ is by construction monotonically increasing so that we can perform the change of variables $u\rightarrow z(u)$ and introduce $f$ such that $u=f(z)$. Since $z(u_A)=-\varepsilon$ and $z(u_A)=\varepsilon$, the integral becomes
\beq
\Theta = \sqrt{2} \Lambda \int_{u_A}^{u_P} \frac{ f^{\prime}(z) \ud z}{\sqrt{\varepsilon^2 - z^2}} = \sqrt{2} \Lambda \int_{-\pi/2}^{\pi/2} f^{\prime}(\varepsilon \sin \phi) \ud \phi \, ,
\eeq
where the last equality follows from the change of variables $z\rightarrow \varepsilon \sin \phi$, with $\phi$ varying between $-\pi/2$ and $\pi/2$ when $z\in[-\varepsilon,\varepsilon]$. Now we assume for $f^{\prime}$ a Taylor expansion around $0$ of the form $f(z)=a_0 + \sum_{n\geq1} a_n z^n$, and integrate term by term to get
\beq \label{Thetasum}
\Theta = \sqrt{2} \pi \Lambda a_0 + 2\sqrt{2} \Lambda \sum_{n\geq1} a_{2n} W_{2n} \varepsilon^{2n}  \, ,
\eeq
with $W_n$ the Wallis integral as given in Eq.~\eqref{Tsum}, and the odd terms vanishing by integration over the symmetric interval $[-\varepsilon,\varepsilon]$. Now if $\Theta$ is to be independent of $\xi$, then it must also be independent of $\varepsilon$, since $\xi_C$ depends only on $\Lambda$. Therefore, we must have $a_{2n}=0$ for all $n\geq1$. We thus obtain the formula $\Theta = \sqrt{2} \pi \Lambda a_0$ and the Taylor expansion of $f'$ therefore writes
\beq
f'(z) = \frac{\Theta}{\sqrt{2}\pi\Lambda} + \sum_{n \geq 1} a_{2n+1} z^{2n+1} \, .
\eeq
Integrating this equation over the interval $[z(u_A),z(u_P)]=[-\varepsilon,\varepsilon]$, we can make the same remarks as we did in the paragraph below Eq.~\eqref{fprime}, except that in this case $f(z_A)=u_A$ and $f(z_P)=u_P$. In the end, we obtain Eq.~\eqref{HenonTheta}, which is the equivalent to Eq.~\eqref{HenonT} for $T$. The right-hand side of that equation is independent of $\xi$, even though the quantities $u_P,u_A$ depend explicitly on $\xi$.

\section{Analysis of \texorpdfstring{$\Theta(\Lambda)$}{Theta(Lambda)} } \label{app:ThetaofLambda}
Let $\psi_i$ be a potential in one of the four families $i\in\{1,2,3,4\}$ as defined in Sec.~\ref{sec:isopara}. In this appendix, we study the properties of the function $\Theta_i(\Lambda)$ defined in Eq.~\eqref{KepTheta1}. These formulae are used in Sec.~\ref{sec:classorbits} to classify the orbits in each isochrone potential $\psi_i$. The claims of Sec.~\ref{sec:classorbits} regarding each function $\Lambda\mapsto\Theta_i(\Lambda)$ follow easily from the mathematical analysis detailed here, with the Latin parameters $(a,b,c,d,e)$ replaced by the Greek ones $(\omega,\varepsilon,\lambda,\mu,\beta)$.
\subsubsection{Harmonic and Kepler family, $\Theta_1(\Lambda)$ and $\Theta_4(\Lambda)$}
For the Harmonic potentials $\psi_1$, the analysis of $\Theta_1(\Lambda)=\pi\Lambda(\Lambda^2+\lambda)^{-1/2}$ is straightforward. For $\lambda<0$, it is strictly decreasing and $\Theta$ varies in $]\pi,+\infty[$ when $\Lambda\in ]\sqrt{-\lambda},+\infty[$. For $\lambda=0$, $\Theta=\pi$ for all $\Lambda\in\mathbb{R}$. For $\lambda>0$, it is strictly increasing and $\Theta$ varies in $]0,\pi[$ when $\Lambda\in ]0,+\infty[$. For the Kepler family $\psi_4$ the analysis is also straightforward since we have for any $\Lambda$ and $\lambda$ the identity $\Theta_4(\Lambda)=2\Theta_1(\Lambda)$ by direct examination of Eq.~\eqref{KepTheta1} when $b=0$ (harmonic) and $d^2=4b^2e$ (Kepler).
\subsubsection{Bounded family, $\Theta_2(\Lambda)$}
For Bounded potentials $\psi_2$, the analysis is more involved. For any $\mu>0$ and $\beta>0$ we write $\alpha:=\lambda/(\lambda+4\mu\beta)$. We also define a function $f(\Lambda,\lambda)$ of the real variables $\Lambda,\lambda$ by the formula
\beq \label{lambda_2}
f(\Lambda,\lambda):=\frac{\Lambda}{(\Lambda^2+\lambda)^{1/2}} - \frac{\Lambda}{(\Lambda^2+\lambda+4\mu\beta)^{1/2}} \, .
\eeq
With these notations, we have $\Theta_2(\Lambda)=\pi f(\Lambda,\lambda)$ (cf. Eq.~\eqref{resumeborne}). We want to study the three cases $\lambda>0$, $\lambda=0$ and $\lambda<0$, used to classify the orbits in Sec.~\ref{sec:classorbits}. \\

\textbullet \, Case $\lambda=0$. In this case, we simply plug $\lambda=0$ in Eq.~\eqref{lambda_2} and we see that $f(\Lambda,0)\in[0,1]$. Moreover, we have easily $\partial_\Lambda f<0$. Therefore, $\Theta(\Lambda)$ is strictly decreasing and varies $[0,\pi]$. \\

\textbullet \, Case $\lambda>0$. In this case, $0<\alpha<1$ and for a fixed $\lambda$, we have $\partial_\Lambda f(\Lambda,\lambda) = \lambda(\Lambda^2+\lambda)^{-3/2} - (\lambda+4\mu\beta)(\Lambda^2+\lambda+4\mu\beta)^{-3/2}$. Then, a few algebraic manipulation show that $\partial_\Lambda f(\Lambda,\lambda)$ vanishes for a value $\Lambda_o$ given by
\beq \label{Lambdao}
\Lambda_o^2 = \lambda \frac{\alpha^{1/3}-1}{\alpha-\alpha^{1/3}} \quad \Rightarrow \quad f(\Lambda_o,\lambda)=\frac{\Lambda_o(1-\alpha^{1/3})}{(\Lambda_o^2+\lambda)^{1/2}} \, .
\eeq
Since $0<\alpha<1$ and $0<\Lambda_o<(\Lambda_o+4\mu\beta)^{1/2}$, we have readily $0<f(\Lambda_o,\lambda)<1$. Now, for any fixed $\lambda>0$, $\Lambda\mapsto f(\Lambda,\lambda)$ is continuous, $\partial_\Lambda f$ vanishes only once at $\Lambda_o$ and furthermore $0<f(\Lambda_o,\lambda)<1$. Furthermore, it is clear that $f(\Lambda,\lambda)$ goes to $0$ as $\Lambda\rightarrow 0$ and $\Lambda\rightarrow +\infty$. With all these results, the general shape of the curve $\Lambda \mapsto f(\Lambda,\lambda)$ can be easily inferred. \\

\textbullet \, Case $\lambda<0$. In this case, $\Lambda\mapsto f(\Lambda;\lambda)$ is defined only when $\Lambda^2>-\lambda$. First subcase: $\lambda<0$ and $\lambda+4\mu\beta<0$. Then, this is the same as in the $\lambda>0$ case, where we saw that $\partial_\Lambda f(\Lambda,\lambda)>0$. Second subcase: $\lambda<0$ but $\lambda+4\mu\beta\geq0$, then setting $g(\Lambda,\lambda) := \lambda(\Lambda^2+\lambda)^{-3/2}$, we have for any $\Lambda^2>-\lambda$
\beq \label{gprimeTheta}
\frac{\partial g}{\partial \lambda} (\Lambda,\lambda) = \frac{2\Lambda^2-\lambda}{2(\Lambda^2+\lambda)^{5/2}} \, .
\eeq
Now, since $\Lambda^2>-\lambda$, the right-hand side of Eq.~\eqref{gprimeTheta} is strictly positive, and therefore $g$ is an increasing function of $\lambda$. In particular, we have $\lambda+4\mu\beta>\lambda \Rightarrow g(\Lambda,\lambda+4\mu\beta) > g(\Lambda,\lambda)$ and by definition of $g$, the latter is exactly $\partial_\Lambda f(\Lambda,\lambda)>0$. To conclude, in the $\lambda<0$ case, $\Lambda\mapsto f(\Lambda,\lambda)$ is strictly decreasing. Furthermore, it is clear that $f(\Lambda,\lambda)$ goes to $+\infty$ as $\Lambda\rightarrow (-\lambda)^{1/2}$, and to $0$ as $\Lambda\rightarrow +\infty$. With all these results, the general shape of the curve $\Lambda \mapsto f(\Lambda,\lambda)$ can be easily inferred.
\subsubsection{H\'enon family, $\Theta_3(\Lambda)$}
For H\'enon potentials $\psi_3$, the analysis is similarly more involved. As for the Bounded potentials we fix $\mu>0$ and $\beta>0$ and write $\alpha:=\lambda/(\lambda+4\mu\beta)$. This time we define a function $h(\Lambda,\lambda)$ of the real variables $\Lambda,\lambda$ by the formula
\beq \label{lambda_3}
h(\Lambda,\lambda):=\frac{\Lambda}{(\Lambda^2+\lambda)^{1/2}} + \frac{\Lambda}{(\Lambda^2+\lambda+4\mu\beta)^{1/2}} \, .
\eeq
With these notations, we have $\Theta_3(\Lambda)=\pi h(\Lambda,\lambda-2\mu\beta)$ (cf. Eq.~\eqref{resumeHenon}). The analysis follows the same lines as what was done for $\Theta_2(\Lambda)$. We want to study the three cases $\lambda>0$, $\lambda=0$ and $\lambda<0$, used to classify the orbits in Sec.~\ref{sec:classorbits}. \\

\textbullet \, Case $\lambda>0$. Then we have $0<\alpha<1$ and there is no problem in showing that $\Lambda\mapsto h(\Lambda,\lambda)$ is strictly increasing and that $0<h(\Lambda,\lambda)<1$. \\

\textbullet \, Case $\lambda=0$. Once again, there is no problem in showing that $\Lambda\mapsto h(\Lambda,\lambda)$ is strictly increasing and that $0<h(\Lambda,\lambda)<2$. \\

\textbullet \, Case $\lambda<0$. In this case, $h$ is only defined when $\Lambda^2>-\lambda$. For any such $(\Lambda,\lambda)$, we have
\beq \label{hprimeTheta}
\frac{\partial h}{\partial \Lambda} (\Lambda,\lambda) = \frac{\lambda}{(\Lambda^2+\lambda)^{3/2}} + \frac{\lambda+4\mu\beta}{(\Lambda^2+\lambda+4\mu\beta)^{3/2}} \, .
\eeq
There are two subcases. First subcase: $\lambda<0$ and $\lambda+4\mu\beta<0$. Then from Eq.~\eqref{hprimeTheta}, $\partial_\Lambda h(\Lambda,\lambda)<0$. Furthermore, $h(\Lambda,\lambda)$ goes to $+\infty$ as $\Lambda\rightarrow (-\lambda)^{1/2}$, and to $2$ as $\Lambda\rightarrow +\infty$. Second subcase: $\lambda<0$ and $\lambda+4\mu\beta<0$. If $|\alpha|<1$, then there is a value $\Lambda_o$ that makes $\partial_\Lambda h(\Lambda,\lambda)$ vanish. It is given by
\beq \label{Lambdao_2}
\Lambda_o^2 = \lambda \frac{|\alpha|^{1/3}+1}{|\alpha|-|\alpha|^{1/3}} \quad \Rightarrow \quad f(\Lambda_o,\lambda)=\frac{\Lambda_o(1+|\alpha|^{1/3})}{(\Lambda_o^2+\lambda)^{1/2}} \, .
\eeq
In this case, the function $\Lambda\mapsto h(\Lambda,\lambda)$ decreases on $[(-\lambda)^{1/2},\Lambda_o]$ and increases on $[\Lambda_o,+\infty[$. The value $h(\Lambda_o,\lambda)$ is always strictly between $1$ and $2$. If $|\alpha|\geq1$, then the function $f$ is strictly decreasing. (It can be seen as the limit $\Lambda_o\rightarrow+\infty$.) The value $h(\Lambda_o,\lambda)$ is in this case always above $2$.

\section{Proof that \texorpdfstring{$c+d\xi<0$}{c+dxi} for isochrone orbits around finite central mass} \label{app:c+dxi}

In Sec.~\ref{sec:param} we used the fact that $a+b\xi<0$ and $c+d\xi<0$ for isochrone orbits in order to prove that our formula \eqref{parameterizationfinal} covers all isochrone orbits. The former identity follows from the generalized Kepler's third law, and here, we prove the latter identity. By assumption, we have a particle $(\xi,\Lambda)$ on an isochrone orbit in a potential with finite central mass whose parabola $\scP:(ax+by)^2+cx+dy=0$ verifies all hypotheses $(H_i)$ of Sec.~\ref{sec:portion}. First we can check easily that $cx+dy=0$ is an equation for the tangent to $\scP$ at the origin. geometrically, since two intersections exist between $\scL$ and $\scP$, the slope of $\scL$ must be bigger than that of this tangent, i.e., we must have $\xi>-c/d$. We just have to show that $d\leq0$ and the result will follow. First, if $b=0$ (harmonic case), then we necessarily have $d<0$ (top-oriented parabola). Second, if $b\neq0$, then since $\lambda=0$ ($\scP$ crosses the origin) we have by Eq.~\eqref{lambda} the equality $-d=\sqrt{d^2}$, which implies $d\leq0$. Therefore, we always have $d\leq0$ and thus $c+d\xi<0$.

\section{Peaks of orbits in Bounded potentials} \label{app:pointy}

Let an arbitrary orbit be given by a polar equation $r(\theta)$, and compute the value of $|\ud r/\ud \theta|$. The latter is a measure of the change of $\ud r$ when moving from $\theta$ to $\theta+\ud\theta$. It vanishes for circles $r=\text{cst}$ and is \textit{infinite} for straight lines $\theta=\text{cst}$. With the help of Eq.~\eqref{eomr} and $\Lambda=r^2\dot{\theta}$, we obtain easily $|\ud r/ \ud \theta|^2 = 2r^4(\xi-\psi_e(r))/ \Lambda^2$.
%
%
Using a Taylor expansion of $\psi(r)$ and Eq.~\eqref{rarp}, we can linearize this equation around the apoapsis $r_A$. We then obtain
\beq \label{drdtheta_lin}
\biggl|\frac{\ud r}{\ud \theta}\biggr|^2 = \frac{2r_A^4}{\Lambda^2}  \biggl( \psi^\prime(r_A)-\frac{\Lambda^2}{r_A^3} \biggr) (r_A-r) + o(r_A-r) \, .
\eeq

Examining Eq.~\eqref{drdtheta_lin}, we see that as $r\rightarrow r_A$ the right-hand side goes to zero as every term is finite in front of $(r_A-r)$. The orbit is therefore smooth and differentiable around the apoapsis. However, the quantity $\psi^\prime(r_A)$ turns out to be very large for the Bounded family, in general. This is because the slope of a Bounded potential increases to infinity as $r$ grows toward $\beta$ from below, as can be seen readily on Eq.~\eqref{resumepotborne}. Therefore, a line $\scL$ can intersect $\scC$ such that $r_A$ is very close to $\beta$, and it is clear from Eq.~\eqref{resumepotborne} that $\psi_2^\prime(r) \rightarrow \infty$ as $r \rightarrow \beta$. As a conclusion, before the apoapsis, the term $(r_A-r)$ does not yet compensate the $\psi^\prime(r_A)$ which is large for the Bounded potential, making $\frac{\ud r}{\ud \theta}$ large and the curve resembles a $\theta=\text{cst}$ line. This is why we see such abrupt and pointy turns in Fig.~\ref{fig:orbitborne}.

\bibliography{main.bib}

\end{document}